\documentclass[]{aa}

\usepackage{natbib,amssymb,color}
\usepackage[breaklinks=true]{hyperref}
\usepackage{amsmath}

\usepackage{graphicx}
\usepackage{txfonts}
\newcommand{\clustera}{\object{RCS2\thinspace$J$232727.7$-$020437}}
\newcommand{\clusteras}{RCS2\thinspace$J$2327}
\newcommand{\ha}[1]{#1}
\newcommand{\hb}[1]{#1}
\newcommand{\reva}[1]{#1}
\newcommand{\revb}[1]{#1}
\newcommand{\revc}[1]{#1}
\newcommand{\revd}[1]{#1}

\def\Bonn{1}
\def\GeminiS{2}
\def\Sarclay{3}
\def\ParisDiderot{4}
\def\Leiden{5}
\def\KICPChicago{6}
\def\Davis{7}
\def\JPL{8}
\def\Ohio{9}
\def\ChicagoAstro{10}
\def\Stonybrook{11}
\def\Tufts{12}
\def\Toronto{13}
\def\Michigan{14}

\def\postagesize{3.0cm}
\begin{document}

\defcitealias{schrabback18}{S18}

\title{Precise  weak lensing constraints from  deep
high-resolution
$K_\mathrm{s}$
    images:
VLT/HAWK-I analysis of the  super-massive  galaxy cluster  \clustera \, at \mbox{$z=0.70$}\thanks{Based on observations conducted with the
ESO Very Large Telescope,  the Large Binocular Telescope, and the NASA/ESA
{\it Hubble} Space Telescope, as detailed in the acknowledgements.}}

\titlerunning{HAWK-I $K_\mathrm{s}$ weak lensing analysis of \clustera.}

   \author{Tim Schrabback\inst{\Bonn}
        \and
          Mischa Schirmer\inst{\GeminiS}
          \and
          Remco~F.~J.~van der Burg\inst{\Sarclay,\ParisDiderot}
          \and
          Henk Hoekstra\inst{\Leiden}
          \and
          Axel Buddendiek\inst{\Bonn}
          \and
          Douglas Applegate\inst{\Bonn,\KICPChicago}
          \and
Maru{\v s}a Brada{\v c}\inst{\Davis}
       \and
       Tim Eifler\inst{\JPL,\Ohio}
      \and
      Thomas Erben\inst{\Bonn}
      \and
      Michael D.~Gladders\inst{\ChicagoAstro,\KICPChicago}
          \and
        Beatriz Hern\'andez-Mart\'in\inst{\Bonn}
          \and
          Hendrik Hildebrandt\inst{\Bonn}
          \and
          Austin Hoag\inst{\Davis}
            \and
            Dominik Klaes\inst{\Bonn}
            \and
            Anja von der Linden\inst{\Stonybrook}
              \and
       Danilo Marchesini\inst{\Tufts}
           \and
        Adam Muzzin\inst{\Toronto}
         \and
        Keren Sharon\inst{\Michigan}
        \and
        Mauro Stefanon\inst{\Leiden}
          }
   \institute{Argelander-Institut f\"{u}r Astronomie, Universit\"{a}t Bonn, Auf dem
     H\"{u}gel 71, 53121, Bonn, Germany\\
               \email{schrabba@astro.uni-bonn.de}
             \and
             Gemini Observatory, Southern Operations Center, Casilla 603, La
             Serena, Chile
            \and
IRFU, CEA, Universit\'e Paris-Saclay, F-91191 Gif-sur-Yvette, France
           \and
 Universit\'e Paris Diderot, AIM, Sorbonne Paris Cit\'e, CEA, CNRS, F-91191 Gif-sur-Yvette, France
\and
              Leiden Observatory, Leiden University, Niels Bohrweg
2, NL-2300 CA Leiden, The Netherlands
             \and Kavli Institute for Cosmological Physics, University of
             Chicago, 5640 South Ellis Avenue, Chicago, IL 60637
              \and Department of Physics, University of California, Davis, CA
              95616, USA
              \and
              Jet Propulsion Laboratory, California Institute of Technology, 4800 Oak Grove Dr., Pasadena, CA 91109, USA
              \and
              Center for Cosmology and Astro-Particle Physics, The Ohio
              State University, 191 W. Woodruff Ave, Columbus, 43210 OH,
              USA
              \and
              Department  of  Astronomy  and  Astrophysics,  University  of
              Chicago, 5640 South Ellis Avenue, Chicago, IL 60637, USA
              \and
              Department of Physics and Astronomy, Stony Brook University,
              Stony Brook, NY 11794, USA
              \and
              Department of Physics and Astronomy, Tufts University, 574 Boston Avenue, Medford, MA 02155, USA
              \and
              Department of Physics and Astronomy, York University, 4700 Keele St., Toronto, Ontario, Canada, MJ3 1P3
              \and
              Department of Astronomy, University of Michigan, 1085 S.
University Ave, Ann Arbor, MI 48109, USA
             }

   \date{Received August 07, 2017; accepted October 27, 2017}

   \abstract{
We demonstrate that deep good-seeing VLT/HAWK-I  $K_\mathrm{s}$ images
complemented with $g$+$z$-band photometry can yield a  \revb{sensitivity for} weak
lensing studies of massive galaxy clusters at redshifts \mbox{$0.7\lesssim z
  \lesssim 1.1,$} which is almost identical to the sensitivity of HST/ACS mosaics of single-orbit depth.
Key reasons for this good performance are the excellent image quality frequently
achievable for $K_\mathrm{s}$ imaging from the ground, a highly effective photometric selection of background galaxies,
and a galaxy ellipticity dispersion that is noticeably lower than
for optically observed high-redshift galaxy samples.
Incorporating results from \revc{the} 3D-HST and UltraVISTA \revc{surveys} we also obtained  a more
accurate calibration of the source redshift distribution than previously
achieved for similar optical weak lensing data sets.
Here we studied the extremely massive galaxy cluster \clustera\,
(\mbox{$z=0.699$}), combining deep VLT/\mbox{HAWK-I} $K_\mathrm{s}$ images
\revc{(point spread function with a 0\farcs35 full width at half maximum)}
with
  LBT/LBC photometry.
The resulting weak lensing mass reconstruction suggests that the cluster
consists of a single overdensity, which is detected with a peak significance of
$10.1\sigma$.
We constrained the cluster mass to
\mbox{$M_\mathrm{200c}/(10^{15} \mathrm{M}_\odot) =2.06^{+0.28}_{-0.26}(\mathrm{stat.})\pm 0.12 (\mathrm{sys.})$}
assuming a spherical
\revc{Navarro, Frenk \& White}
model and simulation-based priors on the concentration,
making it one of the most massive galaxy clusters known in the
\mbox{$z\gtrsim 0.7$} Universe.
We also cross-checked the HAWK-I measurements through an analysis of
overlapping HST/ACS images, yielding fully consistent estimates of the
lensing signal.
}


   \maketitle

   \section{Introduction}

Light bundles from distant galaxies are distorted by the tidal
gravitational field
of foreground structures.
   These weak
   lensing distortions can be constrained statistically from the observed
   shapes of background galaxies, providing information about the differential projected
   mass distribution of the foreground objects,
free of assumptions
\ha{about their dynamical state}
 \citep[e.g.][]{bartelmann01}.
   To conduct such measurements, sufficiently unbiased estimates of galaxy
   shapes have to be obtained, corrected for the impact of the image
   point spread function (PSF).
This is only possible if the observed galaxy images are sufficiently
resolved, as the
\hb{blurring}
 PSF otherwise erases the shape information.
  Weak lensing observations therefore benefit from good image quality, which
  boosts the number density of sufficiently resolved galaxies and thus
  the
 signal-to-noise ratio, while simultaneously reducing the
  required level of PSF corrections and therefore systematic uncertainties
  \citep[e.g.][]{massey13}.

For studies targeting
more distant
lenses
it is
vital to employ deep observations  with superb image quality
to measure the shapes of  the typically faint and small distant
background galaxies carrying the signal.
In red optical filters, queue-scheduled ground-based
observations from the best sites
achieve
a
\ha{stellar}
PSF  full width at half maximum
($\mathrm{FWHM}^*$)
 \mbox{$\simeq 0\farcs6$--$0\farcs7$}
in good conditions
 \citep[e.g.][]{kuijken15,mandelbaum18},
which provides a
\ha{good}
weak lensing sensitivity out to lens redshifts \ha{
\mbox{$z\sim
  0.6$}}
in the case of deep integrations.
Much higher resolution (\mbox{$\mathrm{FWHM}^*\simeq 0\farcs10$}) can be achieved with the {\it Hubble} Space
Telescope (HST),
which has been used
 to probe the weak lensing signatures out
to significantly higher redshifts when targeting
 galaxies \citep[][]{leauthaud12}, galaxy clusters
 \citep[e.g.][\citetalias{schrabback18} henceforth]{jee11,schrabback18}, or the statistical properties of the
large-scale structure itself \citep{massey07b,schrabback10}.
However, HST has a relatively small field of view
of \mbox{$3\farcm3\times 3\farcm3$} for its ACS/WFC detector,
raising the need for
\hb{time-consuming}
mosaics
\ha{in order to cover} a wider area on
the sky.
In particular, studies that aim to
obtain accurate weak lensing mass measurements  for massive galaxy clusters at moderately high redshifts (\mbox{$0.7\lesssim
  z\lesssim 1.1$})
 have so far required mosaic ACS images  to probe the  lensing
signal out to approximately the cluster virial radius
\citep[e.g.~\citetalias{schrabback18};][]{jee09,thoelken18}.

In this paper we
\ha{demonstrate that}
deep ground-based
imaging obtained in the HAWK-I $K_\mathrm{s}$ filter (\mbox{$1.98\mu\mathrm{m}\lesssim \lambda \lesssim
  2.30\mu\mathrm{m}$})
under good seeing conditions can provide a
viable alternative to mosaic HST observations for moderately deep weak lensing
measurements.
\hb{The observational set-up}
\ha{we describe provides several advantages for weak
  lensing studies. First, for} an 8 m class telescope and typical conditions,
the measured
atmospheric
PSF FWHM
is reduced by \mbox{$\simeq 40\%$} at
such long wavelengths   compared to the $V$ band \citep{martinez10}.
As a result, delivered image qualities of \mbox{$\mathrm{FWHM}^*\simeq
  0\farcs3$--$0\farcs4$}
are  achieved in  $K_\mathrm{s}$ in good conditions without having to request the very best seeing quantile.
While not quite reaching an HST-like resolution, this still  provides a major
advantage for weak lensing measurements compared to optical seeing-limited
observations.
\ha{The second advantage is the efficiency of selecting distant background
sources in $K$ (or $K_\mathrm{s}$)-detected galaxy samples,} using the ``$BzK$ selection’’
technique \citep{daddi04} with
observations taken in
only three bands.
\ha{As a third advantage, excellent \revd{deep} reference samples selected in the  near-infrared (NIR)\ have
recently become available to infer the redshift distribution of the weak
lensing source galaxies, including photometric redshifts from UltraVISTA
(\citealt{mccracken12,muzzin13}; Muzzin et al.~in prep.)
and
HST slitless spectroscopy from the 3D-HST programme \citep{momcheva16}.
Finally, at \mbox{$z\sim 2$} $K_\mathrm{s}$ imaging probes the light
distribution of the smoother
stellar component exhibiting lower shape noise,
an advantage over optical imaging, which mostly maps}
the clumpy distribution of star forming regions seen at rest-frame UV wavelengths.

In this study we  analyse new deep VLT/HAWK-I  $K_\mathrm{s}$ observations
of the galaxy cluster \clustera\, \citep[hereafter: \clusteras; \mbox{$z=0.699$,}][]{sharon15} discovered in the Second Red-Sequence
Cluster Survey \citep[RCS2;][]{gilbank11}.
Optical, Sunyaev-Zel'dovich, X-ray, dynamical, strong lensing, and initial weak lensing measurements of the cluster
are consistent with an extremely high mass of
 \mbox{$M_\mathrm{200c}\simeq 2$--$3\times 10^{15}\mathrm{M}_\odot$}
 \citep[][]{menanteau13,sharon15,buddendiek15,hoag15},
where $M_{\Delta\mathrm{c}}$ indicates the mass within the sphere
containing  an average density that exceeds the
critical density of the Universe at the cluster
redshift by a factor $\Delta$.
Hence, this is one of the most massive clusters known
at a comparable or higher redshift.

\citet{king02} presented the first and previously only weak
lensing analysis based on shape measurements in $K_\mathrm{s}$ images.
Their analysis targeting a massive low-redshift cluster  is based
on  imaging obtained with SofI on the 3.6 m
ESO-NTT with an image resolution of 0\farcs73.
\ha{Our analysis
exploits much  deeper $K_\mathrm{s}$ imaging with a resolution that is
better by a factor two, as needed for high-redshift weak lensing
constraints.
}
\ha{We explicitly compare the weak lensing performance
  achieved
with these new  $K_\mathrm{s}$ data to the weak lensing analysis of galaxy clusters at
similar redshift from \citetalias{schrabback18}.
These authors employed \mbox{$2\times
  2$} HST/ACS mosaics of single-orbit depth taken in the F606W  filter for
shape measurements, and a photometric source selection based on
\mbox{$V_{606}-I_{814}$} colour to remove cluster galaxies and
preferentially select distant background galaxies.
}

This paper is organised as follows:
We summarise relevant weak lensing theory and notation in Sect.\thinspace\ref{se:theory}.
 Sect.\thinspace\ref{se:data} describes the analysed data sets and data
 reduction. Sect.\thinspace\ref{se:ana} provides details on the  shape and
 colour measurements,
the background selection, an estimation of the source redshift
distribution, an analysis of the galaxy ellipticity dispersion, and a comparison to shear estimates from HST measurements.
Sect.\thinspace\ref{se:results} presents the
cluster mass reconstruction, the derived cluster mass constraints, and the comparison to previous studies of the cluster.
We compare the weak lensing performance of
the HAWK-I data and
\ha{the previously employed} ACS mosaics in
Sect.\thinspace\ref{se:dis:perf}
and conclude in Sect.\thinspace\ref{se:conclusions}.

     Throughout this paper we assume a  flat $\Lambda$CDM cosmology characterised through
     \mbox{$\Omega_\mathrm{m}=0.3$},
    \mbox{$\Omega_\mathrm{\Lambda}=0.7$},
    \mbox{$H_{0}=70 \, h_{70}\, \mathrm{km\,s^{-1}}$} and
    \mbox{$h_{70}=1$}, as approximately consistent with recent
    constraints from the cosmic microwave background \citep[e.g.][]{hinshaw13,planck15cosmo},
\reva{unless explicitly stated otherwise}.
    At the cluster redshift of \mbox{$z=0.699$},
    $1^{\prime\prime}$ on the sky corresponds to
a physical separation of 7.141 kpc in this cosmology.
All magnitudes are in the AB system.

\section{Summary of relevant weak lensing theory}
\label{se:theory}

In the weak lensing regime, the gravitational lensing effect of a lens at
 redshift $z_\mathrm{l}$ (assumed to be fixed here) onto the shape of
a background galaxy
at redshift $z_\mathrm{s}$ and an observed position
$\vec{\theta}$
can be described through the  anisotropic reduced shear
\begin{equation}
g(\vec{\theta},z_\mathrm{s})=\frac{\gamma(\vec{\theta},z_\mathrm{s})}{1-\kappa(\vec{\theta},z_\mathrm{s})}\,,
\end{equation}
which is a rescaled version of the unobservable shear $\gamma(\vec{\theta},z_\mathrm{s})$,
and
the isotropic convergence
\begin{equation}
\kappa(\vec{\theta},z_\mathrm{s})=\Sigma(\vec{\theta})/\Sigma_\mathrm{crit}(z_\mathrm{l},z_\mathrm{s})
\end{equation}
\hb{(see e.g.~\citealt{bartelmann01} for a general review and
  \citealt{hoekstra13} for applications to clusters)}.
The latter is defined as the ratio of the
surface mass density
$\Sigma(\vec{\theta})$
and the
critical surface mass
density
\begin{equation}
  \label{eqn:sigmacrit}
  \Sigma_{\mathrm{crit}}(z_\mathrm{l},z_\mathrm{s}) = \frac{c^2}{4\pi G}\frac{1}{D_{\mathrm{l}}(z_\mathrm{l}) \beta(z_\mathrm{l},z_\mathrm{s}) },
\end{equation}
where $c$ and  $G$ are the  speed of light and the gravitational
constant, respectively, while  $D_\mathrm{l}$ denotes the angular diameter
distance to the lens.
The geometric lensing efficiency
\begin{equation}
\label{eqn:beta}
\beta(z_\mathrm{l},z_\mathrm{s})=\mathrm{max}\left[0,\frac{D_\mathrm{ls}(z_\mathrm{l},z_\mathrm{s})}{D_\mathrm{s}(z_\mathrm{s})}\right] \,
\end{equation}
is defined in terms of the angular diameter distances  from the observer to
the source  $D_\mathrm{s}$, and from the lens to the source
$D_\mathrm{ls}$.

Given that they are both computed from second-order derivatives of
the  lensing potential,
the weak lensing shear $\gamma$ and convergence $\kappa$ are linked.
The spatial distribution of the convergence  can therefore be reconstructed
from the shear field up to an integration constant \citep{kaiser93}, which represents the mass-sheet degeneracy
\citep{schneider95}.

Weak lensing shape measurement algorithms aim to obtain unbiased estimates of the
complex galaxy ellipticity
\begin{equation}
\epsilon=\epsilon_1+\mathrm{i}\epsilon_2=|\epsilon|\mathrm{e}^{2\mathrm{i}\varphi}\,.
\end{equation}
In the  idealised case of an object that has concentric elliptical isophotes
with a  constant position angle $\varphi$ and constant ratios of the semi-major
and semi-minor axes $a$ and $b$, these are related to the ellipticity as \mbox{$|\epsilon|=(a-b)/(a+b)$}.
The ellipticity transforms under
weak reduced shears (\mbox{$|g|\ll 1$}) as
\begin{equation}
\label{eq:eofges}
\epsilon\simeq
  \epsilon_\mathrm{s}+g
\end{equation}
\citep[for the general case see][]{seitz97,bartelmann01}.
The intrinsic source ellipticity
$\epsilon_\mathrm{s}$
is expected to have a random orientation, yielding an expectation  value \mbox{$\langle \epsilon_\mathrm{s} \rangle=0$}.
Hence,  ellipticity measurements provide noisy estimates for the local reduced
shear, where the noise level is given by the dispersion
\begin{equation}
\sigma_\epsilon=\sigma\left(\epsilon-g\right)
\simeq \sqrt{\sigma_\mathrm{int}^2+\sigma_\mathrm{m}^2}\, ,
\label{eq:sigmae}
\end{equation}
which has contributions from both
the intrinsic ellipticity dispersion
$\sigma_\mathrm{int}=\sigma\left(\epsilon_\mathrm{s}\right)$  of the galaxy
sample\footnote{We absorb the effective broadening of the observed
  ellipticity distribution due to cosmological weak lensing by uncorrelated
  large-scale structure  in $\sigma_\mathrm{int}$. \ha{In
  Eq.\thinspace\ref{eq:sigmae} $g$ refers to the reduced shear caused by
the targeted cluster.}}
 and measurement noise $\sigma_\mathrm{m}$
\citep[e.g.][\citetalias{schrabback18}]{leauthaud07}.
\ha{Assuming dominant shape noise,}
the signal-to-noise ratio of the
detection of the weak
lensing reduced shear signal scales as
\begin{equation}
\label{eq:snwl}
\left(\frac{S}{N}\right)_\mathrm{WL} \propto f \equiv
\frac{
\sqrt{n_\mathrm{gal}}
  \langle\beta\rangle}{\sigma_{\epsilon,\mathrm{eff}}}\, ,
\end{equation}
where $n_\mathrm{gal}$ indicates the weak lensing source density on the sky
\reva{and $\sigma_{\epsilon,\mathrm{eff}}$ corresponds to the effective
  value of $\sigma_\epsilon$ computed taking possible shape weights into account}. The weak lensing signal-to-noise ratio also depends on the
mass, mass
distribution, and radial fitting range \citep[e.g.][]{bartelmann01}.
Shape weights $w_i$ \reva{also} need to be taken into account when computing
$\langle\beta\rangle$,
where we employ magnitude-dependent weights
\begin{equation}
\label{eq:weights}
w_i(\mathrm{mag}_i)=\sigma_\epsilon^{-2}(\mathrm{mag}_i) \, ,
\end{equation}
which are directly related to the expected noise in the reduced shear
estimate for galaxy $i$. In this case the effective ellipticity dispersion
for the sample
from Eq.\thinspace\ref{eq:snwl} \revb{becomes}
\begin{equation}
\sigma_{\epsilon,\mathrm{eff}}=\left(N^{-1}\sum_{i=1}^N w_i\right)^{-\frac{1}{2}}\, .
\end{equation}

For cluster weak lensing analyses it is useful to decompose the ellipticity
(and likewise the reduced shear) into a tangential component
carrying the signal
\begin{equation}
\label{eq:et}
\epsilon_\mathrm{t}  =  - \epsilon_1 \cos{2 \phi} - \epsilon_2
\sin{2\phi}\, ,
\end{equation}
where $\phi$ denotes the azimuthal angle with respect to the cluster centre
and the 45 degrees-rotated cross-component
\begin{equation}
\label{eq:ex}
\epsilon_\times  =  + \epsilon_1 \sin{2\phi} - \epsilon_2 \cos{2 \phi}  \,.
\end{equation}
The averaged tangential ellipticity  profile provides an estimate for the
tangential reduced shear profile $g_\mathrm{t}(r)$ of the cluster, which we fit using  model predictions from \citet{wright00} that assume a
spherical NFW density profile \citep{navarro97}.

\section{Data and data reduction}
\label{se:data}

\ha{In our analysis we make use of high-resolution VLT/HAWK-I $K_\mathrm{s}$ images for the weak lensing shape measurements, which we complement with LBT/LBC imaging for a colour selection. We additionally analyse overlapping HST/ACS data to cross-check the VLT/HAWK-I weak lensing constraints.}

\subsection{VLT/HAWK-I data}
\label{se:data:hawki}
\clusteras\, was observed with VLT UT4 using HAWK-I
under programme 087.A-0933 (PI: Schrabback).
HAWK-I is a high-throughput NIR imager equipped with a \mbox{$2\times 2$}
mosaic of \mbox{$2048\times 2048$} Rockwell HgCdTe
MBE
HAWAII
2 RG arrays, with a plate scale of 0\farcs106 pixel$^{-1}$ and a \mbox{$7\farcm5\times
  7\farcm5$} field of view \citep[see][for details]{kissler-patig08}.
Here we analyse $K_\mathrm{s}$ band images observed
using large dither steps to cover the \mbox{$\sim 15^{\prime\prime}$}
gaps between the detectors.
In total, $326\times 80$\,s exposures were obtained \reva{(total exposure time 7.2h)}, some of which were
repetitions because the seeing constraint ($K_\mathrm{s}$ band image quality
\mbox{$\le 0\farcs4$}) was not fulfilled. Each $80$\,s exposure was
constructed from $8\times 10$\,s \ha{internal} sub-exposures \ha{to avoid background saturation}, averaged using on-detector
arithmetics.

The data were reduced using {\tt THELI} \citep{erben05,schirmer13}
following
standard procedures, including dark subtraction and flat fielding.
A dynamic two-pass background subtraction including object masking was
employed to remove the sky background from individual exposures. The
background models were calculated from a floating median of the eight closest
images \ha{in time}, corresponding to a time window of $13-15$ minutes.
An accurate astrometric reference catalogue is required to align the images
on sky. The 2MASS catalogue has insufficient source density for this purpose,
as RCS2 $J$2327 is located at high galactic latitude of $-58^\circ$. Thus,
 we first
processed and co-added CFHT Megaprime $i$-band data (PI: H. Hoekstra),
\ha{for which an astrometric calibration was possible using
  2MASS  thanks to the  larger field of view.
 We then extracted a deep astrometric reference catalogue from the CFHT data,
which was used both for the HAWK-I reduction and  the reductions described in Sections \ref{se:red:lbt} and \ref{se:data:acs}.}
The astrometry for the HAWK-I data was determined by {\tt THELI} via {\tt Scamp}
\citep{bertin06}. The relative positions of the detectors were accurately fixed
using the dithered exposures and a fixed third-order distortion polynomial
was used to describe the non-linear terms. In total, relative image
registration is accurate to $\sim 1/10$-th of a pixel, which is well sufficient for
our shear analysis. Image co-addition and resampling in {\tt THELI}
was performed with {\tt SWarp} \citep{bertin02}, using a Lanczos3 kernel
matched to the well-sampled PSF.

Given the variation in seeing we created two separate stacks.
The first stack is generated from all exposures for
\reva{photometric measurements},
yielding a total integration time
of 26.1\thinspace ks and a median stellar \mbox{$\mathrm{FWHM}^*=0\farcs40$}
as measured by \texttt{SExtractor} \citep{bertin1996}.
The second stack is  used for the shape measurements. Here we exclude
exposures with
 poorer image quality, yielding a shorter total integration time
of 17.1\thinspace ks (4.8\thinspace h, or \mbox{$\sim 7$}\thinspace h including overheads), but a better image quality with a median
\mbox{$\mathrm{FWHM}^*=0\farcs35$}.
To simplify the comparison to the weak lensing literature we also report
the median stellar FLUX\_RADIUS parameter from \texttt{SExtractor} \mbox{$r_\mathrm{f}^*=0\farcs22$}
and the median stellar half-light radius from  \texttt{analyseldac}
\citep{erben2001}  \mbox{$r_\mathrm{h}^*=0\farcs19$}.

\subsection{LBT/LBC data}
\label{se:red:lbt}
\clusteras\, was also observed by the Large Binocular Telescope (LBT) on Oct 02, 2010 (PI: Eifler) under good seeing conditions (\mbox{$\simeq 0\farcs7$}),
where we make use of $g$-band observations obtained with LBC\_BLUE  \citep{giallongo08} and $z$-band observations obtained with  LBC\_RED.
The data were reduced using \texttt{THELI}
following standard procedures, yielding
co-added total integration times of
2.4\thinspace ks in the $g$ band and 3.0\thinspace ks in the $z$ band.

\subsection{HST/ACS data}
\label{se:data:acs}
To cross-check our HAWK-I shape measurements we also reduced and analysed HST/ACS
observations (HST-GO 13177, PI: Brada{\v c}) of \clusteras\, conducted with the F814W filter as part of
the
        Spitzer Ultra Faint SUrvey Program \citep{bradac14}.
This includes a central pointing (integration time 5.6\thinspace ks) and four
parallel fields (integration times 3.6--5.5\thinspace ks) that overlap with the
outskirts of our HAWK-I observations.
In order to generate a colour image we also \revd{processed} central ACS
observations conduced in the F435W filter (integration time 4.2\thinspace ks) as part of the HST-GO programme
10846 (PI: Gladders).

\revb{Following \citetalias{schrabback18}} we reduced these data employing the pixel-level correction for
charge-transfer inefficiency from \citet{massey2014}, the standard ACS calibration pipeline \texttt{CALACS} for
further basic reduction steps, \texttt{MultiDrizzle} \citep{koekemoer2003}
for the cosmic ray removal and stacking\footnote{\revb{We used the
  \texttt{lanczos3} kernel with the native pixel scale 0\farcs05 and a
  \texttt{pixfrac} of 1.0. These settings minimise the impact of noise
  correlations while introducing only a low level of aliasing for the ellipticity measurements \citep[][]{jee2007}.}}, and scripts from
\citet{schrabback10} for the image registration and optimisation of masks
and weights.

\section{Analysis}
\label{se:ana}
\subsection{HAWK-I shape measurements}
\label{se:hawkshapes}

\begin{figure}
 \includegraphics[width=1\columnwidth]{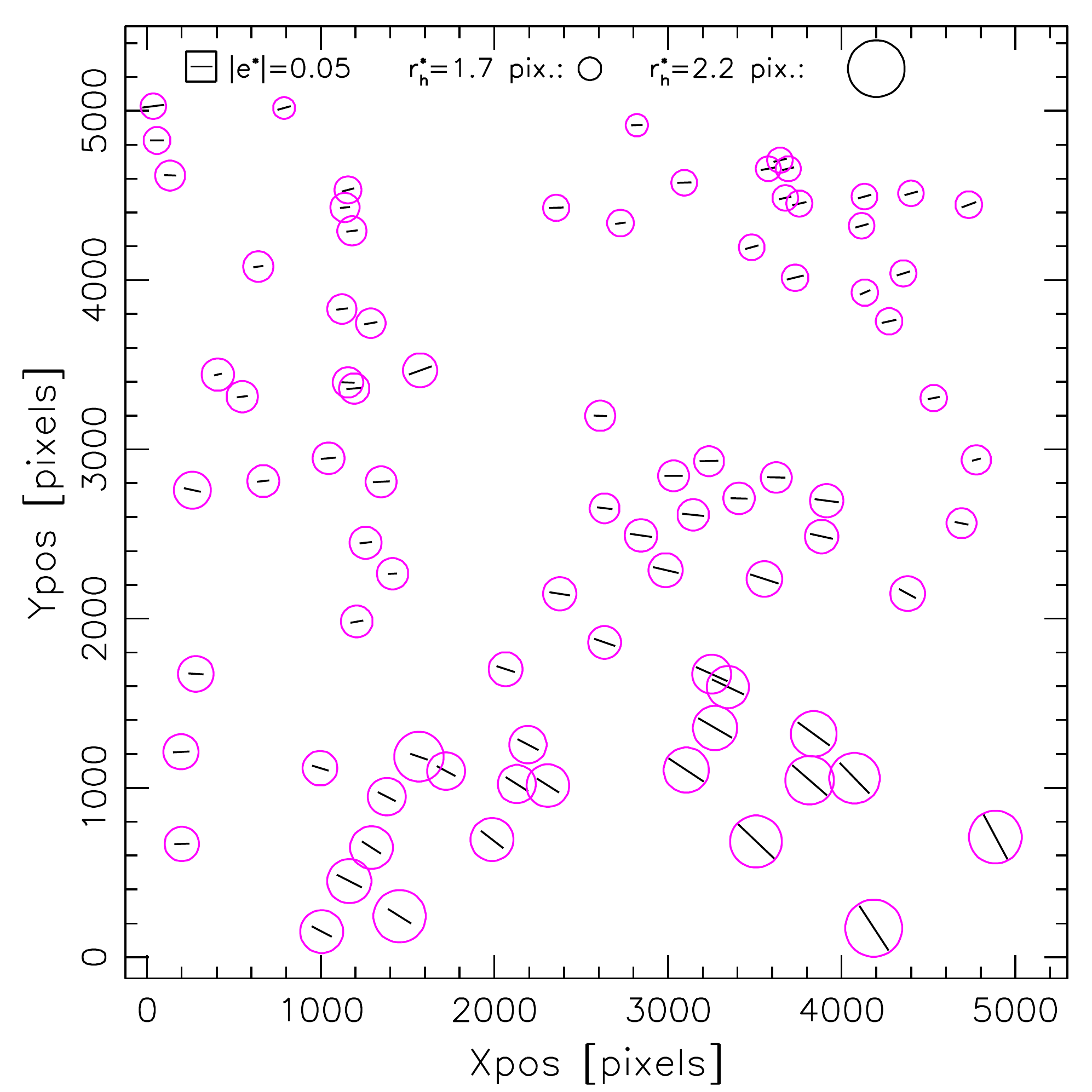}
 \caption{Spatial variation of the PSF in our best-seeing stack of the HAWK-I
   $K_\mathrm{s}$ observations of \clusteras: Each whisker indicates the
  measured polarisation $e^*$ of a star, while the circle indicates its
  half-light radius $r_\mathrm{h}^*$ from \texttt{analyseldac} (see the reference whisker and circles at
  the top for the absolute scale). In this stack north is up and east is left, matching the orientation of the input frames (observations obtained with a default $0^\circ$ position angle).
\label{fig:psf}}
\end{figure}

We detected objects with \texttt{SExtractor}
and measure weak lensing shapes using the \texttt{analyseldac} \citep{erben2001} implementation of the  KSB+ formalism
\citep{kaiser1995,luppino1997,hoekstra1998}
as detailed in \citet{schrabback07}, employing the correction
for multiplicative noise bias as a function of the \texttt{analyseldac} signal-to-noise ratio from
\citet{schrabback10}.
Analysing the measured shapes of stellar images in our  $K_\mathrm{s}$
best-seeing stack we find that the HAWK-I PSF is well behaved in the
majority of the field of view with PSF polarisation amplitudes
\mbox{$|e^*|\lesssim 0.05$}, where
\begin{equation}
\label{eq:elli_e}
   e = e_1 + \textrm{i} e_2 = \frac{Q_{11} - Q_{22} + 2 \textrm{i} Q_{12}}{Q_{11} + Q_{22}} \,
\end{equation}
is defined via weighted second-order brightness moments $Q_{ij}$
as detailed in \citet[][]{schrabback07}.
However, the PSF degrades noticeably towards lower $y$ positions
with larger stellar polarisations and half-light radii $r_\mathrm{h}^*$ as computed by \texttt{analyseldac}  (see Figure
\ref{fig:psf}).
We find that the spatial variations of the KSB+ PSF parameters can be interpolated well using
third-order polynomials combining stars from all chips.
For the weak lensing analysis we required galaxies to be
sufficiently
resolved with half-light radii \mbox{$r_\mathrm{h}>1.2
  r_\mathrm{h,mod}^*(x,y)$}, where $r_\mathrm{h,mod}^*(x,y)$ indicates the
polynomial interpolation of the measured stellar half-light radii at the
position of the galaxy.
We selected galaxies with a flux signal-to-noise ratio defined via
the auto flux from  \texttt{SExtractor}  of
\mbox{$(S/N)_\mathrm{flux}=\mathrm{FLUX\_AUTO}/\mathrm{FLUXERR\_AUTO}>10$}.
Shape selections were also applied according to the trace of the
``pre-seeing'' shear polarisability tensor \mbox{$\mathrm{Tr}P^g/2>0.1$} and
PSF-corrected ellipticity estimate \mbox{$|\epsilon|<1.4$}.
We masked regions around bright foreground objects and reject galaxies that
are flagged by \texttt{SExtractor} or \texttt{analyseldac}, for example owing
to the presence of a nearby object.
Prior to the photometric background selection our catalogue of galaxies
with weak lensing shape estimates has a source number density of
\mbox{$45\,\mathrm{arcmin}^{-2}$}.

Analysing ACS-like image simulations containing weak simulated shears ($|g|<0.06$), \citet{schrabback10} estimated that the basic shape measurement
algorithm also employed in this work leads to residual multiplicative
shear biases \mbox{$|m|<2\%$}.
However, these authors neither tested the performance in the stronger shear regime of
clusters nor the sensitivity to the assumed input ellipticity distribution of galaxies, which
can
affect measured noise biases \citep{viola14,hoekstra15}.
We therefore conducted additional tests with new simulations created with
\texttt{galsim} \citep{rowe15}.
The details of these tests will be described in Hern\'andez-Mart\'in et al.~(in
prep.).
For our current work, the most relevant result from these
simulations is that
 multiplicative biases are limited to  \mbox{$|m|\lesssim
  3\%$} for reduced shears \mbox{$|g|<0.2$} and variations in the intrinsic
ellipticity dispersion in the range \mbox{$0.2\le \sigma_\mathrm{int}\le
  0.3$}.
For stronger shear \mbox{$|g|<0.4$} biases are limited to  \mbox{$|m|\lesssim
  5\%$}, still without \revd{recalibration} compared to the work from
\citet{schrabback10}.
Given that most of the weak lensing mass constraints for \clusteras\,
originate from scales with  \mbox{$|g|<0.2$}, while the innermost radial
bins that are included have \mbox{$|g|<0.4$} (see
Sect.\thinspace\ref{se:results}), we assume an intermediate 4\% systematic
uncertainty on the shear calibration for our systematic error budget.
\reva{Based on the  analysis from Hern\'andez-Mart\'in et al.~(in
prep.) we conclude that this shear calibration uncertainty results from
a combination of limitations in the noise bias correction and a slight
  non-linear response of our KSB+ implementation for stronger shears, both}
\reva{of which can be fixed with a \revb{recalibration} for potential future studies requiring a tighter systematic
error control.}

\subsection{Photometry}
\begin{figure}
 \includegraphics[width=1\columnwidth]{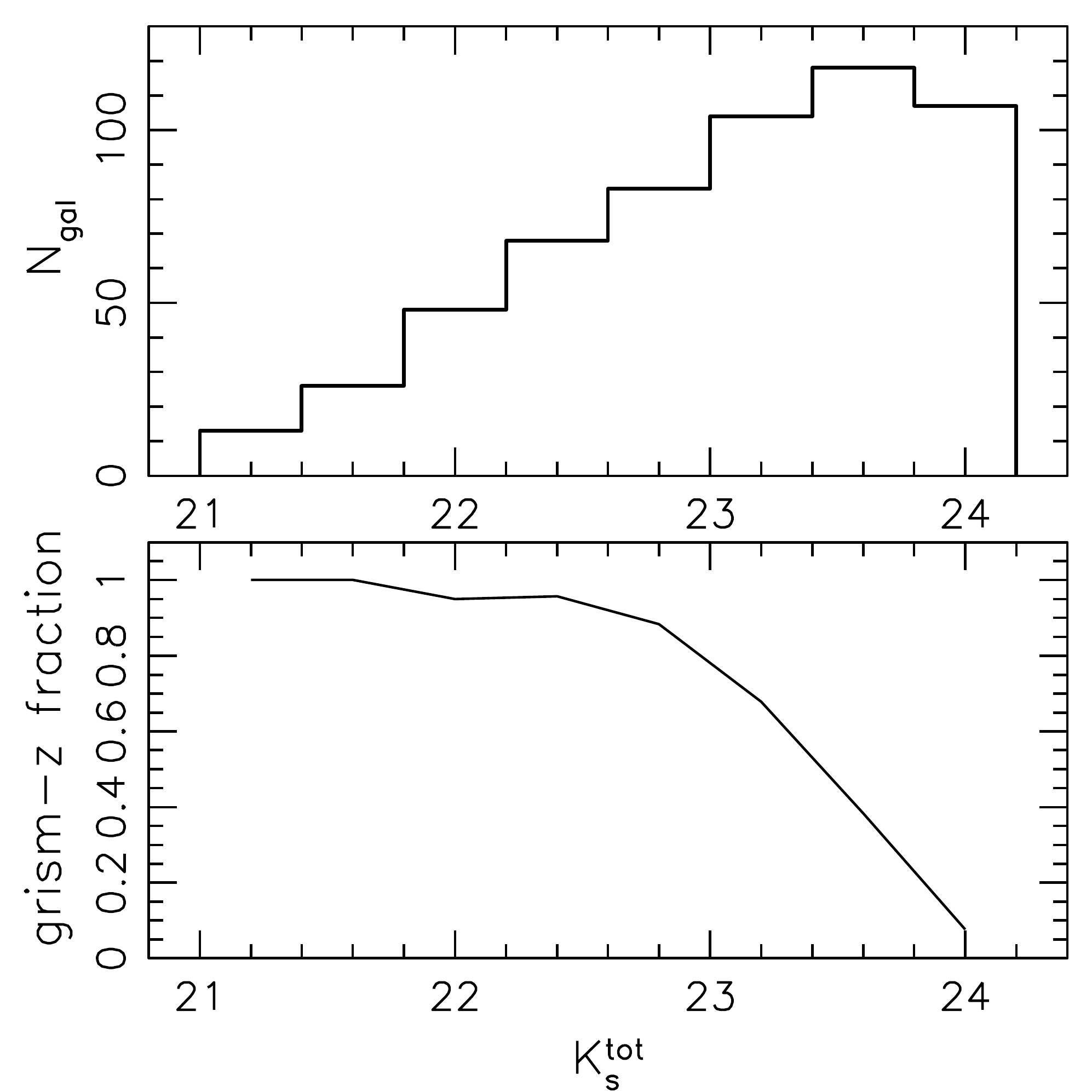}
 \caption{{\it Top:}
   Histogram of the number of colour-selected galaxies in our HAWK-I weak
   lensing shape catalogue
\hb{(covering
a non-masked area of 52.4 arcmin$^2$)}
 as a function of the total $K_\mathrm{s}$ magnitude.
   {\it Bottom:} Fraction of  colour-selected galaxies within the CANDELS/COSMOS 3D-HST grism area
   with a robust HST grism redshift or spectroscopic redshift as a function of   the total $K_\mathrm{s}$ magnitude from UltraVISTA.
   \label{fig:magdist}}
\end{figure}

\reva{For the HAWK-I $K_\mathrm{s}$ data all photometric measurements were conducted on the stack  derived from all available exposures (see Sect.\thinspace\ref{se:data:hawki}).}
We homogenised the PSF between the VLT and LBT
stacks using spatially varying kernels constructed using \texttt{PSFEx}
\citep{bertin11} and measured colours between these PSF-homogenised images employing
2\farcs0
diameter  circular
 apertures.
 We used 2MASS \citep{skrutskie06} $K_\mathrm{s}$ magnitudes for the absolute
 photometric calibration of the HAWK-I data.
 For the $g$ and $z$ bands we initially estimated zero points
with respect to $K_\mathrm{s}$ using stellar locus regression.
We then applied residual zero-point offsets to
 optimise the overlap of the
galaxy colour distributions in $g-z$ versus $z-K_\mathrm{s}$ colour space
between
our catalogue and the UltraVISTA-detected reference catalogue used to
estimate the redshift distribution
(see Sect.\thinspace\ref{color_select_zdist})\footnote{\ha{This is necessary for two
reasons. First, differences in the effective filter curves between our
HAWK-I+LBC data and the VISTA+Subaru data used for \revd{the} UltraVISTA reference catalogue
lead to small differences in the colour calibration for stars and
galaxies. Second, small zero-point offsets have already been applied to
the UltraVISTA reference catalogue to improve the photo-$z$ performance
\citep[see][]{muzzin13}.}}.
Photometric errors were estimated from the flux fluctuations when placing
apertures
at random locations that do not contain detected objects.
For the
2\farcs0 diameter
apertures we computed median $5\sigma$ limiting
magnitudes\footnote{We
  quote limiting
  magnitudes without aperture correction.} of
\mbox{$(26.6,25.9,25.0)$} in the
\mbox{$(g,z,K_\mathrm{s})$} bands.
For the subsequent analysis we excluded regions near the edges of the HAWK-I
mosaic and the LBT chip gaps as they have a significantly
reduced depth in some of the bands.
We also limited the subsequent analysis to galaxies with   \texttt{SExtractor} ``auto'' magnitudes  in the range \mbox{$21<K_\mathrm{s}^\mathrm{tot}<24.2$},
given that brighter \ha{magnitude bins} contain very few background galaxies, while the sample becomes highly incomplete at fainter magnitudes given the shape cuts (compare to the top panel of Fig.\thinspace\ref{fig:magdist}).

\subsection{Reference samples to estimate the source redshift distribution}
\label{color_select_zdist}
\ha{For unbiased mass measurements we  have to accurately estimate
the
weighted-average geometric lensing efficiency $\langle\beta\rangle$
(see Eq.\thinspace\ref{eqn:beta}) of the selected source sample.
Here,
a photometric selection of the lensed background galaxies helps to
increase the measurement sensitivity, while reducing systematic uncertainties
arising from cluster member contamination.}
Similar to the strategy from \citetalias{schrabback18} we employed a colour
selection (see Sect.\thinspace\ref{se:colorselect}) that is designed to yield negligible residual contamination by cluster members
and applied a consistent selection to well-calibrated reference data \ha{from deep fields}   to
estimate the redshift distribution and $\langle\beta\rangle$  (see Sect.\thinspace\ref{se:beta}).

\subsubsection{UltraVISTA reference catalogue}
\label{se:ultravista}
The UltraVISTA Survey \citep{mccracken12} has obtained very deep NIR imaging
in the COSMOS field \citep{scoville07}.
By design the greatest depth is achieved in the ``ultra-deep'' stripes \citep{mccracken12},
reaching  a $5\sigma$ limiting
$K_\mathrm{s}$ magnitude in
2\farcs0
apertures of 25.2 in the
latest DR3 release,
which exceeds even the depth of our HAWK-I imaging by 0.2 mag.
COSMOS/UltraVISTA
allows us to investigate galaxy colour and redshift distributions for our
weak lensing analysis;
\revb{the area of this survey (\mbox{$\sim 0.75$} deg$^2$), which is 50 times larger than the HAWK-I field of view,}
greatly reduces uncertainties from sampling variance (see Sect.\thinspace\ref{se:beta}).
In particular, we employed an updated version of the $K_\mathrm{s}$-selected photometric redshift
catalogue from \citet{muzzin13}, which makes use of the deeper UltraVISTA DR3 data (see Muzzin et al.~in
prep. for details).
In addition to the PSF-matched aperture magnitudes in  $g,z$, and $K_\mathrm{s}$ used for colour measurements, we made use of the   \texttt{SExtractor} ``auto'' magnitudes  $K_\mathrm{s}^\mathrm{tot}$.
For our study we limited the analysis to objects that are photometrically classified as
galaxies,
located in non-masked areas of  the ``ultra-deep''
stripes, and that are not flagged as blends by \texttt{SExtractor}.

While our  HAWK-I+LBC catalogue and the UltraVISTA-detected catalogue have
the same median depth in $g$ (within 0.05 mag), the UltraVISTA-detected catalogue is deeper by 0.2 mag in
$K_\mathrm{s}$ and shallower  by 0.5 mag in $z$.
We expect that the \ha{small} \ha{difference in $K_\mathrm{s}$ depth would be} negligible
for our analysis,
but to further improve the matching in
the source selection between the two catalogues, we added
Gaussian noise to the UltraVISTA $K_\mathrm{s}^\mathrm{tot}$ magnitudes  to
have identical limiting magnitudes; \ha{we also explicitly account
for the incompleteness of the lensing catalogue when computing  $\langle\beta\rangle$ in Sect.\thinspace\ref{se:beta:best}. The impact of differences in
the noise in the  colour measurement is investigated in Sect.\thinspace\ref{se:betauncer}.}

\subsubsection{3D-HST reference catalogue}
\label{se:3dhst}

\begin{figure}
 \includegraphics[width=1\columnwidth]{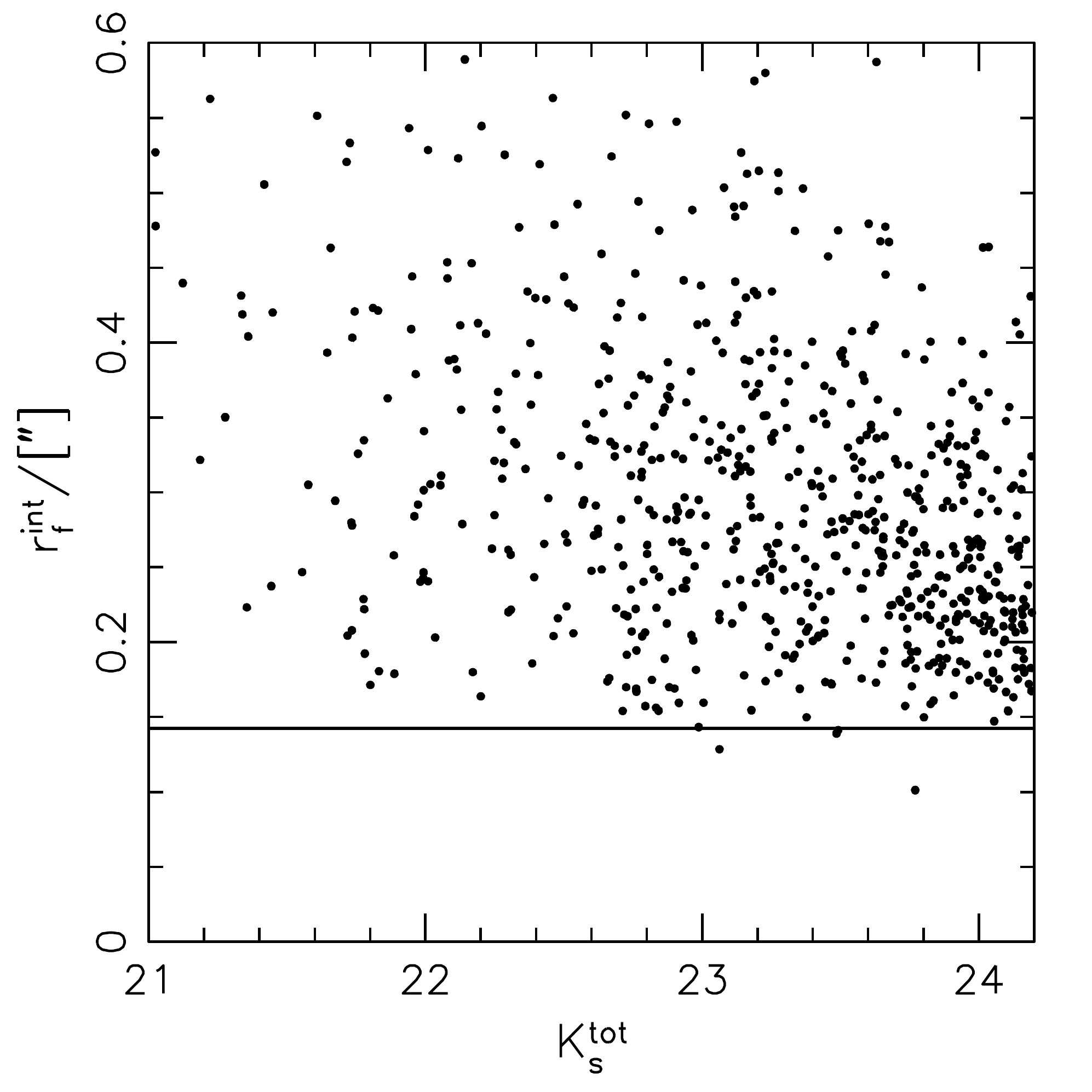}
 \caption{Intrinsic flux radius
   \ha{$r_\mathrm{f}^\mathrm{int}$ as measured in HST/WFC3  $H$-band data for} galaxies in the  CANDELS/COSMOS 3D-HST
   grism area passing our colour selection as a function of  $K_\mathrm{s}^\mathrm{tot}$.
   The horizontal line corresponds to the mean size cut in our HAWK-I weak lensing analysis.
\label{fig:resolved}}
\end{figure}

As a second reference data set to infer the source redshift distribution we employed
redshifts computed by the 3D-HST team for galaxies in the  CANDELS \revb{\citep{grogin2011,koekemoer11}} area
within the COSMOS field.
This includes HST/NIR-selected photometric redshifts based on a total of 44
different photometric data sets  \citep{skelton14} and
``grism''-redshift estimates from WFC3/IR slitless spectroscopy
\citep[][]{momcheva16}, where we also include ground-based spectroscopic redshifts compiled in the 3D-HST catalogue.
Given the deeper NIR photometry and the deep grism spectra, these redshifts are expected to be highly robust,
allowing us to conduct important cross-checks for our analysis.
After applying our magnitude and colour selection  (explained in Sect.\thinspace\ref{se:colorselect})
we find that
99.4\%
of the galaxies in the   UltraVISTA-detected catalogue within the area
covered by the grism spectra have a match in the 3D-HST
catalogue\footnote{The
non-matching galaxies
 can be explained through differences in the deblending and have no relevant impact on our analysis.}.
The bottom panel of Fig.\thinspace\ref{fig:magdist} shows the fraction of
these galaxies that have a spectroscopic redshift or a 3D-HST  grism redshift classified as robust  by \citet{momcheva16} as a function of $K_\mathrm{s}^\mathrm{tot}$ from UltraVISTA.
Most galaxies at \mbox{$K_\mathrm{s}^\mathrm{tot}\lesssim 23$} have a
grism/spec-$z$, but this fraction drops at fainter \revd{magnitudes} because of a
combination of the magnitude limit \mbox{$[JH]<24$} employed by
\revb{\citet{momcheva16}, who used a $J+H$ band stack for detection and
selection, and}
 increased incompleteness at fainter
magnitudes due to contamination by other objects.
Nevertheless, when accounting for the  $K_\mathrm{s}^\mathrm{tot}$
distribution of our HAWK-I data and taking lensing weights into account (see
Sect.\thinspace\ref{se:shapenoise}), we find that effectively
\hb{\mbox{$\simeq 71\%$}}
 of the relevant galaxies in the 3D-HST grism area have a robust
 grism/spec-$z$.
\ha{For comparison, the corresponding  fraction amounts to only 21\% for optically selected
weak lensing source galaxies as employed in  \citetalias{schrabback18}, with shape measurements from ACS F606W data of single-orbit depth and a full-depth \mbox{$V_{606}-I_{814}$} colour selection.
Given the much higher fraction of grism/spec-$z$ in the current study, we have to rely less on the accuracy of photometric redshift
reference samples,}
leading to lower systematic uncertainties in the lensing analysis from the calibration of the redshift distribution (see Sect.\thinspace\ref{se:betauncer}).
For our analysis we define a ``best'' redshift $z_\mathrm{best}$ from the 3D-HST catalogue, which is the spectroscopic or grism redshift of a galaxy when available and its photometric redshift otherwise.

\citet{skelton14} also provided HST/WFC3-measured $H$-band size estimates of CANDELS galaxies, allowing us to check if the galaxy size selection applied in our HAWK-I analysis has a  relevant impact on the estimation  of the redshift distribution.
Fig.\thinspace\ref{fig:resolved} shows the distribution of the
intrinsic flux radius
$r_\mathrm{f}^\mathrm{int}=\sqrt{r_\mathrm{f}^2-r_\mathrm{f,PSF}^2}$,
defined via the  flux radius parameter of the galaxies and stars from  \texttt{SExtractor}, for
the colour-selected CANDELS galaxies as a function of  $K_\mathrm{s}^\mathrm{tot}$.
This shows that
\mbox{$\sim 99.4\%$}
 of the galaxies are sufficiently resolved for shape measurements at the resolution of our HAWK-I data (limit illustrated as horizontal line in Fig.\thinspace\ref{fig:resolved}).
As a result, the application of the size selection has a
 negligible
impact on the estimated average geometric lensing efficiency.
\ha{However, we stress that many of the galaxies are only slightly more extended than required for the shape analysis (see Fig.\thinspace\ref{fig:resolved}). We therefore recommend that similar programmes in the future do not relax the seeing requirements, compared to our study,
  in order to not suffer from a reduced weak lensing source density.
}

\begin{figure}
 \includegraphics[width=1\columnwidth]{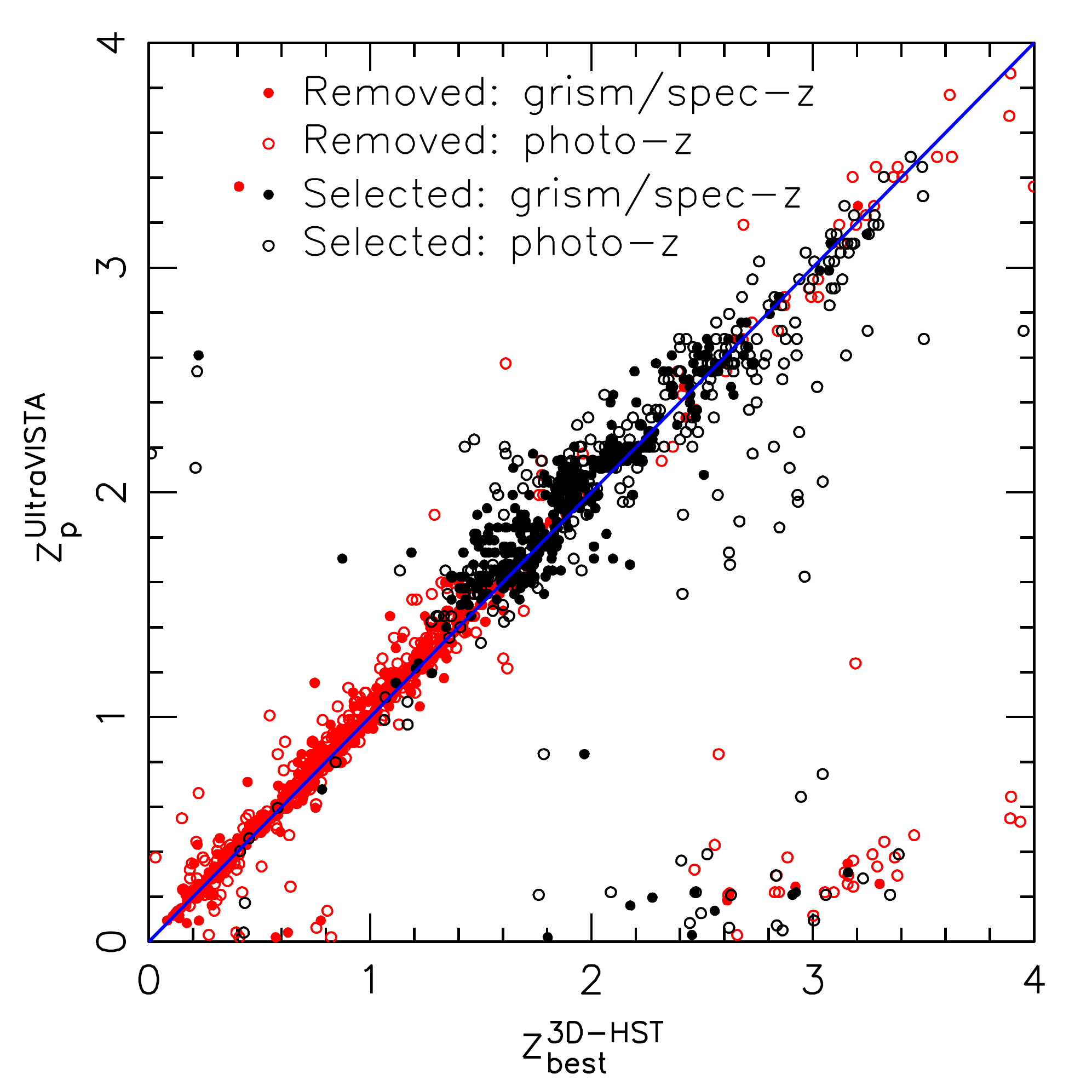}
 \caption{Comparison of the best redshift estimate \mbox{$z_\mathrm{best}$} from 3D-HST and the peak photometric redshift  \mbox{$z_\mathrm{p}$} in the UltraVISTA-detected catalogue for galaxies located in the area covered by the grism observations with  \mbox{$21<K_\mathrm{s}^\mathrm{tot}<24.2$}.
   Galaxies with a spectroscopic or grism redshift in the 3D-HST catalogue are indicated as filled circles, while the galaxies having a  photometric redshift  in the 3D-HST catalogue only are shown as open circles.
   Black symbols correspond to galaxies passing our colour selection, while red symbols show galaxies removed by the colour selection. The blue line shows the one-to-one relation.
\label{fig:zz}}
\end{figure}

\begin{figure*}
  \includegraphics[width=1\columnwidth]{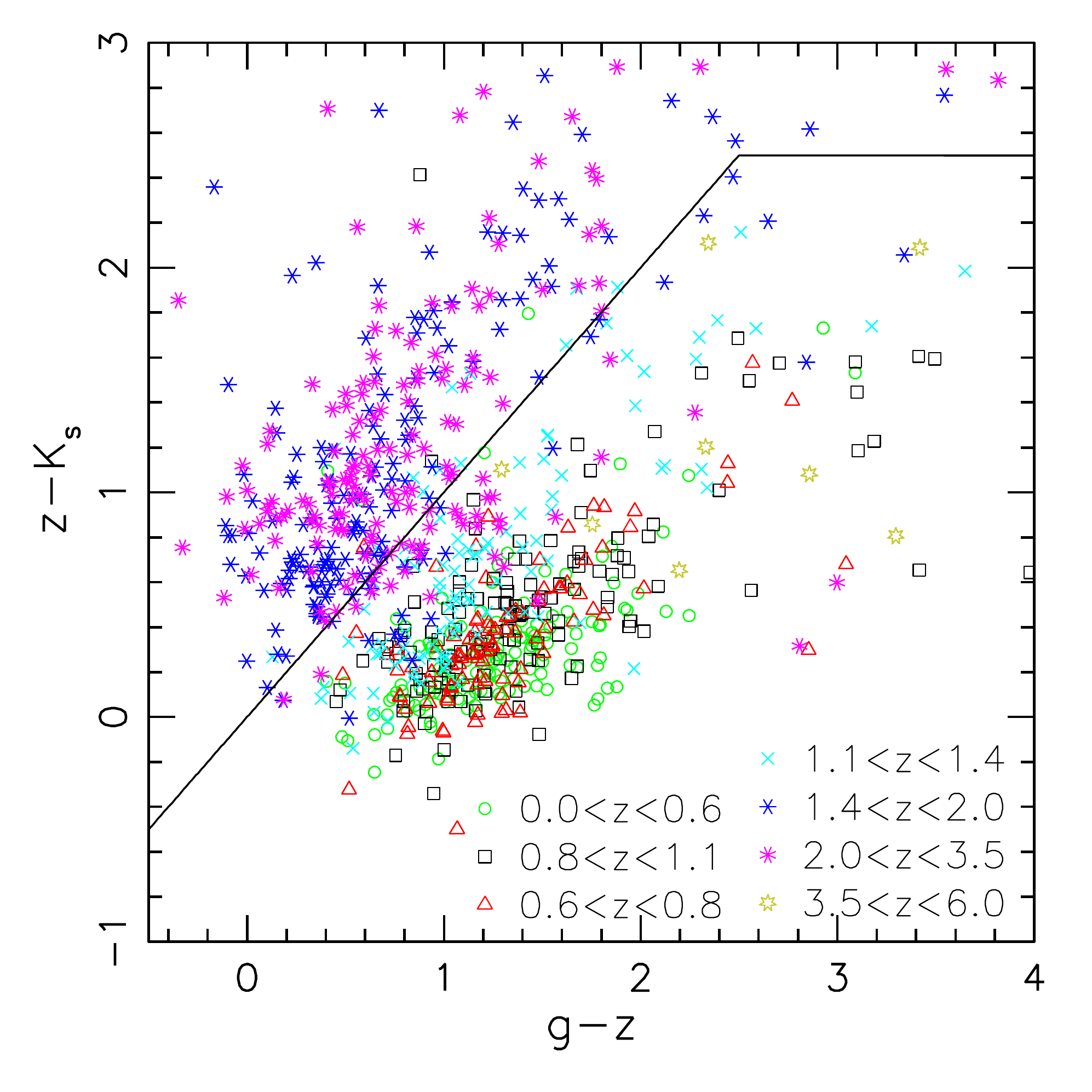}
  \includegraphics[width=1\columnwidth]{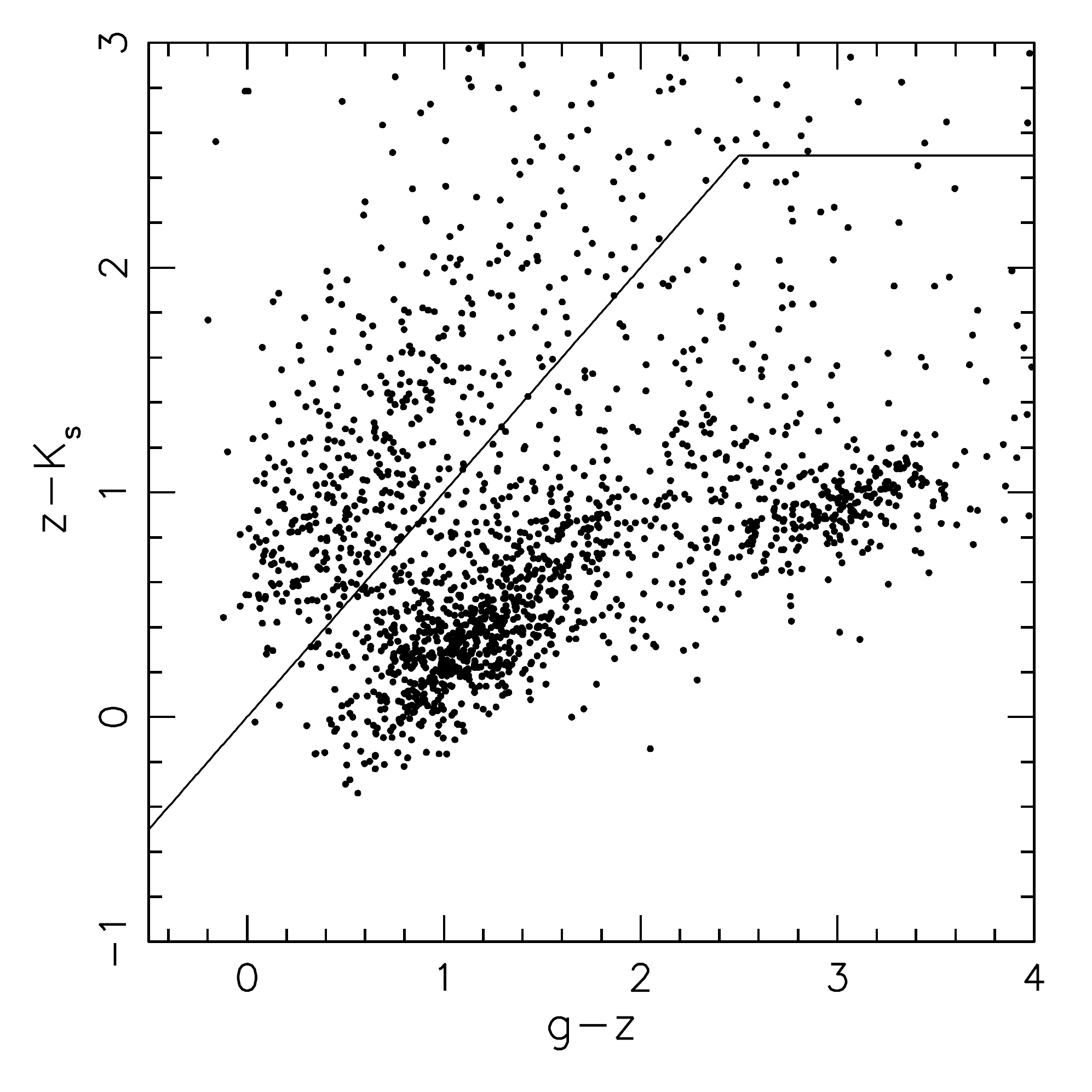}
  \caption{Distribution of
 \revb{galaxies with \mbox{$21<K_{\mathrm{s}}^\mathrm{tot}<24.2$}
   passing our size selection}
in \mbox{$g-z$} vs.   \mbox{$z-K_\mathrm{s}$} colour space.
 The black line indicates the colour selection
 \mbox{$z-K_\mathrm{s}>\mathrm{min}[g-z,2.5]$} employed in our analysis.
The {\it left} panel shows a random 50\% fraction of the galaxies in the
CANDELS/COSMOS 3D-HST grism area, with colours and symbols indicating
various ranges in the best redshift estimate from 3D-HST.
The {\it right} panel shows the galaxies
passing the shape selection in
our
catalogue for   \clusteras.
The excess of galaxies around
\mbox{$g-z\simeq 3$} and \mbox{$z-K_\mathrm{s}\simeq 1$} corresponds to the
cluster red sequence, \ha{which is efficiently removed from our background
  sample, along with bluer cluster members located near \mbox{$g-z\simeq 1.2$} and \mbox{$z-K_\mathrm{s}\simeq 0.3$}.}
\label{fig:colordist_ultravista}}
\end{figure*}

\subsubsection{Redshift comparison}
\label{se:zcomp}
We compared the 3D-HST \mbox{$z_\mathrm{best}$} redshifts to the peak photometric
redshifts $z_\mathrm{p}$ from the UltraVISTA-detected catalogue in
Fig.\thinspace\ref{fig:zz}.
While most galaxies closely follow the one-to-one
relation\footnote{\reva{When defining catastrophic redshift outliers as
    \mbox{$\Delta z=|z_\mathrm{best}-z_\mathrm{p}|>1$}, 5.5\% of the colour-selected galaxies shown in
    Fig.\thinspace\ref{fig:zz} are catastrophic redshift outliers. Excluding these catastrophic outliers, the
    redshift scatter} \reva{of the  remaining galaxies can be quantified via the
    root mean square \mbox{$\mathrm{r.m.s.}(\Delta z/[1+z_\mathrm{best}])=0.07$}.}}, there are some noticeable systematic features visible.
Here we  \revd{focus}
 on those galaxies that pass our colour selection shown in black.
In particular, galaxies close to the one-to-one relation with \mbox{$1.4\lesssim z_\mathrm{best}\lesssim 2.2$} appear to have a peak photometric redshift $z_\mathrm{p}$ in the UltraVISTA-detected catalogue that is  slightly biased high on average.
For galaxies with  \mbox{$2.2\lesssim z_\mathrm{best}\lesssim 3.4$} this
bias  disappears for the galaxies close to the one-to-one relation, but
there is a noticeable fraction of outliers with a  $z_\mathrm{p}$  biased
low, in some cases catastrophically with \mbox{$z_\mathrm{p}\lesssim 0.4$}.
Given that these biases are in opposite directions, their impact partially
cancels when computing the average geometric lensing efficiency (see Sect.\thinspace\ref{se:beta}).

 Indications for similar outliers
have already been noted by \citet{schrabback10}
and \citetalias{schrabback18}.
In particular,  \citetalias{schrabback18} compare
3D-HST \revd{photo-$z$s} to extremely
deep photometric and grism redshifts available in the HUDF.
While \citetalias{schrabback18} conclude that  the 3D-HST photo-$z$s are biased low in
this case, this is not in contradiction with our results given that the
\citetalias{schrabback18}
analysis is based on blue optically selected samples, which are on average significantly
fainter in the NIR compared to the galaxies studied here.
We interpret the various results such that a
noticeable fraction of catastrophic redshift outliers, in the form of
high-$z$ galaxies incorrectly assigned a low photo-$z$, can be present even
if NIR photometry is available, unless that has a  high signal-to-noise ratio.
We expect that
accounting for this effect  will also be relevant when calibrating redshift
distributions for wide-area weak lensing surveys, for example employing the
approach from  \citet{masters17}.
As the catastrophic outliers lead to a  bimodality of the colour-redshift
relation, highly complete spectroscopic redshift measurements will be needed
in the relevant parts of colour-colour space  to adequately map out this bimodality.

\subsection{Colour selection}
\label{se:colorselect}

The left panel of Fig.\thinspace\ref{fig:colordist_ultravista} shows the distribution of resolved galaxies with  \mbox{$21<K_\mathrm{s}^\mathrm{tot}<24.2$}  within the CANDELS/COSMOS 3D-HST grism area in $g-z$ versus $z-K_\mathrm{s}$ colour space, with different symbols indicating different ranges in $z_\mathrm{best}$.
The solid lines indicate our colour selection scheme, where we select
background galaxies that have
\begin{equation}
z-K_\mathrm{s}> \mathrm{min}[g-z,2.5] \, .
\end{equation}
This  selection is similar to the $BzK_\mathrm{s}$
selection introduced by \citet{daddi04}, but is slightly more conservative for
the exclusion of galaxies around the cluster redshift.
It is highly effective in selecting
most of the background galaxies at \mbox{$z_\mathrm{best}>1.4$}, while
efficiently removing galaxies at \mbox{$z_\mathrm{best}<1.1$} (see
Fig.\thinspace\ref{fig:zdist}).
In particular, \ha{98.1\% of the colour-selected galaxies are in the
  background at \mbox{$z_\mathrm{best}>1.1$}.
At the same time, 98.9\% of the galaxies in the parent catalogue at
relevant cluster redshifts
\mbox{$0.6<z_\mathrm{best}<1.1$} are removed by this colour
selection, providing an efficient suppression of cluster member contamination.
}

The right panel of Fig.\thinspace\ref{fig:colordist_ultravista} shows  the distribution
of galaxies in our HAWK-I+LBC shear catalogue in  $g-z$ versus $z-K_\mathrm{s}$ colour space prior to the colour selection.
In addition to the galaxy populations visible in the UltraVISTA-detected catalogue,
this prominently  displays the
population of cluster red-sequence galaxies around
\mbox{$g-z\simeq 3$} and \mbox{$z-K_\mathrm{s}\simeq 1$}.

\subsection{Average geometric lensing efficiency}
\label{se:beta}
\subsubsection{Best estimate}
\label{se:beta:best}
For the
mass measurements we need to
estimate the weighted-average geometric lensing efficiency (see Eq.\thinspace\ref{eqn:beta}) of our source sample.
We
started with the  colour- and  size-selected galaxies in the 3D-HST grism area
and computed
$\langle\beta\rangle_i$ from the  3D-HST \mbox{$z_\mathrm{best}$} redshifts
in
magnitude bins
 of width
  0.4
mag
within  \mbox{$21<K_\mathrm{s}^\mathrm{tot}<24.2$},
taking  the $K_\mathrm{s}^\mathrm{tot}$-dependent shape
weights into account (see Sect.\thinspace\ref{se:shapenoise}).
We then computed a joint estimate
\mbox{$\langle\beta\rangle_\mathrm{3D-HST}^\mathrm{grism-area}=
\left(\sum_i
  \langle\beta\rangle_i
\sum_{j(i)}w_j\right)/
\left(\sum_i
\sum_{j(i)}w_j\right)
=0.501
$}
according to the shape weights $w_j$ of the galaxies in magnitude bin $i$ in
our HAWK-I catalogue.
This procedure accounts for the greater incompleteness
of the HAWK-I catalogue given the lensing $S/N$ cut.

\begin{figure}
 \includegraphics[width=1\columnwidth]{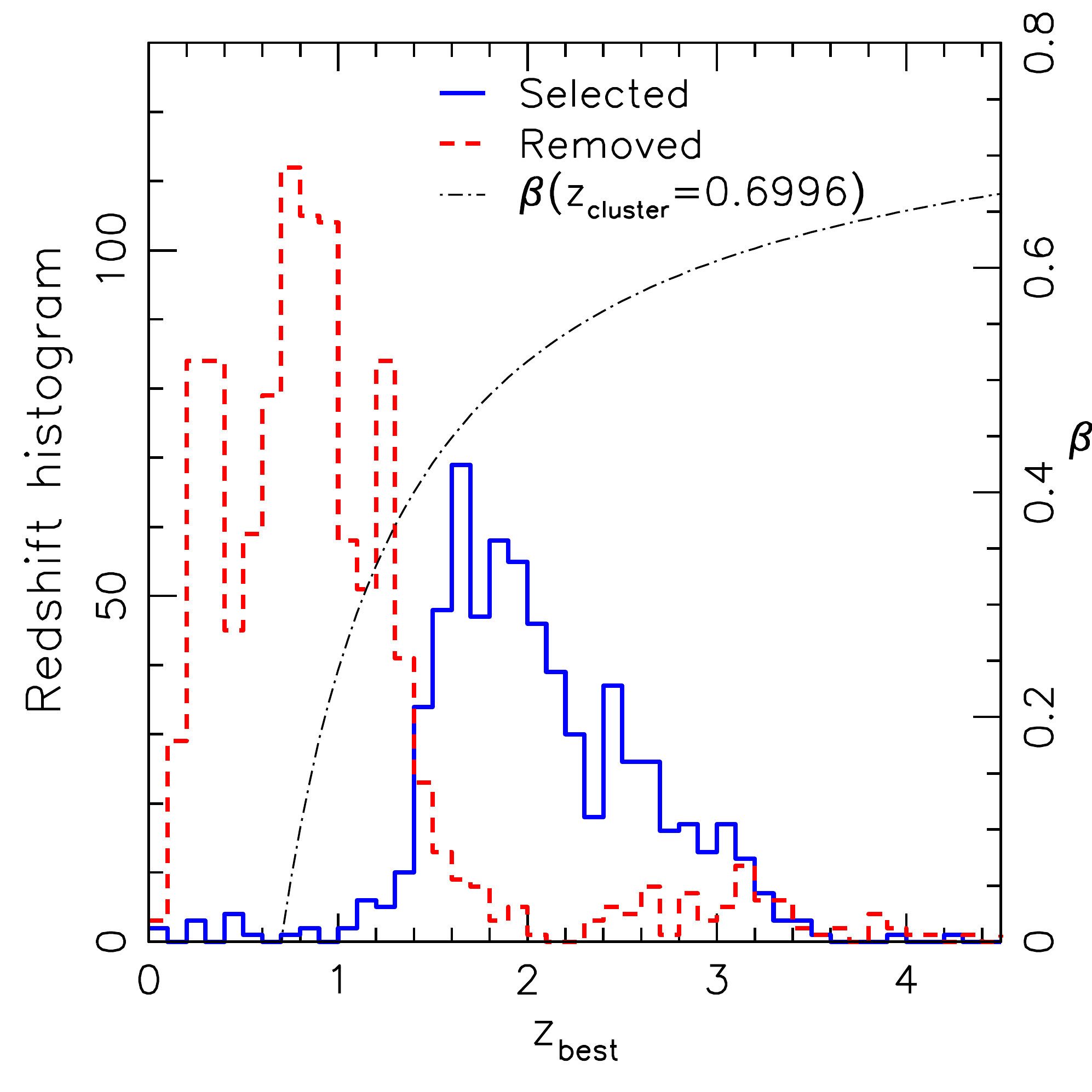}
 \caption{Histogram of the best 3D-HST redshift estimate for \ha{sufficiently} resolved
   galaxies with  \mbox{$21<K_\mathrm{s}^\mathrm{tot}<24.2$}  within the
   CANDELS/COSMOS 3D-HST grism area, split between galaxies selected and
   removed by our $gzK_\mathrm{s}$ selection. The dash-dotted curve shows
   the geometric lensing efficiency $\beta$ as a function of source redshift.
\label{fig:zdist}}
\end{figure}

We quantified and minimised the impact of sampling variance using  the  UltraVISTA-detected catalogue.
For this we
employed the same colour selection and weighting scheme as for the 3D-HST
catalogue, but this time we used the peak photometric
redshift $z_\mathrm{p}$ and dropped the size selection due to the lack of
HST NIR-measured sizes in COSMOS outside the CANDELS footprint.
We then computed estimates both for the full UltraVISTA ultra-deep area
(\mbox{$\langle\beta\rangle_\mathrm{UltraVISTA}^\mathrm{full}=
0.470
 $}) and
the 3D-HST grism area
(\mbox{$\langle\beta\rangle_\mathrm{UltraVISTA}^\mathrm{grism-area}=
0.490
  $}).
The latter covers the same area that was used for the analysis employing the
3D-HST \mbox{$z_\mathrm{best}$} redshifts. Accordingly, the
ratio
\mbox{$r_\mathrm{sys}=\langle\beta\rangle_\mathrm{UltraVISTA}^\mathrm{grism-area}/\langle\beta\rangle_\mathrm{3D-HST}^\mathrm{grism-area}=
0.978
  $}
provides us with a correction factor \ha{$r_\mathrm{sys}^{-1}$} to account for the impact of the
systematic redshift errors in the  UltraVISTA-detected catalogue discussed in Sect.\thinspace\ref{se:zcomp}.
This can be combined with the  estimate from the full UltraVISTA ultra-deep
area, which suffers less from sampling variance, to obtain our best
estimate of the cosmic mean geometric lensing efficiency given our selection
criteria of
\mbox{$\langle\beta\rangle_\mathrm{cor}=\langle\beta\rangle_\mathrm{UltraVISTA}^\mathrm{full}/r_\mathrm{sys}=
  0.481
  $}.

\subsubsection{Line-of-sight variations and \mbox{$\langle\beta^2\rangle$}}

The redshift distribution within the sky patch covered by our
HAWK-I
observation
\revb{likely
deviates} from the cosmic mean distribution because of
sampling variance. To obtain an estimate for this effect we placed 12 tiles of the same
area widely distributed over the  area of the UltraVISTA ultra-deep stripes.
From the variation between the \mbox{$\langle\beta\rangle$} estimates computed
\ha{from} these tiles, we estimated a relative uncertainty of
\hb{\mbox{$\Delta \langle\beta\rangle/\langle\beta\rangle=2.2\%$}}
for our analysis (for a single cluster\footnote{A potential future
scaling relation analysis that incorporates observations from a large number
of clusters would have a systematic uncertainty arising from line-of-sight
variations in the redshift distribution that is approximately reduced by a
factor \mbox{$1/\sqrt{12}\simeq 0.29$}, assuming large-scale structure at
high redshifts is sufficiently uncorrelated between the 12 tiles.}) arising from
line-of-sight
variations in the redshift distribution.

We accounted for \revd{the} impact of the  finite width of the source redshift
distribution in the lensing analysis following
\citet{hoekstra00}, for which we also require an estimate of the weighted
\mbox{$\langle\beta^2\rangle=0.237$},
which we computed based on the 3D-HST $z_\mathrm{best}$ redshifts (given the $z_\mathrm{p}$ outliers),
but rescaled with
the
factor \mbox{$\left(\langle\beta\rangle_\mathrm{UltraVISTA}^\mathrm{full}/\langle\beta\rangle_\mathrm{UltraVISTA}^\mathrm{grism-area}\right)^2$}
to account for the impact of sampling variance.

\subsubsection{Systematic uncertainties}
\label{se:betauncer}
\ha{The  3D-HST-derived $\langle\beta\rangle$ estimates are
expected to be highly robust, as they are mostly based on
accurate grism or spectroscopic redshifts (to \hb{\mbox{$\sim 71\%$}} when
accounting for our weighting scheme, see  Sect.\thinspace\ref{se:3dhst}).
However,}
we cannot fully exclude the possibility
that the \hb{\mbox{$\sim 29\%$}} contribution from 3D-HST photo-$z$ may introduce
systematic uncertainties because of photo-$z$ biases.
To obtain an approximate estimate for this uncertainty,
we recomputed
\mbox{$\langle\beta\rangle_\mathrm{3D-HST}^\mathrm{grism-area}$} using the
3D-HST photometric redshifts for all galaxies, hence using 100\%
photo-$z$ information instead of \hb{29\%}.
This leads to a very small \ha{relative} increase in \mbox{$\langle\beta\rangle$} by
0.4\%.
The expected systematic uncertainty associated with the use of \hb{\mbox{$\sim
  29\%$}} photo-$z$ uncertainty, on the one hand, would be lower than this
number given the smaller fraction of employed photo-$z$s, but, on the other hand, would be larger given
that these galaxies are typically fainter. Considering both aspects, we expect that
0.4\% likely corresponds to a reasonably realistic estimate of the resulting residual uncertainty.

Additional systematic biases in  \mbox{$\langle\beta\rangle$}  may arise
from mismatches in the photometric calibration or \ha{matching} of noise
properties.
To quantify the impact of the former, we tested the sensitivity to systematic errors in
the colour measurements.
We find that a systematic
error in  $g-z$ or $z-K_\mathrm{s}$ colour of 0.1 mag, which provides a conservative
estimate for  the
uncertainty in the
colour calibration,
leads to a relative bias
in  \mbox{$\langle\beta\rangle$} of only
0.5\%.

The matching of noise properties is complicated by the fact
that our HAWK-I+LBC observations are slightly shallower in the
$K_\mathrm{s}$ band than the
reference catalogue, but  deeper in
the $z$ band (see
Sect.\thinspace\ref{se:ultravista}).
Hence, we cannot simply add noise to the colours in the reference
catalogue as performed for $K_\mathrm{s}^\mathrm{tot}$.
However, since the colour selection already achieves an excellent
selection of background galaxies  at the depth of the UltraVISTA-detected
catalogue (Fig.\thinspace\ref{fig:zdist}), we expect that this is also the case
for colour estimates with slightly higher signal-to-noise ratio.
In order to roughly estimate the sensitivity of our analysis to noise in the
colour measurements, we randomly added Gaussian scatter corresponding to a
depth difference of 0.3 mag separately to the $g$, $z$, and $K_\mathrm{s}$
fluxes of the UltraVISTA-detected
catalogue, finding that this leads to relative changes in
\mbox{$\langle\beta\rangle$} of
$+0.0\%$, $-0.2\%$, and $-0.1\%$,
respectively.
Biases at these levels are completely negligible compared to the statistical uncertainties  of our
study.
Added in quadrature, the systematic errors for the \mbox{$\langle\beta\rangle$} estimate identified in this subsection
amount to
0.7\%.

\subsection{Choice of centre}
\label{se:centre}

For our weak lensing \ha{shear profile} analysis we have to adopt a centre. This should match the
position of the centre of the projected mass distribution as best as
possible to minimise miscentring uncertainties \citep[see
e.g.][]{schrabback18}.
For \clusteras\, the centre of the inner projected mass distribution is very
well constrained by strong gravitational lensing
to a location
 $1\farcs17^{+0\farcs47}_{-0\farcs24}$ east and
 $7\farcs42^{+1\farcs42}_{-0\farcs63}$ north from the brightest cluster galaxy (BCG),
in the direction  towards the second brightest cluster galaxy
\citep{sharon15}.
This very small positional uncertainty is completely negligible for weak
lensing studies \citep[e.g. compare~to][]{vonderlinden14}. We therefore fix the centre position for our analysis to the
best-fitting
centre position of the strong lensing analysis
from   \citet{sharon15}
at \mbox{$(\alpha,\delta)=(351.865351,-2.074863)$ deg}.
\subsection{Number density profile}
\label{se:nofr}
\begin{figure}
 \includegraphics[width=1\columnwidth]{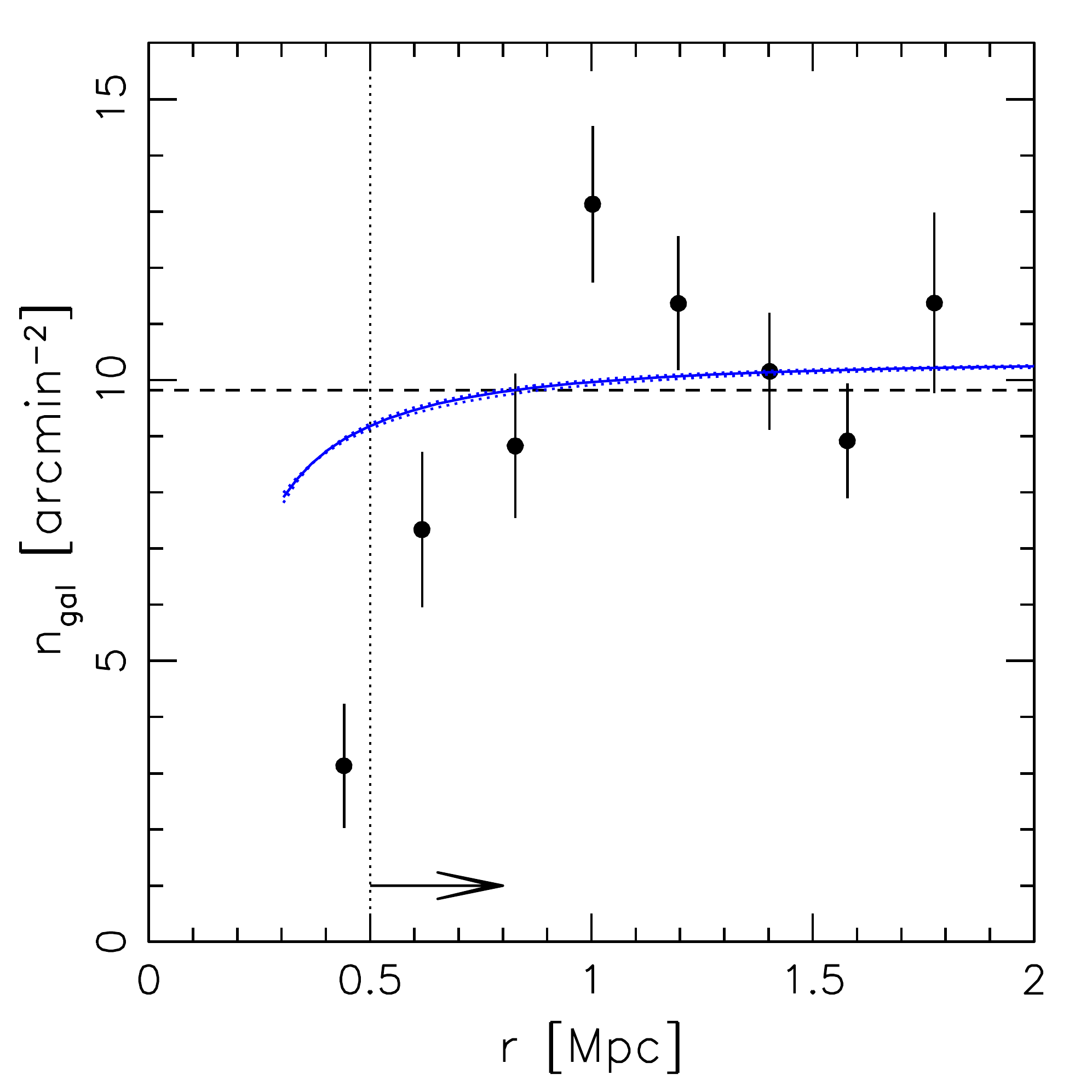}
 \caption{Source density in our colour- and magnitude-selected weak lensing
   source catalogue for \clusteras\,as a function of projected distance from
   the cluster centre, taking field boundaries, manual masks, and a bright
   objects mask
 into account.
\ha{Error bars are underestimated, as they assume Poisson galaxy counts ignoring spatial clustering.
The dashed black line indicates the average density over the whole field of view,
while the
\reva{blue curves indicate} the approximately expected profile due
to lensing magnification \reva{assuming the best-fitting NFW model for \mbox{$c_\mathrm{200c}=5.1$}
    (solid) or  \mbox{$c_\mathrm{200c} \in \left[4.1, 6.1\right]$} (dotted,
    close to the solid curve)}.
 The vertical \reva{black} dotted line and the arrow indicate
the lower radial limit in the weak lensing shear profile fit.}
\label{fig:ngalr}}
\end{figure}

As shown in Sect.\thinspace\ref{se:colorselect}
our colour selection is expected to lead to a negligible residual
contamination by cluster galaxies in the source sample.
As a consistency check for this, we investigated
the radial source number density profile.
Because of the central concentration of cluster galaxies, a substantial
 residual contamination would be detectable as an increase in the source density
 towards the centre.
For our catalogue we do not detect such a central increase.
As shown in Fig.\thinspace\ref{fig:ngalr}, the source
density profile is
\ha{approximately} flat for radii \mbox{$r\gtrsim 0.6$ Mpc} with a \ha{global} mean
density of 9.8 arcmin$^{-2}$.

Further into the cluster core the observed source density drops (see Fig.\thinspace\ref{fig:ngalr}).
We suspect that this may be  due to a
combination of two effects.
First, we cannot detect faint background galaxies behind or close to a bright
foreground cluster galaxy.
In order to account for this effect at least
approximately, we used a  bright  \hb{objects} mask for the sky area
calculation \reva{(already taken into account in
  Fig.\thinspace\ref{fig:ngalr}, causing a \mbox{$\sim 7\%$} correction in
  the inner bins together with the manual masks)}.
We created this by
running \texttt{SExtractor} with a high object detection threshold of
200 \revd{pixels exceeding} the background by $1.5\sigma$ and
then used the ``objects'' check image as a mask.
However, as this mask
\ha{neither accounts for fainter cluster members
nor the
  outer wings of galaxy light profiles or the impact of intra-cluster light}, it  likely still leads
to
an underestimation of the inner source density.

Second, we suspect that lensing magnification may also lead to a net
depletion in the density of faint sources.
This  has the largest impact in the
stronger
magnification regime of cluster cores
\citep[see e.g.][]{fort97}.
Assuming source counts described by a power law and sources at a single redshift, magnification leads to a net depletion in the source counts
if the slope of the logarithmic cumulative number counts is shallow,
\begin{equation}
s=
\frac{\mathrm{d}\log_{10}N(<m)}{\mathrm{d}m}<0.4\,
\end{equation}
\reva{\citep[e.g.][]{broadhurst95,mayen00}}.
We computed this slope for the colour-selected UltraVISTA-detected catalogue
around \mbox{$m=K_\mathrm{s}^\mathrm{tot}\simeq 24$ mag,}
yielding \mbox{$s=0.32\pm
  0.02$} assuming negligible incompleteness,
which is indeed consistent with an expected depletion.
\ha{Making the same simplifying assumptions we plot the expected source density
profile resulting from magnification as solid blue curve in
Fig.\thinspace\ref{fig:ngalr}, employing the best-fit NFW density profile from our reduced shear profile
fit (see Sect.\thinspace\ref{se:masscon}).
This indicates that magnification alone likely cannot explain the very low source
density at \mbox{$r\simeq 0.45$ Mpc}, but that additional effects, such as
the limitations in the bright objects mask may}
dominate.
\reva{In addition, it may just be that the line of sight behind the core of
  \clusteras\, is  noticeably underdense.
In this respect,}
  the error bars shown in Fig.\thinspace\ref{fig:ngalr} assume
Poisson source counts but ignore spatial clustering, which underestimates the
true uncertainty and therefore overestimates the significance of the data
point, \reva{as especially relevant at small radii.}

\subsection{Shape noise and shape weights}
\label{se:shapenoise}

\begin{figure}
 \includegraphics[width=1\columnwidth]{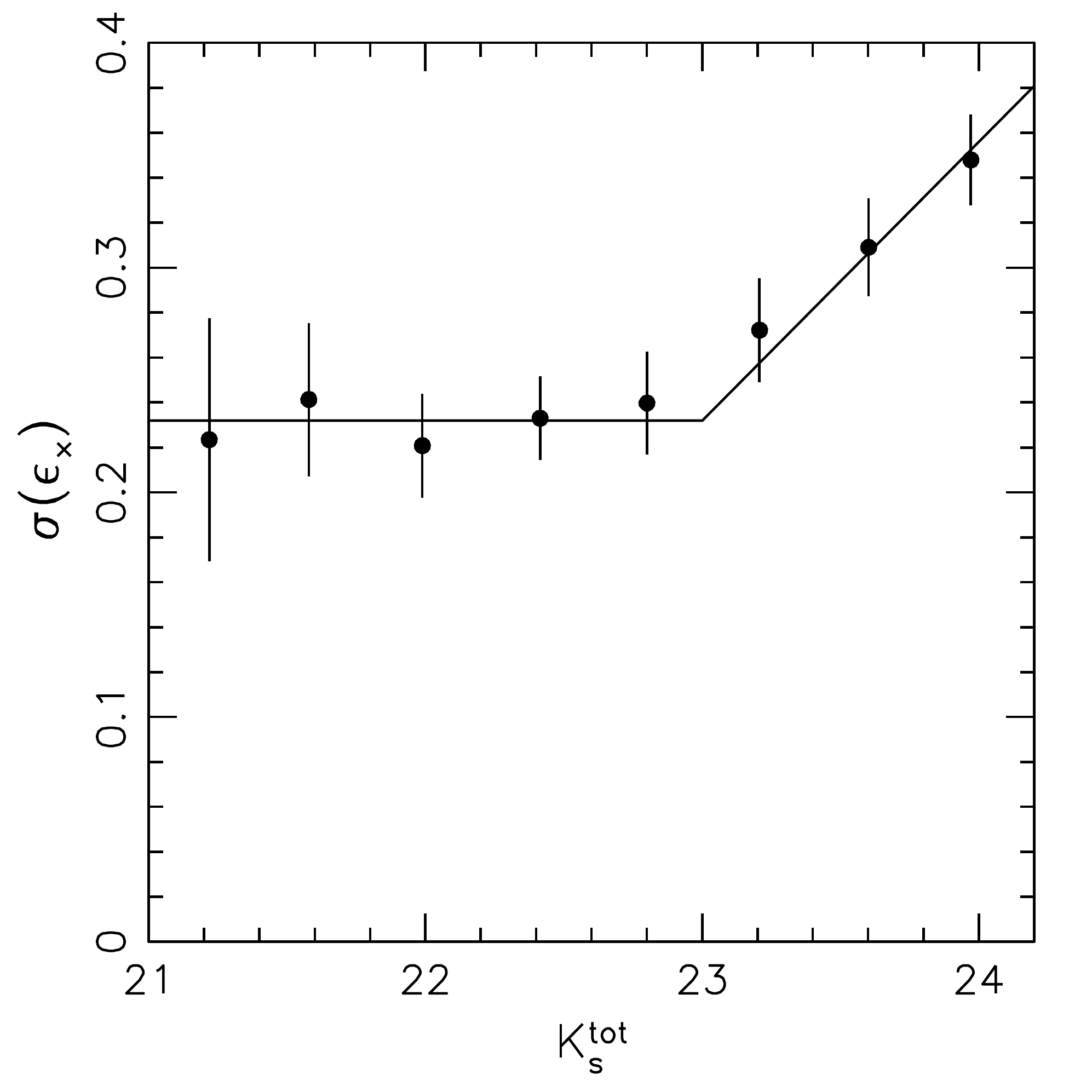}
 \caption{\hb{Dispersion}
of the cross-ellipticity component with respect
   to the cluster centre computed in bins of $K_\mathrm{s}^\mathrm{tot}$
including all lensing and colour-selected galaxies with a
projected separation \mbox{$r>700$ kpc} from the cluster centre. The solid line shows our approximate fit that
is used to define shape weights.
\label{fig:rmsex}}
\end{figure}

\ha{At fixed redshift, fainter sources tend to result in more noisy shear
  estimates than bright sources, for two reasons. First, the higher measurement
  noise leads to more noisy ellipticity measurements. Second, as shown by
  \citetalias{schrabback18}, in optically selected samples the dispersion of
  the \emph{intrinsic} source ellipticity increases at faint magnitudes,
  further increasing the noise in the shear estimate.
As we show below and discuss in Sect.\thinspace\ref{se:dis:perf}, the
$K_\mathrm{s}$ imaging yields shape estimates for high-$z$ galaxies with a lower measured
ellipticity dispersion, indicating a lower intrinsic ellipticity dispersion
than for optical high-$z$ samples.}

\ha{To account for
the more noisy shear estimates at faint magnitudes,
\citetalias{schrabback18} employed}
an empirical weighting scheme according to the
ellipticity dispersion
measured in non-cluster fields as a function of magnitude.
Given the presence of a massive cluster, which significantly shears the background galaxy images, we cannot directly apply the same approach here.
However, as the cluster lensing signature primarily affects the tangential ellipticity component $\epsilon_\mathrm{t}$ with respect to the cluster centre, but not the cross-component  $\epsilon_\times$, we can use the measured
dispersion of the cross-ellipticity component
\hb{\mbox{$\sigma_{\epsilon,\times}= \sigma(\epsilon_\times)$}}
as a function of $K_\mathrm{s}^\mathrm{tot}$
(shown in Fig.\thinspace\ref{fig:rmsex})
to define the weighting scheme.
We find that
 $\sigma_{\epsilon,\times}(K_\mathrm{s}^\mathrm{tot})$
is approximately flat for
\mbox{$K_\mathrm{s}^\mathrm{tot}<23$} with
\hb{
\begin{equation}
\label{eq:seksbright}
  \sigma_{\epsilon,0}\equiv
  \sigma_{\epsilon,\times}\left(21<K_\mathrm{s}^\mathrm{tot}<23\right)
=0.232\pm 0.011
  \end{equation}
}
and increases approximately linearly
as
\hb{
\begin{equation}
\sigma_{\epsilon,\times}\left(K_\mathrm{s}^\mathrm{tot}\right)
 =\sigma_{\epsilon,0}
  +\left(0.124\pm
  0.009\right) \left(K_\mathrm{s}^\mathrm{tot}-23\right)
   \,\,\,\,  \mathrm{for}\,\,  K_\mathrm{s}^\mathrm{tot}>23 \, .
\end{equation}
}
\ha{We expect that this increase is mostly caused by measurement noise, but
  we cannot exclude a possible contribution from an
  increase in the intrinsic ellipticity dispersion at faint magnitudes.}
We  use \mbox{$w\left(K_\mathrm{s}^\mathrm{tot}\right)=\sigma_{\epsilon,\times}^{-2}\left(K_\mathrm{s}^\mathrm{tot}\right)$}
as shape weight.

The
$K_\mathrm{s}$-measured ellipticity dispersion is significantly lower than what has been found by
\citetalias{schrabback18} for galaxies at similar redshifts with a largely identical shape measurement pipeline analysing  optical HST/ACS images of approximately single-orbit depth.
At a relatively bright magnitude \mbox{$V_\mathrm{606,auto}=25$}, where
the contribution from measurement noise is small,
\citetalias{schrabback18} estimate \mbox{$\sigma_\epsilon=0.306$}
for a \mbox{$V_{606}-I_{814}<0.3$} colour-selected sample. This is
significantly larger than the $K_\mathrm{s}$-measured $\sigma_\epsilon$ at
bright magnitudes (Eq.\thinspace\ref{eq:seksbright}).

\subsection{Comparison to HST/ACS weak lensing shear estimates}
\label{se:acscomp}

\begin{figure}
 \includegraphics[width=1\columnwidth]{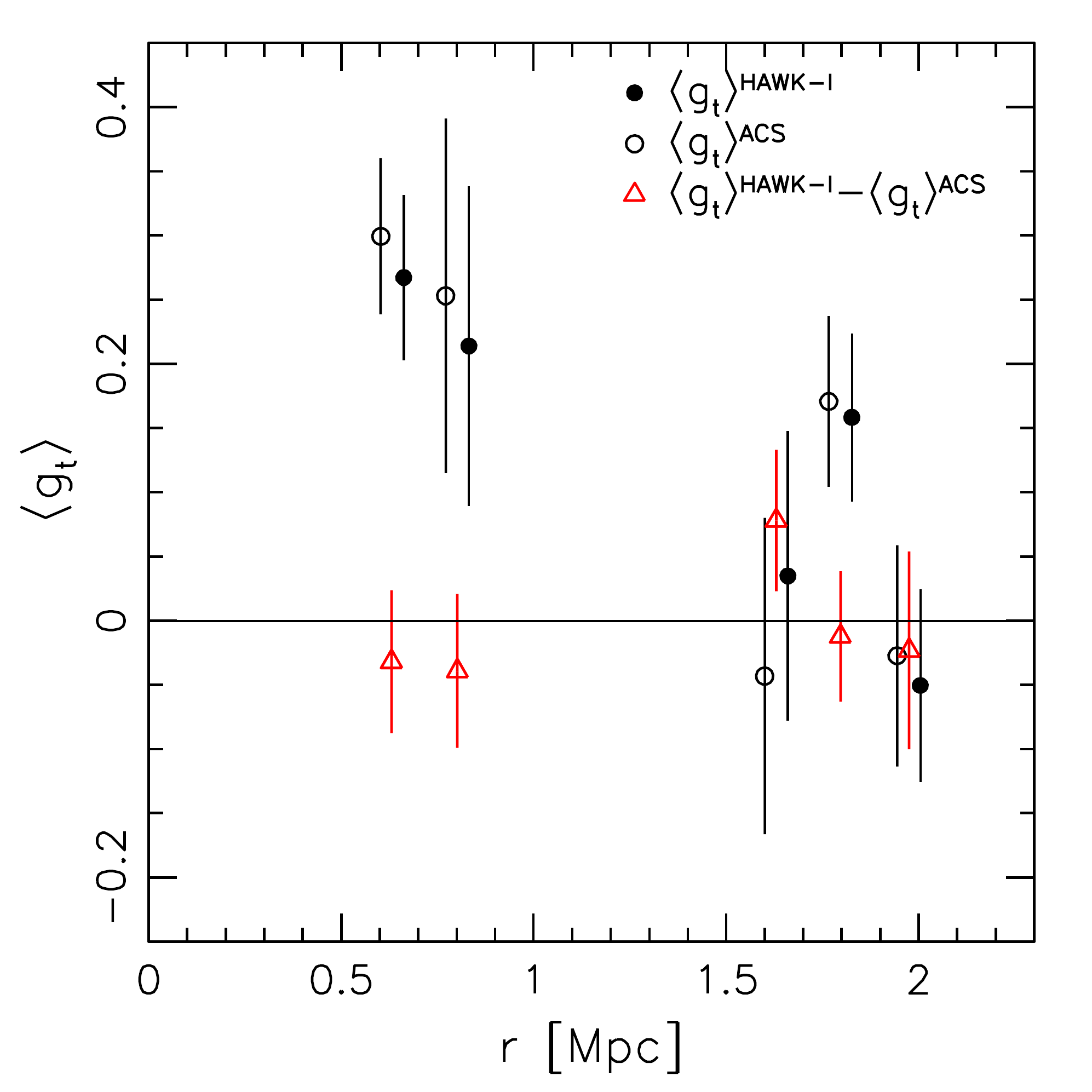}
 \caption{Profile of the
estimated
   tangential  reduced shear for
{\clusteras},
based on the matched HAWK-I and ACS ellipticity catalogue,
employing the HAWK-I+LBC colour selection and uniform weights. \ha{We show all radial bins containing at least five galaxies.
}
The solid
(open) points are based on the HAWK-I (ACS) ellipticity measurements, shown with an offset of $+30$ kpc ($-30$ kpc) for clarity.
\ha{The red open triangles indicate\ the difference between the two estimates with error bars determined by bootstrapping the sample.}
Matched data are only available in the central ACS pointing and near the
corners of the HAWK-I field of view. The resulting smaller area
and lower source density leads to more noisy data
compared to the analysis of the full HAWK-I+LBC-based catalogue (compare
Fig.\thinspace\ref{fig:gtgx})
\ha{and introduces the gap at
intermediate radii}.
\label{fig:acscomp}}
\end{figure}

\begin{figure*}
 \includegraphics[width=2\columnwidth]{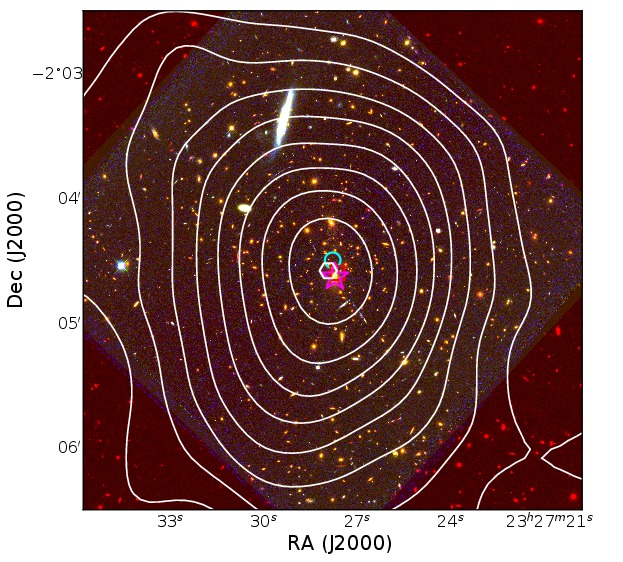}
 \caption{RGB colour image of the central \mbox{$4^\prime\times 4^\prime$}
   of \clusteras\, created from the  VLT/HAWK-I  $K_\mathrm{s}$ best-seeing stack
   and the HST/ACS F814W and F435W images. The contours indicate the weak
   lensing convergence reconstruction  starting at
\mbox{$\kappa_0=0.04$} in steps of \mbox{$\Delta\kappa=0.04$} with the peak
indicated by the white hexagon. The magenta star, red square, and cyan
circle indicate the locations of the BCG, the peak in the X-ray \ha{emission}, and
the strong lensing centre from \citet{sharon15}, respectively.
\label{fig:massrecon}}
\end{figure*}

To
 cross-check our HAWK-I shear estimates we compared these to measurements from overlapping HST/ACS observations (see Sect.\thinspace\ref{se:data:acs}).
For the  ACS catalogue generation we employed the same basic KSB+ implementation as for the
HAWK-I shape measurements (see Sect.\thinspace\ref{se:hawkshapes}), but additionally included the principal component
PSF interpolation from \citet{schrabback10} \citep[building on][]{jarvis04} and the PSF model calibration and shape weighting scheme from \citetalias{schrabback18}.
For the central ACS pointing, the weak lensing catalogue generation has also been described in \citet{hoag15}.

When comparing shape measurements obtained with different resolution and/or
in different band passes, a direct comparison of ellipticity estimates is
not an adequate metric, as the spatial distribution of the light emission may not be
identical and different effective radial weight functions are
used.
This is underlined by the indications we find for a significantly lower
intrinsic ellipticity dispersion for the analysis based on $K_\mathrm{s}$
imaging compared to ACS optical imaging  (see Sect.\thinspace\ref{se:shapenoise}).
Nevertheless, what should be consistent is the estimated reduced tangential
cluster shear profile when a matched catalogue with identical weights is used.
This is shown in Fig.\thinspace\ref{fig:acscomp},
where we employ the HAWK-I+LBC colour selection and uniform weights for the
galaxies in the matched HAWK-I and ACS ellipticity catalogue.
\ha{As the difference in the reduced shear estimates \mbox{$\langle g_\mathrm{t}\rangle^\mathrm{HAWK-I}-\langle g_\mathrm{t}\rangle^\mathrm{ACS}$}
  is consistent with zero, we conclude that the HAWK-I and ACS measurements are fully consistent within the current statistical uncertainty.}

\section{Cluster weak lensing results}
\label{se:results}

\subsection{Mass reconstruction}
\label{se:massrecon}

We reconstructed the convergence ($\kappa$) distribution of \clusteras\, on a grid,
using an improved
version of the \citet{kaiser93}
formalism, which applies a  Wiener filter
as described in
\citet{mcinnes09} and
\citet{simon09}, and as further detailed in
\citetalias{schrabback18}.
\ha{Given the mass-sheet degeneracy we cannot constrain the average
  convergence in the field of view. We fixed it to zero, which is adequate for
  large}
 \hb{fields of view},
\ha{but likely
\revd{leads}
 to an underestimation for our data.}
\reva{This uncertainty is however not a concern for our analysis as we use the mass reconstruction only for illustration and
   consistency checks regarding the location of the cluster centre.}
Given the high cluster mass we apply an iterative reduced-shear correction
\hb{\citep[e.g.][]{seitz96}
based on} the  $\kappa$ distribution from the previous iteration.
Fig.\thinspace\ref{fig:massrecon} shows contours of the resulting
reconstruction starting at
\mbox{$\kappa_0=0.04$} in steps of \mbox{$\Delta\kappa=0.04$}, with a peak value
 \hb{\mbox{$\kappa_\mathrm{max}=0.347$}}.

\begin{figure}
 \includegraphics[width=1\columnwidth]{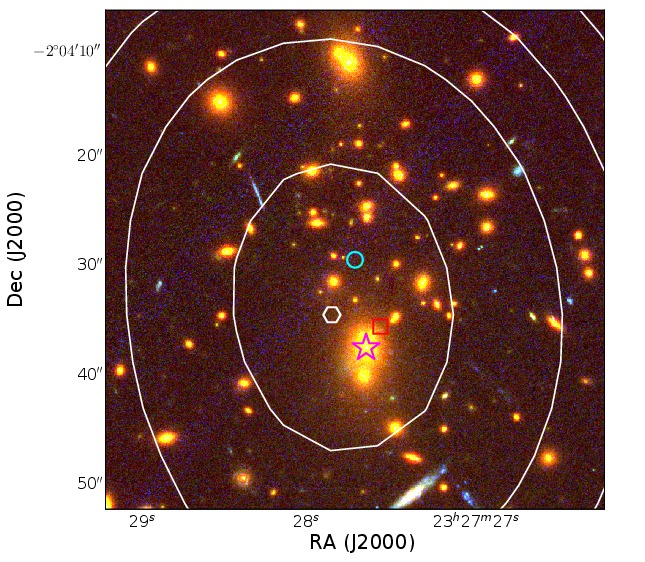}
 \caption{\reva{As Fig.\thinspace\ref{fig:massrecon}, but showing a cut-out
of the central
\mbox{$45^{\prime\prime}\times 45^{\prime\prime}$}
with
contours  in steps of  \mbox{$\Delta\kappa=0.02$}, where the innermost
contour corresponds to \mbox{$\kappa=0.34$}.}
\label{fig:massrecon_zoom}}
\end{figure}

To estimate the peak significance we
apply the same reconstruction algorithm to
 noise catalogues generated  by randomising the ellipticity phases.
Dividing the reconstruction from the real data through the r.m.s. image
of the noise
reconstructions we estimate a
\hb{$10.1\sigma$}
peak significance.
In Fig.\thinspace\ref{fig:massrecon} the contours are overlaid on an RGB colour
image based on the HAWK-I $K_\mathrm{s}$ and the ACS F814W and F435W images,
with indications of the BCG, as well as the  strong lensing centre and the peak
of the X-ray emission from \citet{sharon15}.
The peak of the weak lensing $\kappa$-reconstruction at
\hb{\mbox{$(\alpha,\delta)_\mathrm{peak}=(351.86594,-2.07626)$ deg}}
is  constrained
to
\hb{\mbox{$(\Delta\alpha,\Delta\delta)_\mathrm{peak}=(3\farcs2,5\farcs7)$}}
 as
estimated by bootstrapping the source catalogue,
making it consistent with the locations of the BCG, X-ray centre, and
strong lensing centre
\reva{(see also Fig.\thinspace\ref{fig:massrecon_zoom})}  within \mbox{$\sim 1\sigma$}.

\subsection{\hb{Reduced} shear profile analysis and mass constraints}
\label{se:masscon}

\begin{figure}
 \includegraphics[width=1\columnwidth]{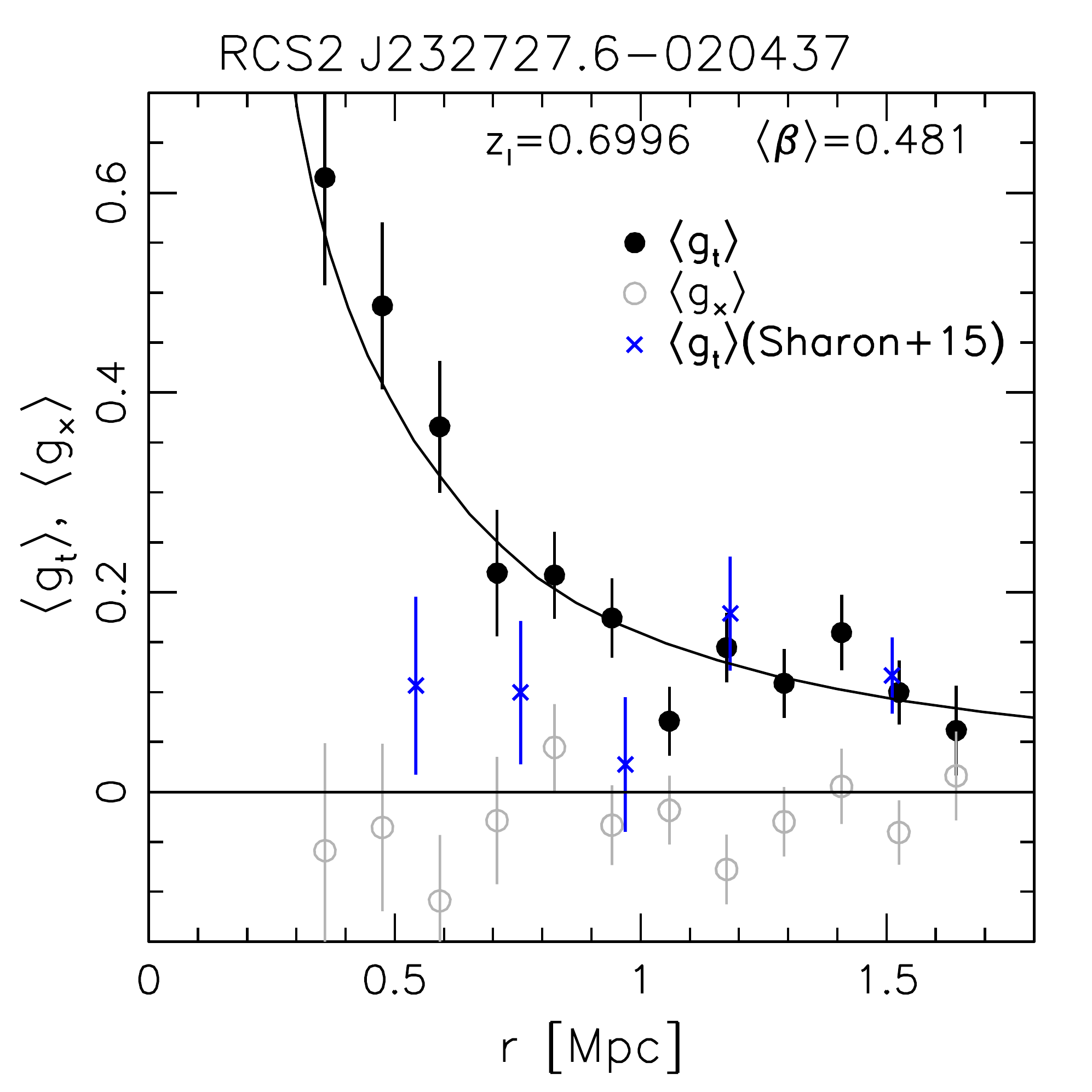}
 \caption{Profile of the tangential reduced shear (filled circles) and the
   45 degrees-rotated cross-component (open circles) for \clusteras\, as
   function of cluster-centric separation.
The solid curve shows the best-fitting NFW
 model prediction for a fixed concentration \mbox{$c_\mathrm{200c}=5.1$}
when considering scales \mbox{$500\,\mathrm{kpc}<r<1.6\,\mathrm{Mpc}$}.
The blue crosses indicate
 tangential  reduced  shear estimates from \citet{sharon15} based on deep
 CFHT weak lensing measurements,  scaled to the
 same $\langle\beta\rangle$ \hb{and excluding points at small radii  that are
   not included in their fit. \citet{sharon15} also incorporate measurements at
   larger radii that are not shown here.}
\label{fig:gtgx}}
\end{figure}

\reva{Weak lensing measurements
can provide non-parametric estimates of projected cluster masses via the
aperture mass statistic \citep[e.g.][]{hoekstra15} if the lensing signal is measured well beyond the
cluster virial radius.
As the HAWK-I field of view does not provide such a large radial coverage
for
 \clusteras , we instead have to rely on model fits of the cluster   tangential reduced shear
  profile  to constrain the cluster} \reva{mass.
This effectively breaks the mass-sheet degeneracy discussed in
Sect.\thinspace\ref{se:massrecon}.
In practise, such idealised mass sheets
are
related to correlated and uncorrelated
large-scale structure projections.
The net impact of such projections
for weak lensing  mass estimates
is additional scatter,
as computed and discussed below.}

We show the tangential reduced shear profile of \clusteras\, \ha{as a function
of separation from the strong lensing centre\footnote{\ha{We do not centre on the peak
of the weak lensing mass reconstruction from Sect.\thinspace\ref{se:massrecon} as this is expected to
yield mass constraints that are biased high \citep[e.g.][]{dietrich12}. However, this would likely be
a minor effect given our very high-significance detection.}} (see Sect.\thinspace\ref{se:centre})} as estimated from our HAWK-I+LBT catalogue in Fig.\thinspace\ref{fig:gtgx}.
We fit these data using reduced shear profile predictions from \citet{wright00} assuming a
spherical NFW density profile \citep{navarro97}.
We only consider  radii in the range
\mbox{$500\,\mathrm{kpc}<r<1.6\,\mathrm{Mpc}$}.
At smaller radii the measured tangential reduced shear
 exceeds the regime tested in the weak lensing image simulations (see
 Sect.\thinspace\ref{se:hawkshapes}).
At larger scales the azimuthal coverage gets increasingly incomplete.

The weak lensing data alone cannot constrain the cluster concentration
\mbox{$c_\mathrm{200c}$}
sufficiently well, which is why we revert to results from numerical
simulations.
Using a suit of simulations, \citet{diemer15} provided a well-calibrated prescription to
compute the expected mean halo concentration as a function of mass, which
would be adequate for a general cluster.
However,  the X-ray analysis from \citet{sharon15} indicated that \clusteras\, is a fairly relaxed cluster,
which is why, on average, a higher  concentration should be expected than for a general cluster.
\cite{neto07} investigated the difference in structural parameters for
relaxed versus general simulated dark matter haloes at redshift \mbox{$z=0$}.
They find that haloes
at the mass-scale of \clusteras\,
have on average larger median concentrations compared to general haloes by a factor 1.16.
Assuming that this factor also holds at higher redshifts, we conducted a two-step fit for \clusteras:
first, we fit the data assuming the concentration--mass relation from
\citet{diemer15}, yielding a best-fit cluster mass that corresponds to a
mean \mbox{$c_\mathrm{200c,D15}=4.4$}.
Based on the results  from \cite{neto07} we then repeated the fit assuming a
larger
concentration
\mbox{$c_\mathrm{200c}=1.16\,c_\mathrm{200c,D15}= 5.1$} yielding
\begin{equation}
  \label{eq:m200}
M_\mathrm{200c}/(10^{15} \mathrm{M}_\odot)
=2.06^{+0.28}_{-0.26}(\mathrm{stat.})\pm 0.12 (\mathrm{sys.})\,,
\end{equation}
where the statistical error contains contributions
added in quadrature
  from
\hb{shape noise ($^{+0.21}_{-0.20}\times 10^{15} \mathrm{M}_\odot$),}
large-scale structure projections
\hb{($\pm 0.12\times 10^{15} \mathrm{M}_\odot$)}
as estimated in \citetalias{schrabback18},
line-of-sight variations in the source redshift distribution ($\pm 0.07\times 10^{15} \mathrm{M}_\odot$; see Sect.\thinspace\ref{se:beta}),
and the impact of the uncertainty in the concentration ($^{+0.12}_{-0.10}\times 10^{15} \mathrm{M}_\odot$).
We derive the latter uncertainty from the estimated scatter in the logarithm of the concentration \mbox{$\sigma(\mathrm{log}_{10}c_\mathrm{200c})=0.061$} for high-mass relaxed haloes as found by \cite{neto07}.
The systematic error in Eq.\thinspace\ref{eq:m200} is dominated by the shear calibration ($\pm 0.12\times 10^{15} \mathrm{M}_\odot$; see Sect.\thinspace\ref{se:hawkshapes}) with a minor contribution from  the systematic uncertainty of the $\langle\beta\rangle$ estimate ($\pm 0.02\times 10^{15} \mathrm{M}_\odot$; see Sect.\thinspace\ref{se:betauncer}).
Based on the \mbox{$M_\mathrm{200c}$} limits and fixed concentration we also
report mass constraints for an overdensity \mbox{$\Delta=500$}
of
\hb{\begin{equation}
  \label{eq:m500}
  M_\mathrm{500c}/(10^{15} \mathrm{M}_\odot) = 1.50^{+0.19}_{-0.17}
  (\mathrm{stat.})\pm 0.09 (\mathrm{sys.}) \,,
\end{equation}
}
taking the same sources of uncertainty into account.
The sensitivity to the uncertainty in the concentration is lower for
$M_\mathrm{500c}$ (3\% relative uncertainty) than for $M_\mathrm{200c}$ (5\% relative uncertainty).
\ha{While the weak lensing data cannot constrain the radii corresponding to
  the considered overdensities $\Delta$ separately, we list the best-fitting values
\mbox{$r_\mathrm{200c}=2.03$ Mpc}  and \mbox{$r_\mathrm{500c}=1.34$ Mpc} given the assumed
concentration to simplify possible mass comparisons in future studies.}

\reva{Our assumptions regarding the concentration--mass relation
  are also  consistent with recent findings from the CLASH project
  \citep{postman12}.
In particular, the constraints derived by \citet{umetsu16}
on the concentration--mass relation of massive clusters
using combined strong lensing, weak lensing, and magnification measurements
are fully consistent with the \citet{diemer15} relation, which we use as a
basis to estimate the mean concentration for a general cluster population as function of mass and redshift.
\citet{meneghetti14} found a higher average
concentration for simulated clusters with regular X-ray morphologies resembling a
subset of the   CLASH clusters, similar to the results from \citet{neto07} for relaxed haloes.
While most CLASH clusters are  at significantly lower
 redshifts compared to \clusteras,  limiting
a direct comparison,
there are two CLASH clusters with  a similar or higher redshift (MACS\thinspace$J0744+39$
and CL\thinspace$J1226+3332$). For these clusters \citet{merten15} estimated
concentrations \mbox{$c_\mathrm{200c}=4.1\pm 1.0$} and
\mbox{$c_\mathrm{200c}=4.0\pm 0.9$}, respectively, in reasonable agreement with the simulation-based priors assumed in our analysis.
}

\ha{For  a pure lensing signal the 45 degrees-rotated  cross-component, shown as the open circles in Fig.\thinspace\ref{fig:gtgx},  should be consistent with zero.
The measured
signal  appears to be
\hb{slightly}
negative,
with a significance at the
\hb{\mbox{$1.9\sigma$}}
level when all data points at \mbox{$r>1$ Mpc}
are considered.
This could possibly indicate the presence of residual systematics, for example from incomplete PSF anisotropy correction, which is typically referred to as  additive shape measurement bias.
While our employed basic KSB+ implementation was among the methods with the lowest additive biases in the blind test analysis from \citet[][]{heymans06}, there are simplifying assumptions
in the KSB+ approach that may break down for complex PSFs \citep[e.g.][]{kaiser00}.}
As a \ha{sensitivity test  to investigate if this can have a significant impact on our analysis, we
artificially  doubled the level of the   PSF anisotropy correction.
This reduces  the  significance of the negative cross-component to
\hb{\mbox{$1.1\sigma$},}
but has only a very minor
\hb{+2.7\%}
 impact on the estimated cluster mass.
Compared to the statistical uncertainty we conclude that possible PSF anisotropy residuals are therefore of no concern for our current study. Potential future investigations with  larger samples will be able to test
for possible residual systematics with a higher sensitivity. If detected, such analyses could revert to alternative shape estimation techniques, which do not rely on simplifying assumptions regarding the PSF \citep[e.g.][]{melchior11}.
}

\subsection{Comparison to results from previous studies}

\begin{figure}
 \includegraphics[width=1\columnwidth]{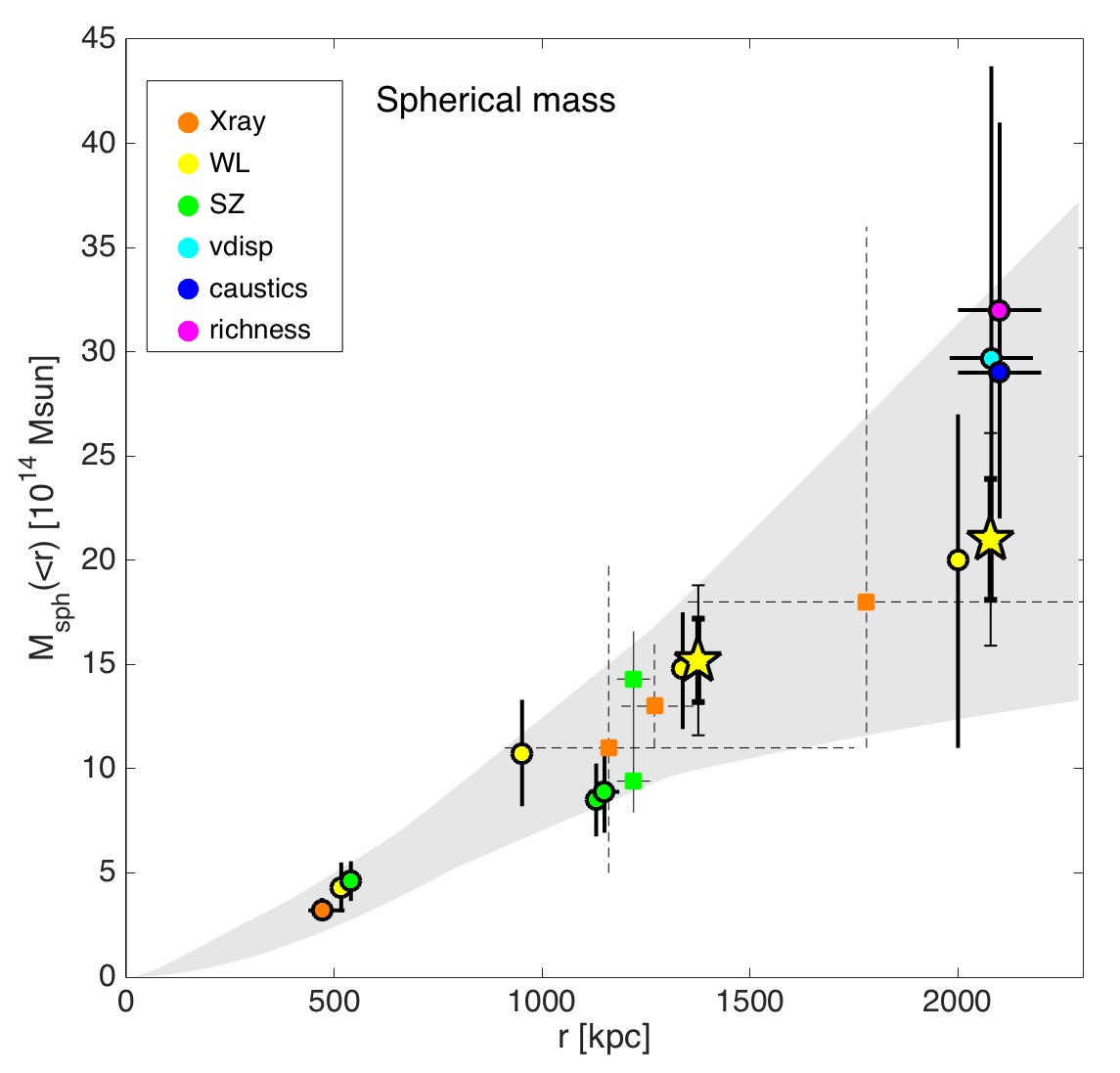}
 \caption{\reva{Updated version of Fig.\thinspace 16 from \citet{sharon15}, showing various estimates for the enclosed spherical mass of \clusteras\, as function of radius. The  stars-shaped data points show our  weak lensing measurements, recomputed for \mbox{$\Omega_\mathrm{m}=0.27$} and  \mbox{$\Omega_\Lambda=0.73$} as assumed by  \citet{sharon15}.
     The thick (thin) error bars correspond to our combined statistical and systematic uncertainty without (with) including an additional \mbox{$\sim 20\%$} intrinsic scatter from cluster triaxiality and correlated large-scale structure.
     The green squares show SZ mass estimates from \citet{hasselfield13}. The other mass measurements are described in  \citet{sharon15} and were derived from Magellan spectroscopic, {\it Chandra} X-ray, SZA  Sunyaev-Zel'dovich, and CFHT wide-field weak lensing observations, as well as richness measurements, where points with dashed error bars indicate extrapolated results.
     The shaded grey region shows the $1\sigma$ range of spherical NFW mass profiles \citet{sharon15} fit
     to the spherical mass estimates indicated with thick circles.
\label{fig:massradius}}}
\end{figure}

\citet{sharon15} presented a first weak lensing analysis of \clusteras\,
based on deep wide-field CFHT/Megacam observations, yielding a mass
constraint
\mbox{$M_\mathrm{200c}=2.0^{+0.9}_{-0.8}\times 10^{15}\mathrm{M}_\odot$}.
\reva{Recomputing our analysis for the cosmology assumed
in their study
  ($\Lambda$CDM with \mbox{$\Omega_\mathrm{m}=0.27$},  \mbox{$\Omega_\Lambda=0.73$}, and \mbox{$h=0.7$}),
our result
\mbox{$M_\mathrm{200c}/(10^{15} \mathrm{M}_\odot) =2.10^{+0.29}_{-0.27}(\mathrm{stat.})\pm 0.12 (\mathrm{sys.})$}}
is fully consistent with this previous measurement, but provides a constraint that is three
times tighter.
The major increase in sensitivity is also visible in Fig.\thinspace\ref{fig:gtgx}, where  the estimated tangential reduced shear profiles of the two studies scaled to the same $\langle\beta\rangle$ are compared. While the CFHT results are noisier, they agree well for scales  \mbox{$1\,\mathrm{Mpc} \lesssim r \lesssim 1.7\,\mathrm{Mpc}$}.
However, at smaller radii the rescaled estimate from \citet{sharon15} is significantly lower than our estimated
reduced shear profile.
This may be a consequence of the colour selection scheme employed in \citet{sharon15}, which yields only a partial removal of cluster galaxies and therefore needs to be complemented with a contamination correction, thereby introducing additional uncertainties especially at smaller radii.
\citet{sharon15} also included measurements at larger radii, which are not probed by our HAWK-I observations.

We can also compare our weak lensing cluster mass \ha{constraints} with mass estimates derived by \citet{sharon15} and \citet{hasselfield13} using other techniques.
In particular, we compare to SZ and dynamical mass estimates, as they probe the cluster mass distribution at similar scales as the weak lensing signal.
The dynamical mass constraints tend to be higher,
for example~\mbox{$M_\mathrm{200c}/(10^{15} \mathrm{M}_\odot) =2.9^{+1.0}_{-0.7}$}
from a caustics analysis, but are still consistent with our measurements.
\ha{There is a noticeable spread in the SZ-derived mass constraints for the cluster.
  \citet{sharon15} estimated a mass \mbox{$M_\mathrm{500c}/(10^{15} \mathrm{M}_\odot) =0.85\pm0.11$}
based on scaling relations from  \citet{andersson11}
or \mbox{$M_\mathrm{500c}/(10^{15} \mathrm{M}_\odot) =0.89\pm0.08$} when employing  the method from \citet{mroczkowski11}.
  \citet{hasselfield13} obtained a similar mass estimate \mbox{$M_\mathrm{500c}/(10^{15} \mathrm{M}_\odot) =0.94\pm0.15$} when assuming universal pressure profiles,  but higher masses when assuming other scaling relations or models,
for example~\mbox{$M_\mathrm{500c}/(10^{15} \mathrm{M}_\odot) =1.49\pm0.30$} based on dynamical masses from \citet{sifon12}.
  Our derived constraint
  \reva{\mbox{$M_\mathrm{500c}/(10^{15} \mathrm{M}_\odot) = 1.52^{+0.19}_{-0.17}
    (\mathrm{stat.})\pm 0.09 (\mathrm{sys.}),$} when assuming the same cosmology as \citealt{sharon15},}
agrees well with the latter SZ results.
We note that our mass constraint assumes a spherical NFW mass model. Cluster
triaxiality and correlated large-scale structure can introduce an additional
\mbox{$\simeq 20\%$} intrinsic scatter in comparison to the 3D halo mass
\citep[compare e.g.][]{becker11}. Likewise, there is intrinsic scatter
between the 3D halo mass and SZ-inferred mass estimates.}
\reva{Fig.\thinspace\ref{fig:massradius} compares our results
  to the mass estimates from \citet{sharon15}
and \citet{hasselfield13}, where we show error bars for our constraints both with and without including intrinsic scatter.}

    Our analysis confirms that \clusteras\, is one of the most massive clusters known in the \mbox{$z\gtrsim 0.7$} Universe.
    Its largest  rival is likely  ACT-CL\,\mbox{$J$0102$-$4915} \citep{menanteau12},
    for which existing weak lensing measurements indicate a possibly higher mass, but
    here the uncertainties are increased because of the complex merger geometry \citep[compare \citetalias{schrabback18};][]{jee14}.
    Comparing our improved mass constraints for \clusteras\, with the analysis from \citet{buddendiek15} we conclude that the existence of   \clusteras\,
    does not pose a significant challenge to standard $\Lambda$CDM predictions.

\section{Weak lensing performance: HAWK-I versus ACS}
\label{se:dis:perf}

\begin{table*}
\caption{Comparison of weak lensing data and performance.
\label{tab:comp}}
\begin{center}
\begin{tabular}{lll}
\hline\hline
            & HAWK-I+LBC analysis    & \citetalias{schrabback18}-like ACS
            analysis (with full-depth colour selection)\\
\hline
Shapes from (total duration) & VLT/HAWK-I $K_\mathrm{s}$
($\simeq 7$\thinspace h) &  HST/ACS $F606W$
$2\times 2$ mosaic (4 orbits $\simeq 6.3$\thinspace h)\\
For colours (total duration) & LBT/LBC  $g$ + $z$ ($\simeq 2$\thinspace h)  &  HST/ACS
$F814W$ mosaic ($\simeq 6.3$\thinspace h) or  8m-class $i$ band ($\simeq 2$\thinspace h)$^1$\\
Useful field of view
&\mbox{$\simeq 7^{\prime}\times 7^{\prime}$}
& \mbox{$\simeq 6\farcm5\times 6\farcm5$} \\
PSF FWHM & $\simeq 0\farcs35 $ & $\simeq 0\farcs1 $ \\
$n_\mathrm{gal}/\mathrm{arcmin^{-2}}$ & 9.8 (for \mbox{$z_\mathrm{l}\le 1.1$}) & 18.1  (for \mbox{$z_\mathrm{l}\le 1.0$})$^2$\\
$\langle\beta\rangle (z_\mathrm{l}=0.7)$ & 0.481&  0.466\\
$\sigma_{\epsilon,\mathrm{eff}}$ &
\hb{0.259}
 &0.322\\
$f/\mathrm{arcmin^{-1}} (z_\mathrm{l}=0.7)$ &
\hb{5.82}
& 6.15 \\
\hline
\end{tabular}
\end{center}
\footnotesize{\flushleft
Notes. ---
$^1$: This corresponds to the F814W/$i$-band imaging that would be needed to apply the colour selection for the full depth of the shape catalogue to reach the source density $n_\mathrm{gal}$. \\
$^2$: \citetalias{schrabback18} reach this average source density for a
colour selection including F814W imaging and clusters at
\mbox{$z_\mathrm{l}\le 1.0$}. At higher cluster redshifts a more stringent colour
selection reduces the source density. \\
}
\end{table*}

\begin{figure*}
\includegraphics[width=\postagesize]{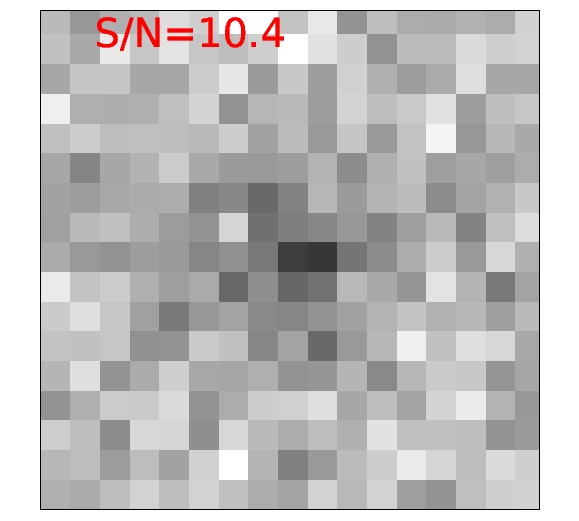}
\includegraphics[width=\postagesize]{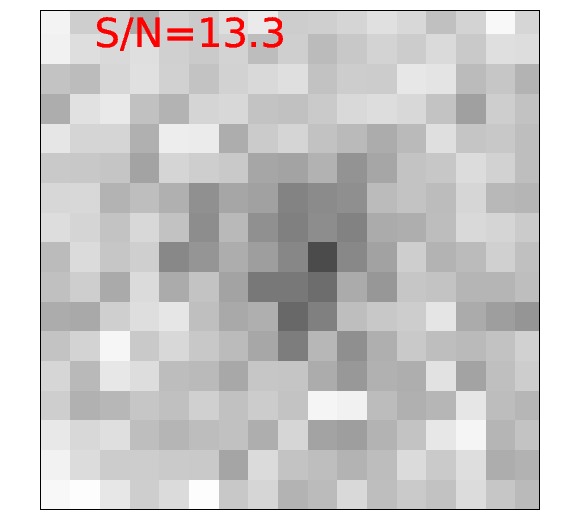}
\includegraphics[width=\postagesize]{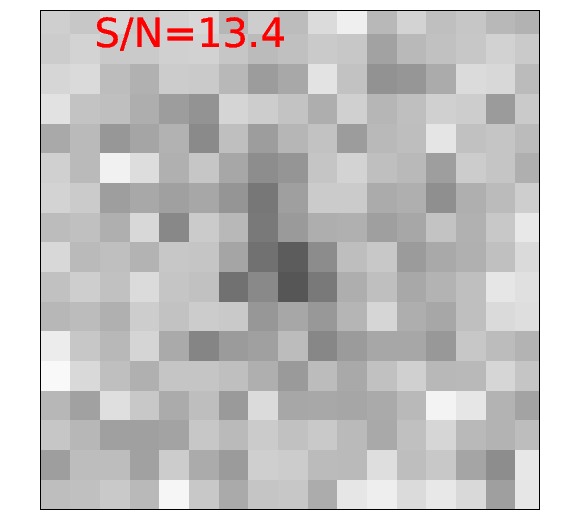}
\includegraphics[width=\postagesize]{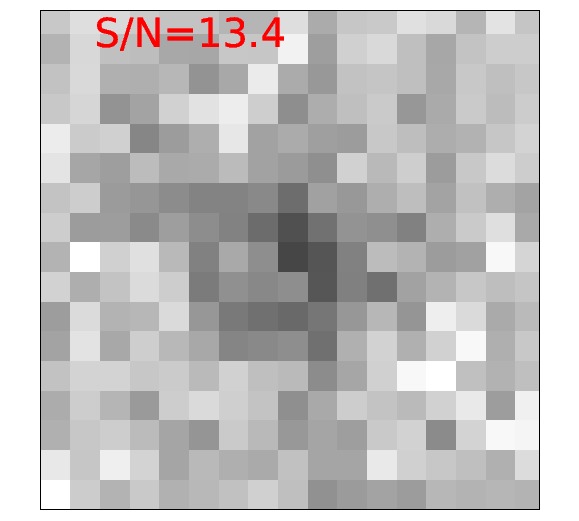}
\includegraphics[width=\postagesize]{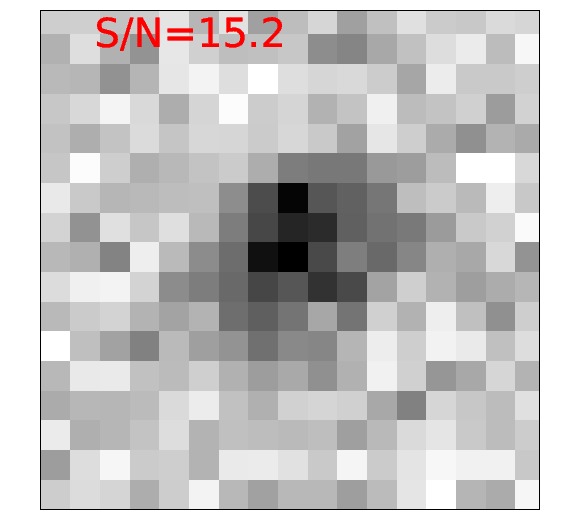}
\includegraphics[width=\postagesize]{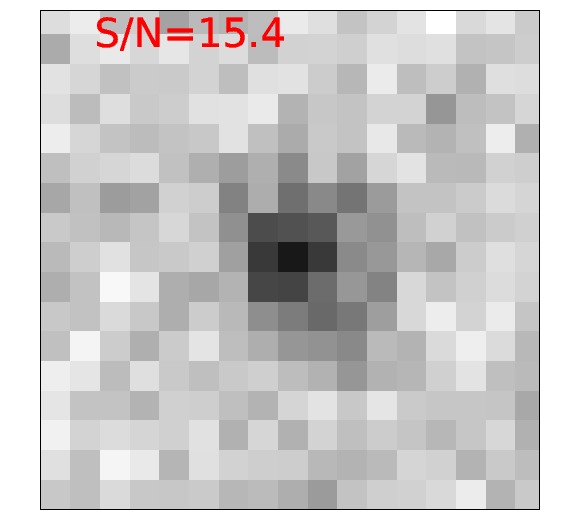}
\includegraphics[width=\postagesize]{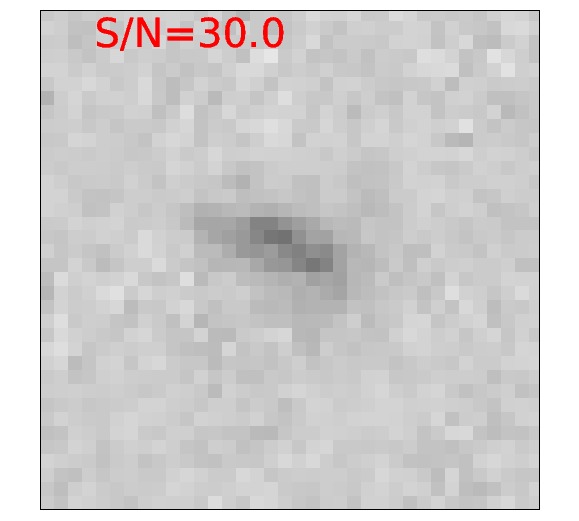}
\includegraphics[width=\postagesize]{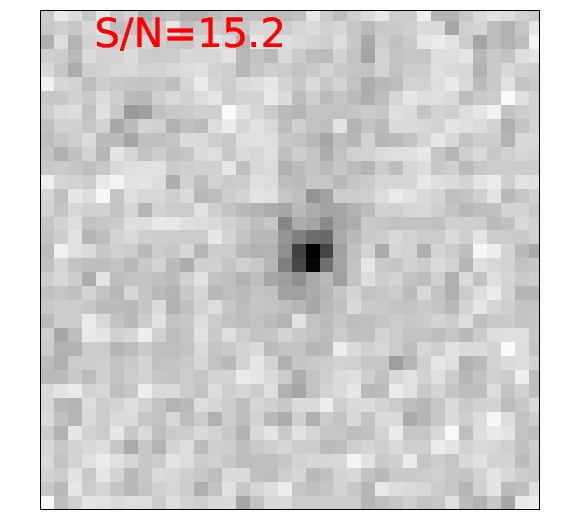}
\includegraphics[width=\postagesize]{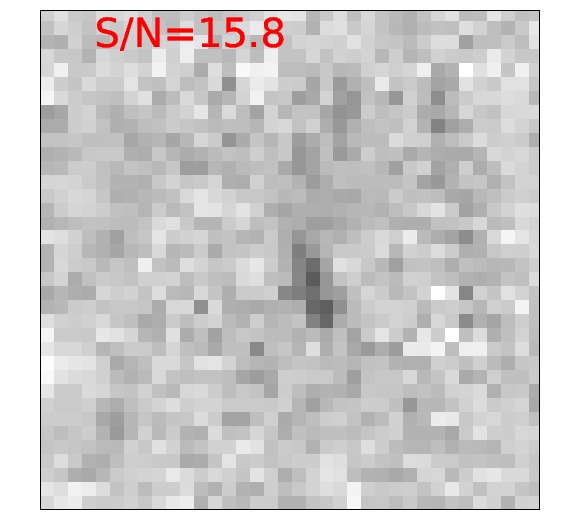}
\includegraphics[width=\postagesize]{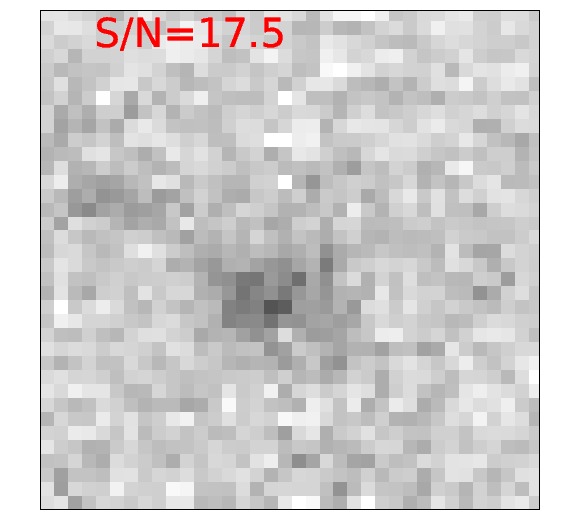}
\includegraphics[width=\postagesize]{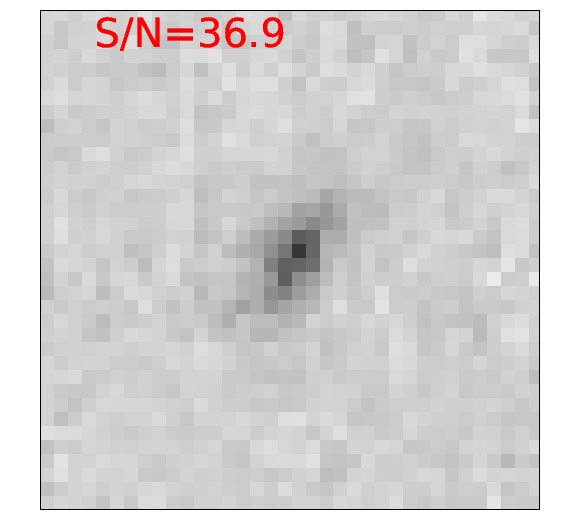}
\includegraphics[width=\postagesize]{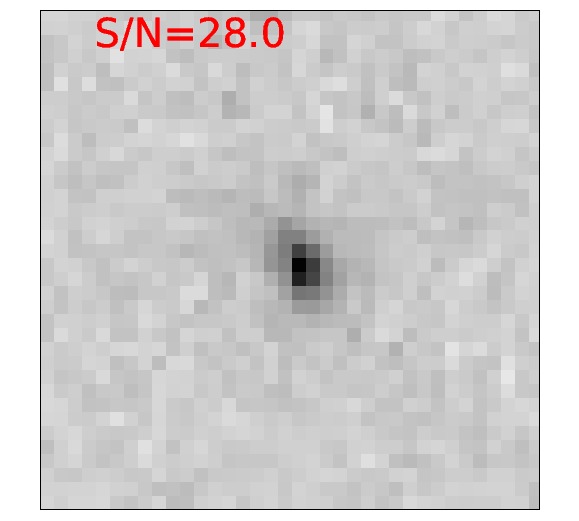}

\vspace{0.2cm}
\includegraphics[width=\postagesize]{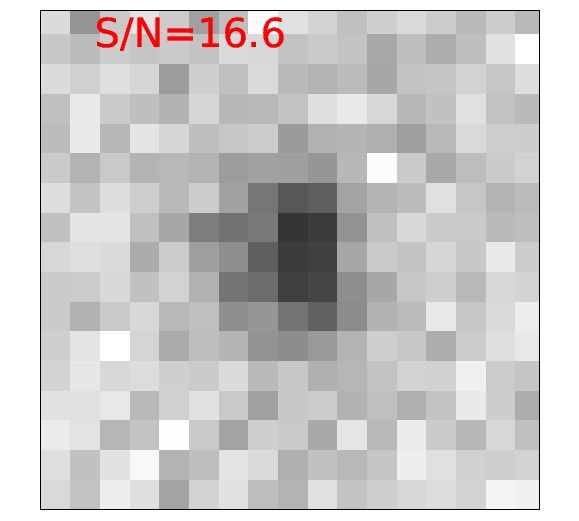}
\includegraphics[width=\postagesize]{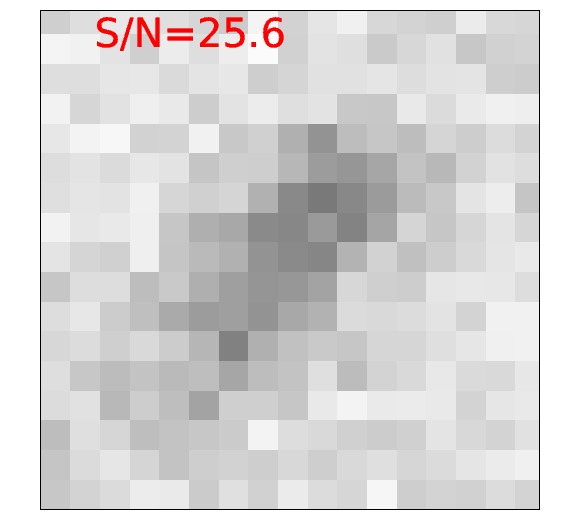}
\includegraphics[width=\postagesize]{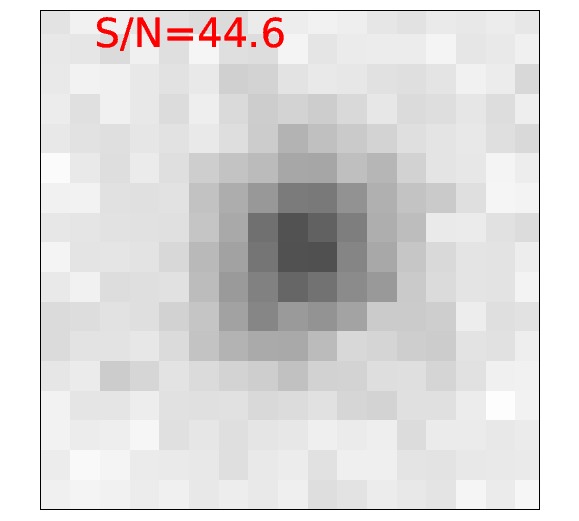}
\includegraphics[width=\postagesize]{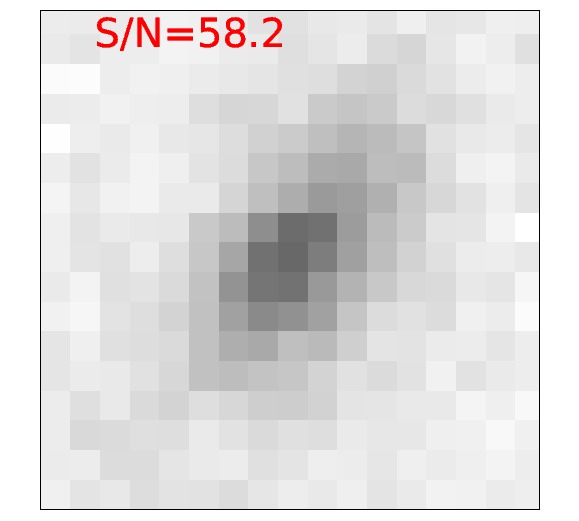}
\includegraphics[width=\postagesize]{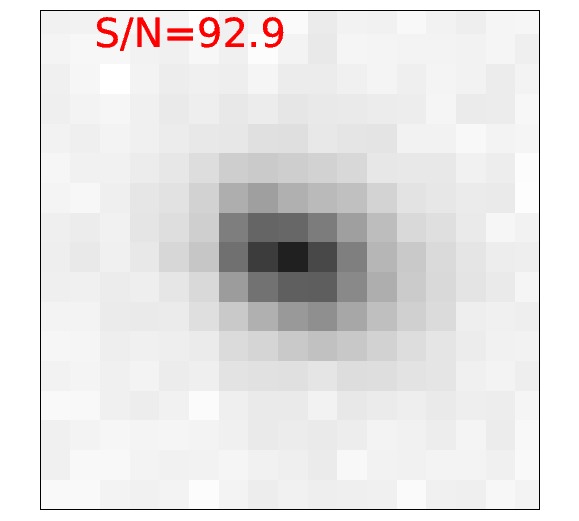}
\includegraphics[width=\postagesize]{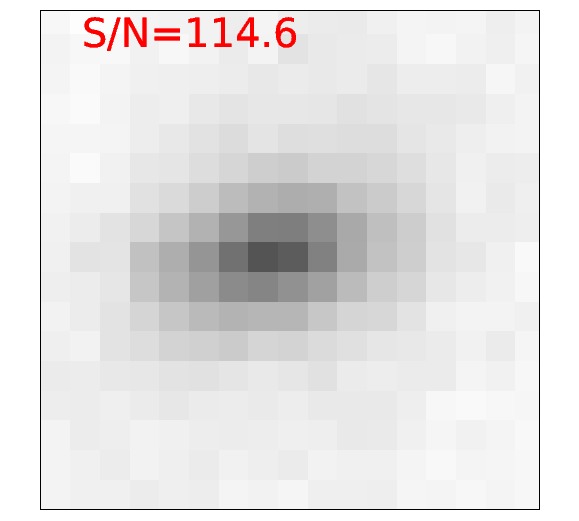}
\includegraphics[width=\postagesize]{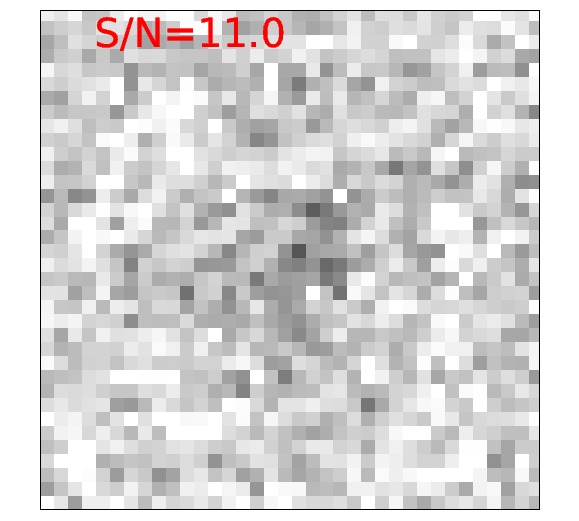}
\includegraphics[width=\postagesize]{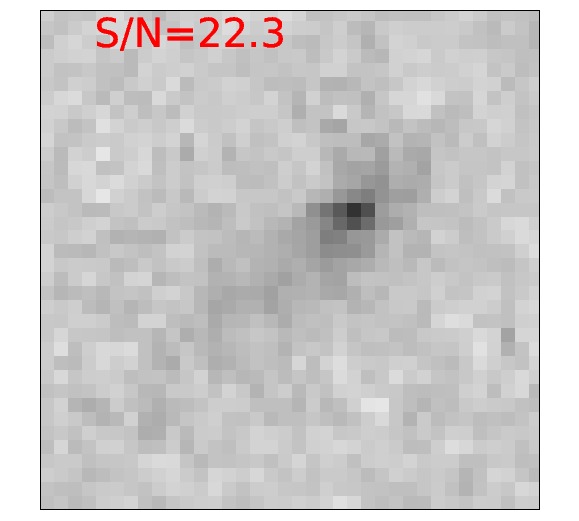}
\includegraphics[width=\postagesize]{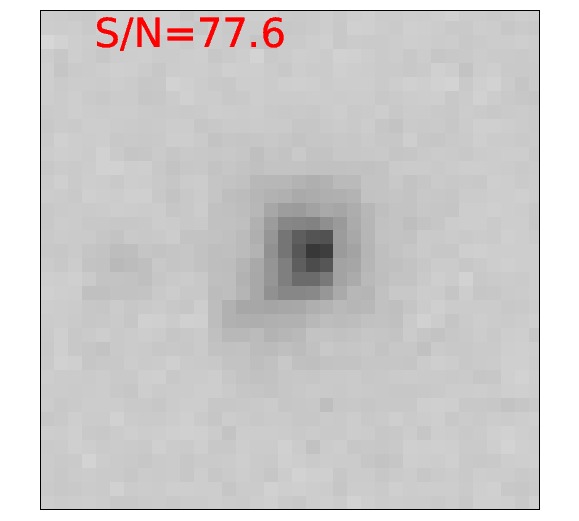}
\includegraphics[width=\postagesize]{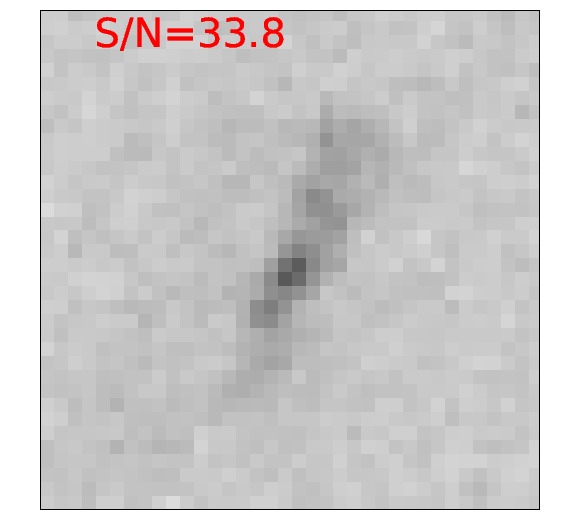}
\includegraphics[width=\postagesize]{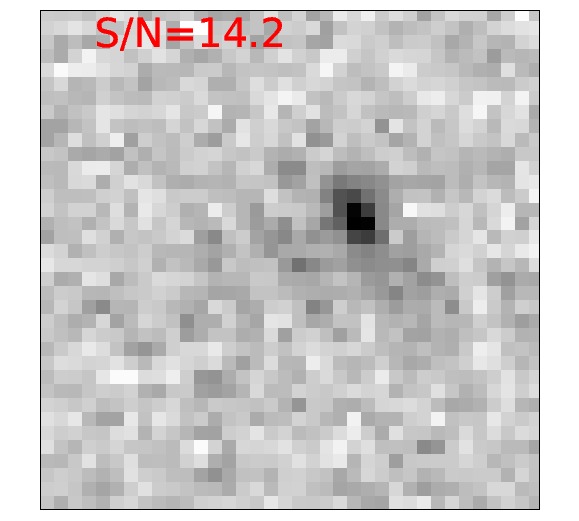}
\includegraphics[width=\postagesize]{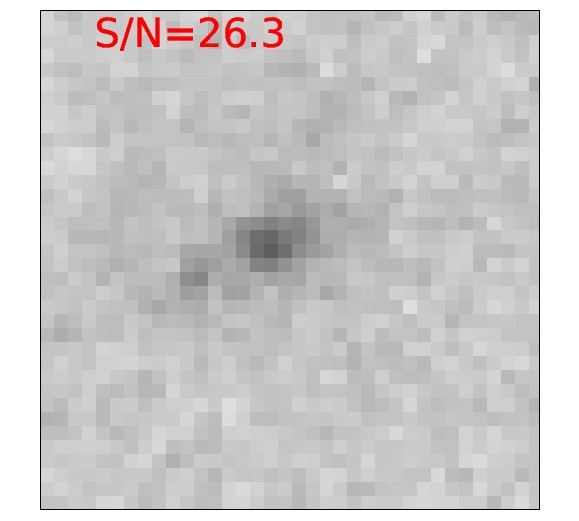}
\caption{\ha{\mbox{$2\farcs0 \times 2\farcs0$} cut-outs
of background-selected \revd{galaxies included} in \emph{both} the weak
lensing  catalogue obtained from the VLT/HAWK-I imaging \emph{and} the weak
lensing catalogue derived from the HST/ACS data.
Rows one and three show the  HAWK-I cut-outs sorted according to the HAWK-I
$(S/N)_\mathrm{flux}$,
while
rows two and four show the corresponding ACS cut-outs of the same galaxies.
All cut-outs are  oriented with north$=$up and east$=$left, and are centred on the HAWK-I galaxy position.
The grey scale is  linear with flux for all cut-outs, but the
range in
flux is adjusted according the individual  $(S/N)_\mathrm{flux}$.}
\label{fig:acsimcom0}}
\end{figure*}

A primary goal of this study is to investigate whether our experimental
set-up, which employs shape measurements in high-resolution ground-based
$K_\mathrm{s}$ images and a $g-z$ versus $z-K_\mathrm{s}$ colour selection,
can
provide a viable alternative to mosaic HST observations for the weak
lensing analysis of massive galaxy clusters at moderately high redshifts.
For this we compare our results to the study from \citetalias{schrabback18},
as summarised in Table \ref{tab:comp}. In their work,
\citetalias{schrabback18}
measured shapes in \mbox{$2\times 2$} ACS F606W mosaics with single-orbit depth per pointing using the same
underlying KSB+ implementation employed here.
These authors applied a
\mbox{$V_{606}-I_{814}<0.3$} colour selection (for clusters at
\mbox{$0.6\lesssim z_\mathrm{l}\lesssim 1.0$}). Here we consider only the case of
adequately deep data for the colour selection as provided e.g.~by the ACS F814W
imaging in
\citetalias{schrabback18}.
While the  ACS \reva{background-selected} source density is higher by a factor 1.85, this advantage is
\hb{almost completely  cancelled  by the larger
  $\sigma_{\epsilon,\mathrm{eff}}$ and slightly lower $\langle\beta\rangle$ for the
ACS  catalogue \hb{(quoted numbers assume a cluster at \mbox{$z_\mathrm{l}=0.7$})},}
yielding
\hb{very similar}
 weak lensing sensitivity \hb{factors} $f$ (see
Eq.\thinspace\ref{eq:snwl})
\hb{with
\mbox{$f_\mathrm{HAWK-I}/f_\mathrm{ACS}=0.95$}}.
\hb{Hence, our HAWK-I+LBC set-up provides a nearly identical weak lensing sensitivity
as the ACS set-up employed by \citetalias{schrabback18}.}

An important reason for the good performance of the HAWK-I+LBC set-up is given by
the
\hb{lower}
effective ellipticity dispersion $\sigma_{\epsilon,\mathrm{eff}}$
found for the colour-selected HAWK-I shear catalogue (see Sect.\thinspace\ref{se:shapenoise}).
\reva{In part this may be due to
differences in the selected galaxy populations.}
But even for galaxies that would be included in both the HAWK-I and the ACS
selection schemes
we expect that the $K_\mathrm{s}$-based shape measurements  yield a lower
intrinsic ellipticity dispersion as they primarily probe the smoother and
typically rounder stellar component.
In contrast, probing  rest-frame UV wavelengths, the optical ACS imaging
primarily
shows clumpy star-forming regions, yielding more irregular shapes with a
larger  ellipticity dispersion.
\revb{For illustration}
 we compare the  HAWK-I  \mbox{$K_\mathrm{s}$} images for some of the galaxies in our
weak lensing catalogue to their counter parts  in ACS F814W images in Fig.\thinspace\ref{fig:acsimcom0}.
For example, the second but last galaxy shown in rows three and four
exhibits a small light-emitting region in the ACS image likely constituting
a compact star-forming region, which is  spatially offset  compared to
the centre of the stellar  light distribution
\hb{visible}
in the \mbox{$K_\mathrm{s}$} image.

In addition to the statistical performance we also have to compare
the systematic uncertainties associated with both approaches, which is
particularly relevant when considering future studies of larger samples.
For this we ignore mass modelling uncertainties, as they
are  essentially identical for both approaches given the similar radial
coverage, and given that they
can be improved via
simulations (e.g.~see the discussion in \citetalias{schrabback18}).
Residual shape measurement biases are in principle expected to be
lower for the
ACS-based analysis given the higher resolution \citep[e.g.][]{massey13}.
\ha{However, we  expect that  shape
measurement biases will not be a limiting systematic for
the analysis of
future large weak lensing follow-up programmes of massive high-$z$ clusters.
Any  such programme that is realistically
conceivable
in the next few years
will have
statistical uncertainties at the several per cent level, which is why systematic
error control at the \mbox{$\sim 1\%$} level  suffices \citep[see
also][]{koehlinger15}.
With advanced shape
measurement techniques, this level of accuracy has already been demonstrated for cosmic shear
measurements \citep[e.g.][]{fenechconti17}, while \citet{bernstein16} even
achieve a further order of magnitude improvement on simplified simulations.
Additionally, \citet{hoekstra15,hoekstra17} demonstrate how image simulations can be
employed to calibrate shape measurement techniques for the impact of
real survey effects for next generation cosmic shear experiments.
What is currently still missing is the calibration of  shape measurement algorithms in the
stronger shear regime of clusters \citep[see e.g.][]{lsst12}, but such
efforts are already well underway (e.g.~Hern\'andez-Mart\'in et al.~in prep.).}

This leaves the final and most relevant source of systematic uncertainty,
which is the calibration of the source redshift distribution and estimation
of $\langle\beta\rangle$.
Combining the various relevant contributors to this uncertainty in
\citetalias{schrabback18},
the current systematic uncertainty on  $\langle\beta\rangle$ amounts to
\mbox{$\sim 2.6\%$} for the ACS-based analysis.
 For comparison, the systematic effects considered in Sect.\thinspace\ref{se:betauncer}
yield a smaller combined  systematic uncertainty  on
$\langle\beta\rangle$ for the HAWK-I-based analysis of \mbox{$\sim 0.7\%$}.
\ha{One of the reasons for this low systematic  uncertainty is
 the availability} of NIR-selected reference samples with deep
high-quality redshift information.
In particular in the
3D-HST reference sample
effectively
\hb{\mbox{$\sim 71\%$}}
of
the colour-selected galaxies at the relevant depth have a spectroscopic or HST/WFC3 grism redshift when
taking our source magnitude distribution and weights into account
(see Sect.\thinspace\ref{se:3dhst}).
\ha{Comparably deep and complete spectroscopic reference samples
do not yet exist for the deep optically selected ACS weak lensing data
sets \citep[but samples are increasing, see e.g.][]{lefevre15}.
In \citetalias{schrabback18} a significant contribution to the systematic uncertainty
related to the $\langle\beta\rangle$ estimate
comes from the correction for catastrophic redshift outliers.
These incorrectly scatter from the high-$z$ source population into a
low-$z$ contamination sample, which cannot be removed with the colour
selection scheme from \citetalias{schrabback18}.
The  $gzK_\mathrm{s}$ selection applied in our current study does not suffer from
such a low-$z$ contamination, and is therefore affected  less by catastrophic redshift outliers.
}

\revb{There are further advantages of the  HAWK-I+LBC-based analysis.
The chosen default colour selection scheme
can be applied
out to a higher maximum cluster redshift \mbox{$z_\mathrm{l,max}=1.1$} (instead of
\mbox{$z_\mathrm{l}=1.0$} for the  \mbox{$V_{606}-I_{814}<0.3$} ACS
colour selection scheme), which can possibly be extended
to
 \mbox{$z_\mathrm{l,max}\simeq 1.2$--$1.3$} (instead of
\mbox{$z_\mathrm{l}=1.15$} for the ACS-based analysis) with more stringent
colour selection criteria (compare
Figures\thinspace\ref{fig:colordist_ultravista} and \ref{fig:zdist}).
The HAWK-I+LBC-based colour selection also yields
 a}
better suppression fraction of galaxies at relevant cluster redshifts
(98.9\%
versus 98.1\%).

Taking all this together we conclude that
the
chosen set-up of the HAWK-I+LBC data yields a weak lensing
performance that is
\hb{similarly}
powerful as the considered ACS-based
analysis scheme.
While the required integration time is significant for the $K_\mathrm{s}$
imaging, this is compensated by the ability to cover a larger field of view
with \revb{imagers such} as HAWK-I.
The $K_\mathrm{s}$-based approach is therefore particularly efficient for
the analysis of high-mass \reva{(\mbox{$M_\mathrm{200c}> 5 \times 10^{14} \mathrm{M}_\odot$}) clusters at redshifts \mbox{$0.7\lesssim
  z_\mathrm{l}\lesssim 1.1$}, for which
mosaics} would be
needed with HST/ACS to probe the weak lensing signal out to approximately
the virial radius\footnote{\reva{The achievable signal-to-noise ratio of the mass constraints naturally increases with cluster mass and decreases with cluster redshift. For example, for an individual \mbox{$M_\mathrm{200c}\simeq 6 \times 10^{14} \mathrm{M}_\odot$} cluster at \mbox{$z\simeq 1.0$} and a set-up similar to our analysis we expect a  \mbox{$\sim 50\%$} statistical mass uncertainty.
}} (see Table \ref{tab:comp} for the approximate total
observing times).
\reva{For \revd{less} massive clusters
and clusters at even higher redshifts
deeper observations} are
needed, while a wide angular coverage  is less important
\citep[e.g.][]{jee11}. In this regime deeper single pointing HST
observations likely provide a more adequate observing strategy, as required $K_\mathrm{s}$
integration times would become prohibitively long, \ha{and the virial radius fits within the ACS field of view}.

\section{Summary and conclusions}
\label{se:conclusions}

We have presented the first weak gravitational lensing analysis
that exploits the superb image resolution
(\mbox{$\mathrm{FWHM}^*=0\farcs35$}) that can
 be achieved in the $K_\mathrm{s}$ band
under good seeing conditions with optimised imagers such as the
employed VLT/HAWK-I
 to measure weak lensing galaxy shapes.
Here we summarise our main conclusions:
\begin{itemize}
\item At the resolution of the  $K_\mathrm{s}$ imaging, nearly all relevant
  background galaxies are sufficiently resolved for weak lensing
  measurements.
\item The employed photometric selection  in $g-z$ versus $z-K_\mathrm{s}$
  colour space is highly effective for the selection of most of  the lensed
  background galaxies and the removal of
diluting foreground and cluster galaxies.
\item \ha{Our analysis indicates that the intrinsic ellipticity dispersion
    is
\hb{noticeably}
lower
for high-$z$ galaxies in
    $K_\mathrm{s}$ weak lensing data compared to high-$z$
    sources studied in the optical, boosting the weak lensing
    sensitivity.}
\item \ha{Despite a lower source density the} analysed data
\hb{therefore yield almost the same weak lensing sensitivity as}
the analysis of mosaic HST/ACS data with single-orbit
  depth per pointing from \citetalias{schrabback18}.
\item The systematic uncertainty regarding the
  calibration of the source redshift distribution is lower for the
  HAWK-I analysis compared to the \citetalias{schrabback18} ACS analysis.
This is   thanks to the use of NIR-selected redshift reference samples from 3D-HST
  and UltraVISTA and the improved removal of contaminating low-$z$ galaxies from the
  source sample, reducing the sensitivity to catastrophic redshift errors.
\item Comparing to  HST/ACS data that overlap with parts of our
  HAWK-I observations  of \clusteras, we find fully consistent estimates
of the  tangential reduced shear profile between the two data sets in a
matched catalogue, providing an important confirmation for the
$K_\mathrm{s}$-based  analysis.
\item Given the larger field of view, good-seeing VLT/HAWK-I $K_\mathrm{s}$
  observations, complemented with $g$ and $z$ (or $B$ and $z$) photometry,
  provide an efficient alternative to mosaic HST/ACS observations for the
  weak lensing analysis of massive galaxy clusters at  redshifts \mbox{$0.7\lesssim
  z_\mathrm{l}\lesssim 1.1$}.
\item \reva{Especially for clusters at higher redshifts}
    significantly deeper
  observations with higher resolution
are required, while a smaller  field of view is \reva{typically} sufficient.
In this regime  deep HST observations with a smaller
  angular coverage provide the most
  effective and efficient observing strategy.
\item We stress that calibrations of the source redshift distribution for
  weak lensing studies have to carefully account for catastrophic redshift
  outliers, which appear to be present even when NIR imaging is available
  (see Sect.\thinspace\ref{se:zcomp}).
\item \ha{While our observations confirm that \clusteras\, is one of the most massive
  galaxy clusters known in the \mbox{$z\gtrsim 0.7$} Universe, its existence
 is not in tension with standard $\Lambda$CDM expectations according to our mass constraints.}
\item \ha{The extreme mass of \clusteras\, leads to the significant weak lensing
    signal we detect, but we stress that our conclusions regarding the
    sensitivity of the HAWK-I weak lensing measurements (hence, the noise
    level) do not depend on its extreme mass. The approach is also directly
    applicable to massive, but less extreme clusters at redshifts \mbox{$0.7\lesssim
  z_\mathrm{l}\lesssim 1.1$} (e.g.~from the \citealt{bleem15} sample).}
\end{itemize}

\begin{acknowledgements}
This work is directly based on observations collected at the European Organisation
for Astronomical Research in the Southern Hemisphere under ESO programme(s)
087.A-0933,
at the Large Binocular Telescope (LBT),
and with the
NASA/ESA Hubble Space Telescope under GO programmes 13177 and 10846.
This work also makes use of catalogues created by the 3D-HST Treasury Program (GO
12177 and 12328) and catalogues derived from the ESO UltraVISTA Programme 179.A-2005.

We thank ESO staff for obtaining the excellent VLT/HAWK-I images and
Paul Martini, David Atlee, Erica Hesselbach, Jeff Blackburne, and Matthias
Dietrich for conducting the LBT/LBC observations.
\ha{We thank Patrick Simon for providing the codes employed in this work to
  reconstruct the cluster mass distribution and to generate Gaussian shear
  field realisations for the estimation of the impact of large-scale structure projections.}
\hb{We thank Peter Schneider for useful discussions and for providing comments on this manuscript.}

TS, DA, BH, and DK acknowledge
support from the German Federal Ministry of Economics and Technology
(BMWi) provided through DLR under projects 50 OR 1210, 50 OR 1308, 50
OR 1407, and 50 OR 1610.
RFJvdB acknowledges support from the European Research Council under FP7 grant number 340519.
 TE is supported by the Deutsche
      Forschungsgemeinschaft in the framework of the TR33 'The Dark
      Universe'.
HHi is supported by an Emmy Noether grant (No. Hi 1495/2-1) of the Deutsche
Forschungsgemeinschaft.
This work was supported in part by the Kavli Institute for Cosmological
Physics at the University of Chicago through grant NSF PHY-1125897 and an
endowment from the Kavli Foundation and its founder Fred Kavli.
\revd{Part of the research was carried out at the Jet Propulsion Laboratory, California Institute of Technology, under a contract with the National Aeronautics and Space Administration.}

HST is operated by the Association of
Universities for Research in Astronomy, Incorporated, under NASA contract
NAS5-26555.
The LBT is an international collaboration among institutions in the United States, Italy and Germany. LBT Corporation partners are The University of Arizona on behalf of the Arizona university system; Istituto Nazionale di Astrofisica, Italy; LBT Beteiligungsgesellschaft, Germany, representing the Max-Planck Society, the Astrophysical Institute Potsdam, and Heidelberg University; The Ohio State University, and The Research Corporation, on behalf of The University of Notre Dame, University of Minnesota and University of Virginia.

This research made use of APLpy, an open-source plotting package for Python \citep{robitaille12}.

\end{acknowledgements}

 \bibliographystyle{aa}
\bibliography{oir}

\begin{thebibliography}{91}
\expandafter\ifx\csname natexlab\endcsname\relax\def\natexlab#1{#1}\fi

\bibitem[{{Andersson} {et~al.}(2011){Andersson}, {Benson}, {Ade}, {Aird},
  {Armstrong}, {Bautz}, {Bleem}, {Brodwin}, {Carlstrom}, {Chang}, {Crawford},
  {Crites}, {de Haan}, {Desai}, {Dobbs}, {Dudley}, {Foley}, {Forman},
  {Garmire}, {George}, {Gladders}, {Halverson}, {High}, {Holder}, {Holzapfel},
  {Hrubes}, {Jones}, {Joy}, {Keisler}, {Knox}, {Lee}, {Leitch}, {Lueker},
  {Marrone}, {McMahon}, {Mehl}, {Meyer}, {Mohr}, {Montroy}, {Murray}, {Padin},
  {Plagge}, {Pryke}, {Reichardt}, {Rest}, {Ruel}, {Ruhl}, {Schaffer}, {Shaw},
  {Shirokoff}, {Song}, {Spieler}, {Stalder}, {Staniszewski}, {Stark}, {Stubbs},
  {Vanderlinde}, {Vieira}, {Vikhlinin}, {Williamson}, {Yang}, {Zahn}, \&
  {Zenteno}}]{andersson11}
{Andersson}, K., {Benson}, B.~A., {Ade}, P.~A.~R., {et~al.} 2011, \apj, 738, 48

\bibitem[{{Bartelmann} \& {Schneider}(2001)}]{bartelmann01}
{Bartelmann}, M. \& {Schneider}, P. 2001, \physrep, 340, 291

\bibitem[{{Becker} \& {Kravtsov}(2011)}]{becker11}
{Becker}, M.~R. \& {Kravtsov}, A.~V. 2011, \apj, 740, 25

\bibitem[{{Bernstein} {et~al.}(2016){Bernstein}, {Armstrong}, {Krawiec}, \&
  {March}}]{bernstein16}
{Bernstein}, G.~M., {Armstrong}, R., {Krawiec}, C., \& {March}, M.~C. 2016,
  \mnras, 459, 4467

\bibitem[{{Bertin}(2006)}]{bertin06}
{Bertin}, E. 2006, in Astronomical Society of the Pacific Conference Series,
  Vol. 351, Astronomical Data Analysis Software and Systems XV, ed.
  C.~{Gabriel}, C.~{Arviset}, D.~{Ponz}, \& S.~{Enrique}, 112

\bibitem[{{Bertin}(2011)}]{bertin11}
{Bertin}, E. 2011, in Astronomical Society of the Pacific Conference Series,
  Vol. 442, Astronomical Data Analysis Software and Systems XX, ed. I.~N.
  {Evans}, A.~{Accomazzi}, D.~J. {Mink}, \& A.~H. {Rots}, 435

\bibitem[{{Bertin} \& {Arnouts}(1996)}]{bertin1996}
{Bertin}, E. \& {Arnouts}, S. 1996, \aaps, 117, 393

\bibitem[{{Bertin} {et~al.}(2002){Bertin}, {Mellier}, {Radovich}, {Missonnier},
  {Didelon}, \& {Morin}}]{bertin02}
{Bertin}, E., {Mellier}, Y., {Radovich}, M., {et~al.} 2002, in Astronomical
  Society of the Pacific Conference Series, Vol. 281, Astronomical Data
  Analysis Software and Systems XI, ed. D.~A. {Bohlender}, D.~{Durand}, \&
  T.~H. {Handley}, 228

\bibitem[{{Bleem} {et~al.}(2015){Bleem}, {Stalder}, {de Haan}, {Aird}, {Allen},
  {Applegate}, {Ashby}, {Bautz}, {Bayliss}, {Benson}, {Bocquet}, {Brodwin},
  {Carlstrom}, {Chang}, {Chiu}, {Cho}, {Clocchiatti}, {Crawford}, {Crites},
  {Desai}, {Dietrich}, {Dobbs}, {Foley}, {Forman}, {George}, {Gladders},
  {Gonzalez}, {Halverson}, {Hennig}, {Hoekstra}, {Holder}, {Holzapfel},
  {Hrubes}, {Jones}, {Keisler}, {Knox}, {Lee}, {Leitch}, {Liu}, {Lueker},
  {Luong-Van}, {Mantz}, {Marrone}, {McDonald}, {McMahon}, {Meyer}, {Mocanu},
  {Mohr}, {Murray}, {Padin}, {Pryke}, {Reichardt}, {Rest}, {Ruel}, {Ruhl},
  {Saliwanchik}, {Saro}, {Sayre}, {Schaffer}, {Schrabback}, {Shirokoff},
  {Song}, {Spieler}, {Stanford}, {Staniszewski}, {Stark}, {Story}, {Stubbs},
  {Vanderlinde}, {Vieira}, {Vikhlinin}, {Williamson}, {Zahn}, \&
  {Zenteno}}]{bleem15}
{Bleem}, L.~E., {Stalder}, B., {de Haan}, T., {et~al.} 2015, \apjs, 216, 27

\bibitem[{{Brada{\v c}} {et~al.}(2014){Brada{\v c}}, {Ryan}, {Casertano},
  {Huang}, {Lemaux}, {Schrabback}, {Gonzalez}, {Allen}, {Cain}, {Gladders},
  {Hall}, {Hildebrandt}, {Hinz}, {von der Linden}, {Lubin}, {Treu}, \&
  {Zaritsky}}]{bradac14}
{Brada{\v c}}, M., {Ryan}, R., {Casertano}, S., {et~al.} 2014, \apj, 785, 108

\bibitem[{{Broadhurst} {et~al.}(1995){Broadhurst}, {Taylor}, \&
  {Peacock}}]{broadhurst95}
{Broadhurst}, T.~J., {Taylor}, A.~N., \& {Peacock}, J.~A. 1995, \apj, 438, 49

\bibitem[{{Buddendiek} {et~al.}(2015){Buddendiek}, {Schrabback}, {Greer},
  {Hoekstra}, {Sommer}, {Eifler}, {Erben}, {Erler}, {Hicks}, {High},
  {Hildebrandt}, {Marrone}, {Morris}, {Muzzin}, {Reiprich}, {Schirmer},
  {Schneider}, \& {von der Linden}}]{buddendiek15}
{Buddendiek}, A., {Schrabback}, T., {Greer}, C.~H., {et~al.} 2015, \mnras, 450,
  4248

\bibitem[{{Daddi} {et~al.}(2004){Daddi}, {Cimatti}, {Renzini}, {Fontana},
  {Mignoli}, {Pozzetti}, {Tozzi}, \& {Zamorani}}]{daddi04}
{Daddi}, E., {Cimatti}, A., {Renzini}, A., {et~al.} 2004, \apj, 617, 746

\bibitem[{{Diemer} \& {Kravtsov}(2015)}]{diemer15}
{Diemer}, B. \& {Kravtsov}, A.~V. 2015, \apj, 799, 108

\bibitem[{{Dietrich} {et~al.}(2012){Dietrich}, {B{\"o}hnert}, {Lombardi},
  {Hilbert}, \& {Hartlap}}]{dietrich12}
{Dietrich}, J.~P., {B{\"o}hnert}, A., {Lombardi}, M., {Hilbert}, S., \&
  {Hartlap}, J. 2012, \mnras, 419, 3547

\bibitem[{{Erben} {et~al.}(2005){Erben}, {Schirmer}, {Dietrich}, {Cordes},
  {Haberzettl}, {Hetterscheidt}, {Hildebrandt}, {Schmithuesen}, {Schneider},
  {Simon}, {Deul}, {Hook}, {Kaiser}, {Radovich}, {Benoist}, {Nonino}, {Olsen},
  {Prandoni}, {Wichmann}, {Zaggia}, {Bomans}, {Dettmar}, \&
  {Miralles}}]{erben05}
{Erben}, T., {Schirmer}, M., {Dietrich}, J.~P., {et~al.} 2005, Astronomische
  Nachrichten, 326, 432

\bibitem[{{Erben} {et~al.}(2001){Erben}, {Van Waerbeke}, {Bertin}, {Mellier},
  \& {Schneider}}]{erben2001}
{Erben}, T., {Van Waerbeke}, L., {Bertin}, E., {Mellier}, Y., \& {Schneider},
  P. 2001, \aap, 366, 717

\bibitem[{{Fenech Conti} {et~al.}(2017){Fenech Conti}, {Herbonnet}, {Hoekstra},
  {Merten}, {Miller}, \& {Viola}}]{fenechconti17}
{Fenech Conti}, I., {Herbonnet}, R., {Hoekstra}, H., {et~al.} 2017, \mnras,
  467, 1627

\bibitem[{{Fort} {et~al.}(1997){Fort}, {Mellier}, \& {Dantel-Fort}}]{fort97}
{Fort}, B., {Mellier}, Y., \& {Dantel-Fort}, M. 1997, \aap, 321, 353

\bibitem[{{Giallongo} {et~al.}(2008){Giallongo}, {Ragazzoni}, {Grazian},
  {Baruffolo}, {Beccari}, {de Santis}, {Diolaiti}, {di Paola}, {Farinato},
  {Fontana}, {Gallozzi}, {Gasparo}, {Gentile}, {Green}, {Hill}, {Kuhn},
  {Pasian}, {Pedichini}, {Radovich}, {Salinari}, {Smareglia}, {Speziali},
  {Testa}, {Thompson}, {Vernet}, \& {Wagner}}]{giallongo08}
{Giallongo}, E., {Ragazzoni}, R., {Grazian}, A., {et~al.} 2008, \aap, 482, 349

\bibitem[{{Gilbank} {et~al.}(2011){Gilbank}, {Gladders}, {Yee}, \&
  {Hsieh}}]{gilbank11}
{Gilbank}, D.~G., {Gladders}, M.~D., {Yee}, H.~K.~C., \& {Hsieh}, B.~C. 2011,
  \aj, 141, 94

\bibitem[{{Grogin} {et~al.}(2011){Grogin}, {Kocevski}, {Faber}, {Ferguson},
  {Koekemoer}, {Riess}, {Acquaviva}, {Alexander}, {Almaini}, {Ashby}, {Barden},
  {Bell}, {Bournaud}, {Brown}, {Caputi}, {Casertano}, {Cassata}, {Castellano},
  {Challis}, {Chary}, {Cheung}, {Cirasuolo}, {Conselice}, {Roshan Cooray},
  {Croton}, {Daddi}, {Dahlen}, {Dav{\'e}}, {de Mello}, {Dekel}, {Dickinson},
  {Dolch}, {Donley}, {Dunlop}, {Dutton}, {Elbaz}, {Fazio}, {Filippenko},
  {Finkelstein}, {Fontana}, {Gardner}, {Garnavich}, {Gawiser}, {Giavalisco},
  {Grazian}, {Guo}, {Hathi}, {H{\"a}ussler}, {Hopkins}, {Huang}, {Huang},
  {Jha}, {Kartaltepe}, {Kirshner}, {Koo}, {Lai}, {Lee}, {Li}, {Lotz}, {Lucas},
  {Madau}, {McCarthy}, {McGrath}, {McIntosh}, {McLure}, {Mobasher},
  {Moustakas}, {Mozena}, {Nandra}, {Newman}, {Niemi}, {Noeske}, {Papovich},
  {Pentericci}, {Pope}, {Primack}, {Rajan}, {Ravindranath}, {Reddy}, {Renzini},
  {Rix}, {Robaina}, {Rodney}, {Rosario}, {Rosati}, {Salimbeni}, {Scarlata},
  {Siana}, {Simard}, {Smidt}, {Somerville}, {Spinrad}, {Straughn}, {Strolger},
  {Telford}, {Teplitz}, {Trump}, {van der Wel}, {Villforth}, {Wechsler},
  {Weiner}, {Wiklind}, {Wild}, {Wilson}, {Wuyts}, {Yan}, \& {Yun}}]{grogin2011}
{Grogin}, N.~A., {Kocevski}, D.~D., {Faber}, S.~M., {et~al.} 2011, \apjs, 197,
  35

\bibitem[{{Hasselfield} {et~al.}(2013){Hasselfield}, {Hilton}, {Marriage},
  {Addison}, {Barrientos}, {Battaglia}, {Battistelli}, {Bond}, {Crichton},
  {Das}, {Devlin}, {Dicker}, {Dunkley}, {D{\"u}nner}, {Fowler}, {Gralla},
  {Hajian}, {Halpern}, {Hincks}, {Hlozek}, {Hughes}, {Infante}, {Irwin},
  {Kosowsky}, {Marsden}, {Menanteau}, {Moodley}, {Niemack}, {Nolta}, {Page},
  {Partridge}, {Reese}, {Schmitt}, {Sehgal}, {Sherwin}, {Sievers}, {Sif{\'o}n},
  {Spergel}, {Staggs}, {Swetz}, {Switzer}, {Thornton}, {Trac}, \&
  {Wollack}}]{hasselfield13}
{Hasselfield}, M., {Hilton}, M., {Marriage}, T.~A., {et~al.} 2013, \jcap, 7, 8

\bibitem[{{Heymans} {et~al.}(2006){Heymans}, {Van Waerbeke}, {Bacon}, {Berge},
  {Bernstein}, {Bertin}, {Bridle}, {Brown}, {Clowe}, {Dahle}, {Erben}, {Gray},
  {Hetterscheidt}, {Hoekstra}, {Hudelot}, {Jarvis}, {Kuijken}, {Margoniner},
  {Massey}, {Mellier}, {Nakajima}, {Refregier}, {Rhodes}, {Schrabback}, \&
  {Wittman}}]{heymans06}
{Heymans}, C., {Van Waerbeke}, L., {Bacon}, D., {et~al.} 2006, \mnras, 368,
  1323

\bibitem[{{Hinshaw} {et~al.}(2013){Hinshaw}, {Larson}, {Komatsu}, {Spergel},
  {Bennett}, {Dunkley}, {Nolta}, {Halpern}, {Hill}, {Odegard}, {Page}, {Smith},
  {Weiland}, {Gold}, {Jarosik}, {Kogut}, {Limon}, {Meyer}, {Tucker}, {Wollack},
  \& {Wright}}]{hinshaw13}
{Hinshaw}, G., {Larson}, D., {Komatsu}, E., {et~al.} 2013, \apjs, 208, 19

\bibitem[{{Hoag} {et~al.}(2015){Hoag}, {Brada{\v c}}, {Huang}, {Ryan},
  {Sharon}, {Schrabback}, {Schmidt}, {Cain}, {Gonzalez}, {Hildebrandt}, {Hinz},
  {Lemaux}, {von der Linden}, {Lubin}, {Treu}, \& {Zaritsky}}]{hoag15}
{Hoag}, A., {Brada{\v c}}, M., {Huang}, K.~H., {et~al.} 2015, \apj, 813, 37

\bibitem[{{Hoekstra} {et~al.}(2013){Hoekstra}, {Bartelmann}, {Dahle}, {Israel},
  {Limousin}, \& {Meneghetti}}]{hoekstra13}
{Hoekstra}, H., {Bartelmann}, M., {Dahle}, H., {et~al.} 2013, \ssr, 177, 75

\bibitem[{{Hoekstra} {et~al.}(2000){Hoekstra}, {Franx}, \&
  {Kuijken}}]{hoekstra00}
{Hoekstra}, H., {Franx}, M., \& {Kuijken}, K. 2000, \apj, 532, 88

\bibitem[{{Hoekstra} {et~al.}(1998){Hoekstra}, {Franx}, {Kuijken}, \&
  {Squires}}]{hoekstra1998}
{Hoekstra}, H., {Franx}, M., {Kuijken}, K., \& {Squires}, G. 1998, \apj, 504,
  636

\bibitem[{{Hoekstra} {et~al.}(2015){Hoekstra}, {Herbonnet}, {Muzzin}, {Babul},
  {Mahdavi}, {Viola}, \& {Cacciato}}]{hoekstra15}
{Hoekstra}, H., {Herbonnet}, R., {Muzzin}, A., {et~al.} 2015, \mnras, 449, 685

\bibitem[{{Hoekstra} {et~al.}(2017){Hoekstra}, {Viola}, \&
  {Herbonnet}}]{hoekstra17}
{Hoekstra}, H., {Viola}, M., \& {Herbonnet}, R. 2017, \mnras, 468, 3295

\bibitem[{{Jarvis} \& {Jain}(2004)}]{jarvis04}
{Jarvis}, M. \& {Jain}, B. 2004, astro-ph/0412234 [\eprint{astro-ph/0412234}]

\bibitem[{{Jee} {et~al.}(2007){Jee}, {Blakeslee}, {Sirianni}, {Martel},
  {White}, \& {Ford}}]{jee2007}
{Jee}, M.~J., {Blakeslee}, J.~P., {Sirianni}, M., {et~al.} 2007, \pasp, 119,
  1403

\bibitem[{{Jee} {et~al.}(2011){Jee}, {Dawson}, {Hoekstra}, {Perlmutter},
  {Rosati}, {Brodwin}, {Suzuki}, {Koester}, {Postman}, {Lubin}, {Meyers},
  {Stanford}, {Barbary}, {Barrientos}, {Eisenhardt}, {Ford}, {Gilbank},
  {Gladders}, {Gonzalez}, {Harris}, {Huang}, {Lidman}, {Rykoff}, {Rubin}, \&
  {Spadafora}}]{jee11}
{Jee}, M.~J., {Dawson}, K.~S., {Hoekstra}, H., {et~al.} 2011, \apj, 737, 59

\bibitem[{{Jee} {et~al.}(2014){Jee}, {Hughes}, {Menanteau}, {Sif{\'o}n},
  {Mandelbaum}, {Barrientos}, {Infante}, \& {Ng}}]{jee14}
{Jee}, M.~J., {Hughes}, J.~P., {Menanteau}, F., {et~al.} 2014, \apj, 785, 20

\bibitem[{{Jee} {et~al.}(2009){Jee}, {Rosati}, {Ford}, {Dawson}, {Lidman},
  {Perlmutter}, {Demarco}, {Strazzullo}, {Mullis}, {B{\"o}hringer}, \&
  {Fassbender}}]{jee09}
{Jee}, M.~J., {Rosati}, P., {Ford}, H.~C., {et~al.} 2009, \apj, 704, 672

\bibitem[{{Kaiser}(2000)}]{kaiser00}
{Kaiser}, N. 2000, \apj, 537, 555

\bibitem[{{Kaiser} \& {Squires}(1993)}]{kaiser93}
{Kaiser}, N. \& {Squires}, G. 1993, \apj, 404, 441

\bibitem[{Kaiser {et~al.}(1995)Kaiser, Squires, \& Broadhurst}]{kaiser1995}
Kaiser, N., Squires, G., \& Broadhurst, T. 1995, ApJ, 449, 460

\bibitem[{{King} {et~al.}(2002){King}, {Clowe}, {Lidman}, {Schneider}, {Erben},
  {Kneib}, \& {Meylan}}]{king02}
{King}, L.~J., {Clowe}, D.~I., {Lidman}, C., {et~al.} 2002, \aap, 385, L5

\bibitem[{{Kissler-Patig} {et~al.}(2008){Kissler-Patig}, {Pirard}, {Casali},
  {Moorwood}, {Ageorges}, {Alves de Oliveira}, {Baksai}, {Bedin}, {Bendek},
  {Biereichel}, {Delabre}, {Dorn}, {Esteves}, {Finger}, {Gojak}, {Huster},
  {Jung}, {Kiekebush}, {Klein}, {Koch}, {Lizon}, {Mehrgan}, {Petr-Gotzens},
  {Pritchard}, {Selman}, \& {Stegmeier}}]{kissler-patig08}
{Kissler-Patig}, M., {Pirard}, J.-F., {Casali}, M., {et~al.} 2008, \aap, 491,
  941

\bibitem[{{Koekemoer} {et~al.}(2011){Koekemoer}, {Faber}, {Ferguson}, {Grogin},
  {Kocevski}, {Koo}, {Lai}, {Lotz}, {Lucas}, {McGrath}, {Ogaz}, {Rajan},
  {Riess}, {Rodney}, {Strolger}, {Casertano}, {Castellano}, {Dahlen},
  {Dickinson}, {Dolch}, {Fontana}, {Giavalisco}, {Grazian}, {Guo}, {Hathi},
  {Huang}, {van der Wel}, {Yan}, {Acquaviva}, {Alexander}, {Almaini}, {Ashby},
  {Barden}, {Bell}, {Bournaud}, {Brown}, {Caputi}, {Cassata}, {Challis},
  {Chary}, {Cheung}, {Cirasuolo}, {Conselice}, {Roshan Cooray}, {Croton},
  {Daddi}, {Dav{\'e}}, {de Mello}, {de Ravel}, {Dekel}, {Donley}, {Dunlop},
  {Dutton}, {Elbaz}, {Fazio}, {Filippenko}, {Finkelstein}, {Frazer}, {Gardner},
  {Garnavich}, {Gawiser}, {Gruetzbauch}, {Hartley}, {H{\"a}ussler},
  {Herrington}, {Hopkins}, {Huang}, {Jha}, {Johnson}, {Kartaltepe},
  {Khostovan}, {Kirshner}, {Lani}, {Lee}, {Li}, {Madau}, {McCarthy},
  {McIntosh}, {McLure}, {McPartland}, {Mobasher}, {Moreira}, {Mortlock},
  {Moustakas}, {Mozena}, {Nandra}, {Newman}, {Nielsen}, {Niemi}, {Noeske},
  {Papovich}, {Pentericci}, {Pope}, {Primack}, {Ravindranath}, {Reddy},
  {Renzini}, {Rix}, {Robaina}, {Rosario}, {Rosati}, {Salimbeni}, {Scarlata},
  {Siana}, {Simard}, {Smidt}, {Snyder}, {Somerville}, {Spinrad}, {Straughn},
  {Telford}, {Teplitz}, {Trump}, {Vargas}, {Villforth}, {Wagner}, {Wandro},
  {Wechsler}, {Weiner}, {Wiklind}, {Wild}, {Wilson}, {Wuyts}, \&
  {Yun}}]{koekemoer11}
{Koekemoer}, A.~M., {Faber}, S.~M., {Ferguson}, H.~C., {et~al.} 2011, \apjs,
  197, 36

\bibitem[{{Koekemoer} {et~al.}(2003){Koekemoer}, {Fruchter}, {Hook}, \&
  {Hack}}]{koekemoer2003}
{Koekemoer}, A.~M., {Fruchter}, A.~S., {Hook}, R.~N., \& {Hack}, W. 2003, in
  HST Calibration Workshop : Hubble after the Installation of the ACS and the
  NICMOS Cooling System, ed. S.~{Arribas}, A.~{Koekemoer}, \& B.~{Whitmore},
  337

\bibitem[{{K{\"o}hlinger} {et~al.}(2015){K{\"o}hlinger}, {Hoekstra}, \&
  {Eriksen}}]{koehlinger15}
{K{\"o}hlinger}, F., {Hoekstra}, H., \& {Eriksen}, M. 2015, \mnras, 453, 3107

\bibitem[{{Kuijken} {et~al.}(2015){Kuijken}, {Heymans}, {Hildebrandt},
  {Nakajima}, {Erben}, {de Jong}, {Viola}, {Choi}, {Hoekstra}, {Miller}, {van
  Uitert}, {Amon}, {Blake}, {Brouwer}, {Buddendiek}, {Conti}, {Eriksen},
  {Grado}, {Harnois-D{\'e}raps}, {Helmich}, {Herbonnet}, {Irisarri},
  {Kitching}, {Klaes}, {La Barbera}, {Napolitano}, {Radovich}, {Schneider},
  {Sif{\'o}n}, {Sikkema}, {Simon}, {Tudorica}, {Valentijn}, {Verdoes Kleijn},
  \& {van Waerbeke}}]{kuijken15}
{Kuijken}, K., {Heymans}, C., {Hildebrandt}, H., {et~al.} 2015, \mnras, 454,
  3500

\bibitem[{{Le F{\`e}vre} {et~al.}(2015){Le F{\`e}vre}, {Tasca}, {Cassata},
  {Garilli}, {Le Brun}, {Maccagni}, {Pentericci}, {Thomas}, {Vanzella},
  {Zamorani}, {Zucca}, {Amorin}, {Bardelli}, {Capak}, {Cassar{\`a}},
  {Castellano}, {Cimatti}, {Cuby}, {Cucciati}, {de la Torre}, {Durkalec},
  {Fontana}, {Giavalisco}, {Grazian}, {Hathi}, {Ilbert}, {Lemaux}, {Moreau},
  {Paltani}, {Ribeiro}, {Salvato}, {Schaerer}, {Scodeggio}, {Sommariva},
  {Talia}, {Taniguchi}, {Tresse}, {Vergani}, {Wang}, {Charlot}, {Contini},
  {Fotopoulou}, {L{\'o}pez-Sanjuan}, {Mellier}, \& {Scoville}}]{lefevre15}
{Le F{\`e}vre}, O., {Tasca}, L.~A.~M., {Cassata}, P., {et~al.} 2015, \aap, 576,
  A79

\bibitem[{{Leauthaud} {et~al.}(2007){Leauthaud}, {Massey}, {Kneib}, {Rhodes},
  {Johnston}, {Capak}, {Heymans}, {Ellis}, {Koekemoer}, {Le F{\`e}vre},
  {Mellier}, {R{\'e}fr{\'e}gier}, {Robin}, {Scoville}, {Tasca}, {Taylor}, \&
  {Van Waerbeke}}]{leauthaud07}
{Leauthaud}, A., {Massey}, R., {Kneib}, J.-P., {et~al.} 2007, \apjs, 172, 219

\bibitem[{{Leauthaud} {et~al.}(2012){Leauthaud}, {Tinker}, {Bundy}, {Behroozi},
  {Massey}, {Rhodes}, {George}, {Kneib}, {Benson}, {Wechsler}, {Busha},
  {Capak}, {Cort{\^e}s}, {Ilbert}, {Koekemoer}, {Le F{\`e}vre}, {Lilly},
  {McCracken}, {Salvato}, {Schrabback}, {Scoville}, {Smith}, \&
  {Taylor}}]{leauthaud12}
{Leauthaud}, A., {Tinker}, J., {Bundy}, K., {et~al.} 2012, \apj, 744, 159

\bibitem[{{LSST Dark Energy Science Collaboration}(2012)}]{lsst12}
{LSST Dark Energy Science Collaboration}. 2012, arXiv:1211.0310
  [\eprint[arXiv]{1211.0310}]

\bibitem[{{Luppino} \& {Kaiser}(1997)}]{luppino1997}
{Luppino}, G.~A. \& {Kaiser}, N. 1997, \apj, 475, 20

\bibitem[{{Mandelbaum} {et~al.}(2018){Mandelbaum}, {Miyatake}, {Hamana},
  {Oguri}, {Simet}, {Armstrong}, {Bosch}, {Murata}, {Lanusse}, {Leauthaud},
  {Coupon}, {More}, {Takada}, {Miyazaki}, {Speagle}, {Shirasaki}, {Sif{\'o}n},
  {Huang}, {Nishizawa}, {Medezinski}, {Okura}, {Okabe}, {Czakon}, {Takahashi},
  {Coulton}, {Hikage}, {Komiyama}, {Lupton}, {Strauss}, {Tanaka}, \&
  {Utsumi}}]{mandelbaum18}
{Mandelbaum}, R., {Miyatake}, H., {Hamana}, T., {et~al.} 2018, \pasj, 70, S25

\bibitem[{{Martinez} {et~al.}(2010){Martinez}, {Kolb}, {Tokovinin}, \&
  {Sarazin}}]{martinez10}
{Martinez}, P., {Kolb}, J., {Tokovinin}, A., \& {Sarazin}, M. 2010, \aap, 516,
  A90

\bibitem[{{Massey} {et~al.}(2013){Massey}, {Hoekstra}, {Kitching}, {Rhodes},
  {Cropper}, {Amiaux}, {Harvey}, {Mellier}, {Meneghetti}, {Miller},
  {Paulin-Henriksson}, {Pires}, {Scaramella}, \& {Schrabback}}]{massey13}
{Massey}, R., {Hoekstra}, H., {Kitching}, T., {et~al.} 2013, \mnras, 429, 661

\bibitem[{{Massey} {et~al.}(2007){Massey}, {Rhodes}, {Leauthaud}, {Capak},
  {Ellis}, {Koekemoer}, {R{\'e}fr{\'e}gier}, {Scoville}, {Taylor}, {Albert},
  {Berg{\'e}}, {Heymans}, {Johnston}, {Kneib}, {Mellier}, {Mobasher},
  {Semboloni}, {Shopbell}, {Tasca}, \& {Van Waerbeke}}]{massey07b}
{Massey}, R., {Rhodes}, J., {Leauthaud}, A., {et~al.} 2007, \apjs, 172, 239

\bibitem[{{Massey} {et~al.}(2014){Massey}, {Schrabback}, {Cordes}, {Marggraf},
  {Israel}, {Miller}, {Hall}, {Cropper}, {Prod'homme}, \& {Matias
  Niemi}}]{massey2014}
{Massey}, R., {Schrabback}, T., {Cordes}, O., {et~al.} 2014, \mnras, 439, 887

\bibitem[{{Masters} {et~al.}(2017){Masters}, {Stern}, {Cohen}, {Capak},
  {Rhodes}, {Castander}, \& {Paltani}}]{masters17}
{Masters}, D.~C., {Stern}, D.~K., {Cohen}, J.~G., {et~al.} 2017, \apj, 841, 111

\bibitem[{{Mayen} \& {Soucail}(2000)}]{mayen00}
{Mayen}, C. \& {Soucail}, G. 2000, \aap, 361, 415

\bibitem[{{McCracken} {et~al.}(2012){McCracken}, {Milvang-Jensen}, {Dunlop},
  {Franx}, {Fynbo}, {Le F{\`e}vre}, {Holt}, {Caputi}, {Goranova}, {Buitrago},
  {Emerson}, {Freudling}, {Hudelot}, {L{\'o}pez-Sanjuan}, {Magnard}, {Mellier},
  {M{\o}ller}, {Nilsson}, {Sutherland}, {Tasca}, \& {Zabl}}]{mccracken12}
{McCracken}, H.~J., {Milvang-Jensen}, B., {Dunlop}, J., {et~al.} 2012, \aap,
  544, A156

\bibitem[{{McInnes} {et~al.}(2009){McInnes}, {Menanteau}, {Heavens}, {Hughes},
  {Jimenez}, {Massey}, {Simon}, \& {Taylor}}]{mcinnes09}
{McInnes}, R.~N., {Menanteau}, F., {Heavens}, A.~F., {et~al.} 2009, \mnras,
  399, L84

\bibitem[{{Melchior} {et~al.}(2011){Melchior}, {Viola}, {Sch{\"a}fer}, \&
  {Bartelmann}}]{melchior11}
{Melchior}, P., {Viola}, M., {Sch{\"a}fer}, B.~M., \& {Bartelmann}, M. 2011,
  \mnras, 412, 1552

\bibitem[{{Menanteau} {et~al.}(2012){Menanteau}, {Hughes}, {Sif{\'o}n},
  {Hilton}, {Gonz{\'a}lez}, {Infante}, {Barrientos}, {Baker}, {Bond}, {Das},
  {Devlin}, {Dunkley}, {Hajian}, {Hincks}, {Kosowsky}, {Marsden}, {Marriage},
  {Moodley}, {Niemack}, {Nolta}, {Page}, {Reese}, {Sehgal}, {Sievers},
  {Spergel}, {Staggs}, \& {Wollack}}]{menanteau12}
{Menanteau}, F., {Hughes}, J.~P., {Sif{\'o}n}, C., {et~al.} 2012, \apj, 748, 7

\bibitem[{{Menanteau} {et~al.}(2013){Menanteau}, {Sif{\'o}n}, {Barrientos},
  {Battaglia}, {Bond}, {Crichton}, {Das}, {Devlin}, {Dicker}, {D{\"u}nner},
  {Gralla}, {Hajian}, {Hasselfield}, {Hilton}, {Hincks}, {Hughes}, {Infante},
  {Kosowsky}, {Marriage}, {Marsden}, {Moodley}, {Niemack}, {Nolta}, {Page},
  {Partridge}, {Reese}, {Schmitt}, {Sievers}, {Spergel}, {Staggs}, {Switzer},
  \& {Wollack}}]{menanteau13}
{Menanteau}, F., {Sif{\'o}n}, C., {Barrientos}, L.~F., {et~al.} 2013, \apj,
  765, 67

\bibitem[{{Meneghetti} {et~al.}(2014){Meneghetti}, {Rasia}, {Vega}, {Merten},
  {Postman}, {Yepes}, {Sembolini}, {Donahue}, {Ettori}, {Umetsu}, {Balestra},
  {Bartelmann}, {Ben{\'{\i}}tez}, {Biviano}, {Bouwens}, {Bradley},
  {Broadhurst}, {Coe}, {Czakon}, {De Petris}, {Ford}, {Giocoli},
  {Gottl{\"o}ber}, {Grillo}, {Infante}, {Jouvel}, {Kelson}, {Koekemoer},
  {Lahav}, {Lemze}, {Medezinski}, {Melchior}, {Mercurio}, {Molino},
  {Moscardini}, {Monna}, {Moustakas}, {Moustakas}, {Nonino}, {Rhodes},
  {Rosati}, {Sayers}, {Seitz}, {Zheng}, \& {Zitrin}}]{meneghetti14}
{Meneghetti}, M., {Rasia}, E., {Vega}, J., {et~al.} 2014, \apj, 797, 34

\bibitem[{{Merten} {et~al.}(2015){Merten}, {Meneghetti}, {Postman}, {Umetsu},
  {Zitrin}, {Medezinski}, {Nonino}, {Koekemoer}, {Melchior}, {Gruen},
  {Moustakas}, {Bartelmann}, {Host}, {Donahue}, {Coe}, {Molino}, {Jouvel},
  {Monna}, {Seitz}, {Czakon}, {Lemze}, {Sayers}, {Balestra}, {Rosati},
  {Ben{\'{\i}}tez}, {Biviano}, {Bouwens}, {Bradley}, {Broadhurst}, {Carrasco},
  {Ford}, {Grillo}, {Infante}, {Kelson}, {Lahav}, {Massey}, {Moustakas},
  {Rasia}, {Rhodes}, {Vega}, \& {Zheng}}]{merten15}
{Merten}, J., {Meneghetti}, M., {Postman}, M., {et~al.} 2015, \apj, 806, 4

\bibitem[{{Momcheva} {et~al.}(2016){Momcheva}, {Brammer}, {van Dokkum},
  {Skelton}, {Whitaker}, {Nelson}, {Fumagalli}, {Maseda}, {Leja}, {Franx},
  {Rix}, {Bezanson}, {Da Cunha}, {Dickey}, {F{\"o}rster Schreiber},
  {Illingworth}, {Kriek}, {Labb{\'e}}, {Ulf Lange}, {Lundgren}, {Magee},
  {Marchesini}, {Oesch}, {Pacifici}, {Patel}, {Price}, {Tal}, {Wake}, {van der
  Wel}, \& {Wuyts}}]{momcheva16}
{Momcheva}, I.~G., {Brammer}, G.~B., {van Dokkum}, P.~G., {et~al.} 2016, \apjs,
  225, 27

\bibitem[{{Mroczkowski}(2011)}]{mroczkowski11}
{Mroczkowski}, T. 2011, \apjl, 728, L35

\bibitem[{{Muzzin} {et~al.}(2013){Muzzin}, {Marchesini}, {Stefanon}, {Franx},
  {Milvang-Jensen}, {Dunlop}, {Fynbo}, {Brammer}, {Labb{\'e}}, \& {van
  Dokkum}}]{muzzin13}
{Muzzin}, A., {Marchesini}, D., {Stefanon}, M., {et~al.} 2013, \apjs, 206, 8

\bibitem[{{Navarro} {et~al.}(1997){Navarro}, {Frenk}, \& {White}}]{navarro97}
{Navarro}, J.~F., {Frenk}, C.~S., \& {White}, S.~D.~M. 1997, \apj, 490, 493

\bibitem[{{Neto} {et~al.}(2007){Neto}, {Gao}, {Bett}, {Cole}, {Navarro},
  {Frenk}, {White}, {Springel}, \& {Jenkins}}]{neto07}
{Neto}, A.~F., {Gao}, L., {Bett}, P., {et~al.} 2007, \mnras, 381, 1450

\bibitem[{{Planck Collaboration} {et~al.}(2016){Planck Collaboration}, {Ade},
  {Aghanim}, {Arnaud}, {Ashdown}, {Aumont}, {Baccigalupi}, {Banday},
  {Barreiro}, {Bartlett}, \& et~al.}]{planck15cosmo}
{Planck Collaboration}, {Ade}, P.~A.~R., {Aghanim}, N., {et~al.} 2016, \aap,
  594, A13

\bibitem[{{Postman} {et~al.}(2012){Postman}, {Coe}, {Ben{\'{\i}}tez},
  {Bradley}, {Broadhurst}, {Donahue}, {Ford}, {Graur}, {Graves}, {Jouvel},
  {Koekemoer}, {Lemze}, {Medezinski}, {Molino}, {Moustakas}, {Ogaz}, {Riess},
  {Rodney}, {Rosati}, {Umetsu}, {Zheng}, {Zitrin}, {Bartelmann}, {Bouwens},
  {Czakon}, {Golwala}, {Host}, {Infante}, {Jha}, {Jimenez-Teja}, {Kelson},
  {Lahav}, {Lazkoz}, {Maoz}, {McCully}, {Melchior}, {Meneghetti}, {Merten},
  {Moustakas}, {Nonino}, {Patel}, {Reg{\"o}s}, {Sayers}, {Seitz}, \& {Van der
  Wel}}]{postman12}
{Postman}, M., {Coe}, D., {Ben{\'{\i}}tez}, N., {et~al.} 2012, \apjs, 199, 25

\bibitem[{{Robitaille} \& {Bressert}(2012)}]{robitaille12}
{Robitaille}, T. \& {Bressert}, E. 2012, {APLpy: Astronomical Plotting Library
  in Python}, Astrophysics Source Code Library

\bibitem[{{Rowe} {et~al.}(2015){Rowe}, {Jarvis}, {Mandelbaum}, {Bernstein},
  {Bosch}, {Simet}, {Meyers}, {Kacprzak}, {Nakajima}, {Zuntz}, {Miyatake},
  {Dietrich}, {Armstrong}, {Melchior}, \& {Gill}}]{rowe15}
{Rowe}, B.~T.~P., {Jarvis}, M., {Mandelbaum}, R., {et~al.} 2015, Astronomy and
  Computing, 10, 121

\bibitem[{{Schirmer}(2013)}]{schirmer13}
{Schirmer}, M. 2013, \apjs, 209, 21

\bibitem[{{Schneider} \& {Seitz}(1995)}]{schneider95}
{Schneider}, P. \& {Seitz}, C. 1995, \aap, 294, 411

\bibitem[{{Schrabback} {et~al.}(2018){Schrabback}, {Applegate}, {Dietrich},
  {Hoekstra}, {Bocquet}, {Gonzalez}, {von der Linden}, {McDonald}, {Morrison},
  {Raihan}, {Allen}, {Bayliss}, {Benson}, {Bleem}, {Chiu}, {Desai}, {Foley},
  {de Haan}, {High}, {Hilbert}, {Mantz}, {Massey}, {Mohr}, {Reichardt}, {Saro},
  {Simon}, {Stern}, {Stubbs}, \& {Zenteno}}]{schrabback18}
{Schrabback}, T., {Applegate}, D., {Dietrich}, J.~P., {et~al.} 2018, \mnras,
  474, 2635

\bibitem[{{Schrabback} {et~al.}(2007){Schrabback}, {Erben}, {Simon},
  {Miralles}, {Schneider}, {Heymans}, {Eifler}, {Fosbury}, {Freudling},
  {Hetterscheidt}, {Hildebrandt}, \& {Pirzkal}}]{schrabback07}
{Schrabback}, T., {Erben}, T., {Simon}, P., {et~al.} 2007, \aap, 468, 823

\bibitem[{{Schrabback} {et~al.}(2010){Schrabback}, {Hartlap}, {Joachimi},
  {Kilbinger}, {Simon}, {Benabed}, {Brada{\v c}}, {Eifler}, {Erben},
  {Fassnacht}, {High}, {Hilbert}, {Hildebrandt}, {Hoekstra}, {Kuijken},
  {Marshall}, {Mellier}, {Morganson}, {Schneider}, {Semboloni}, {van Waerbeke},
  \& {Velander}}]{schrabback10}
{Schrabback}, T., {Hartlap}, J., {Joachimi}, B., {et~al.} 2010, \aap, 516, A63

\bibitem[{{Scoville} {et~al.}(2007){Scoville}, {Aussel}, {Brusa}, {Capak},
  {Carollo}, {Elvis}, {Giavalisco}, {Guzzo}, {Hasinger}, {Impey}, {Kneib},
  {LeFevre}, {Lilly}, {Mobasher}, {Renzini}, {Rich}, {Sanders}, {Schinnerer},
  {Schminovich}, {Shopbell}, {Taniguchi}, \& {Tyson}}]{scoville07}
{Scoville}, N., {Aussel}, H., {Brusa}, M., {et~al.} 2007, \apjs, 172, 1

\bibitem[{{Seitz} \& {Schneider}(1997)}]{seitz97}
{Seitz}, C. \& {Schneider}, P. 1997, \aap, 318, 687

\bibitem[{{Seitz} \& {Schneider}(1996)}]{seitz96}
{Seitz}, S. \& {Schneider}, P. 1996, \aap, 305, 383

\bibitem[{{Sharon} {et~al.}(2015){Sharon}, {Gladders}, {Marrone}, {Hoekstra},
  {Rasia}, {Bourdin}, {Gifford}, {Hicks}, {Greer}, {Mroczkowski}, {Barrientos},
  {Bayliss}, {Carlstrom}, {Gilbank}, {Gralla}, {Hlavacek-Larrondo}, {Leitch},
  {Mazzotta}, {Miller}, {Muchovej}, {Schrabback}, {Yee}, \&
  {RCS-Team}}]{sharon15}
{Sharon}, K., {Gladders}, M.~D., {Marrone}, D.~P., {et~al.} 2015, \apj, 814, 21

\bibitem[{{Sif{\'o}n} {et~al.}(2013){Sif{\'o}n}, {Menanteau}, {Hasselfield},
  {Marriage}, {Hughes}, {Barrientos}, {Gonz{\'a}lez}, {Infante}, {Addison},
  {Baker}, {Battaglia}, {Bond}, {Crichton}, {Das}, {Devlin}, {Dunkley},
  {D{\"u}nner}, {Gralla}, {Hajian}, {Hilton}, {Hincks}, {Kosowsky}, {Marsden},
  {Moodley}, {Niemack}, {Nolta}, {Page}, {Partridge}, {Reese}, {Sehgal},
  {Sievers}, {Spergel}, {Staggs}, {Thornton}, {Trac}, \& {Wollack}}]{sifon12}
{Sif{\'o}n}, C., {Menanteau}, F., {Hasselfield}, M., {et~al.} 2013, \apj, 772,
  25

\bibitem[{{Simon} {et~al.}(2009){Simon}, {Taylor}, \& {Hartlap}}]{simon09}
{Simon}, P., {Taylor}, A.~N., \& {Hartlap}, J. 2009, \mnras, 399, 48

\bibitem[{{Skelton} {et~al.}(2014){Skelton}, {Whitaker}, {Momcheva}, {Brammer},
  {van Dokkum}, {Labb{\'e}}, {Franx}, {van der Wel}, {Bezanson}, {Da Cunha},
  {Fumagalli}, {F{\"o}rster Schreiber}, {Kriek}, {Leja}, {Lundgren}, {Magee},
  {Marchesini}, {Maseda}, {Nelson}, {Oesch}, {Pacifici}, {Patel}, {Price},
  {Rix}, {Tal}, {Wake}, \& {Wuyts}}]{skelton14}
{Skelton}, R.~E., {Whitaker}, K.~E., {Momcheva}, I.~G., {et~al.} 2014, \apjs,
  214, 24

\bibitem[{{Skrutskie} {et~al.}(2006){Skrutskie}, {Cutri}, {Stiening},
  {Weinberg}, {Schneider}, {Carpenter}, {Beichman}, {Capps}, {Chester},
  {Elias}, {Huchra}, {Liebert}, {Lonsdale}, {Monet}, {Price}, {Seitzer},
  {Jarrett}, {Kirkpatrick}, {Gizis}, {Howard}, {Evans}, {Fowler}, {Fullmer},
  {Hurt}, {Light}, {Kopan}, {Marsh}, {McCallon}, {Tam}, {Van Dyk}, \&
  {Wheelock}}]{skrutskie06}
{Skrutskie}, M.~F., {Cutri}, R.~M., {Stiening}, R., {et~al.} 2006, \aj, 131,
  1163

\bibitem[{{Th{\"o}lken} {et~al.}(2018){Th{\"o}lken}, {Schrabback}, {Reiprich},
  {Lovisari}, {Allen}, {Hoekstra}, {Applegate}, {Buddendiek}, \&
  {Hicks}}]{thoelken18}
{Th{\"o}lken}, S., {Schrabback}, T., {Reiprich}, T.~H., {et~al.} 2018, \aap,
  610, A71

\bibitem[{{Umetsu} {et~al.}(2016){Umetsu}, {Zitrin}, {Gruen}, {Merten},
  {Donahue}, \& {Postman}}]{umetsu16}
{Umetsu}, K., {Zitrin}, A., {Gruen}, D., {et~al.} 2016, \apj, 821, 116

\bibitem[{{Viola} {et~al.}(2014){Viola}, {Kitching}, \& {Joachimi}}]{viola14}
{Viola}, M., {Kitching}, T.~D., \& {Joachimi}, B. 2014, \mnras, 439, 1909

\bibitem[{{von der Linden} {et~al.}(2014){von der Linden}, {Allen},
  {Applegate}, {Kelly}, {Allen}, {Ebeling}, {Burchat}, {Burke}, {Donovan},
  {Morris}, {Blandford}, {Erben}, \& {Mantz}}]{vonderlinden14}
{von der Linden}, A., {Allen}, M.~T., {Applegate}, D.~E., {et~al.} 2014,
  \mnras, 439, 2

\bibitem[{{Wright} \& {Brainerd}(2000)}]{wright00}
{Wright}, C.~O. \& {Brainerd}, T.~G. 2000, \apj, 534, 34

\end{thebibliography}

\end{document}